\documentclass[12pt]{iopart}
\usepackage{iopams}
\expandafter\let\csname equation*\endcsname\relax
\expandafter\let\csname endequation*\endcsname\relax
\usepackage{amsmath}
\usepackage{amsfonts,amssymb}
\usepackage{amscd}
\usepackage{ulem}
\usepackage{graphicx,epsf}
\usepackage{caption,subcaption}
\usepackage{color}         
\usepackage{graphicx}
\usepackage{epsfig}
\usepackage{soul}
\usepackage{hyperref}

\begin{document}
\newcommand{\gae}{\lower 2pt \hbox{$\, \buildrel {\scriptstyle >}\over {\scriptstyle
\sim}\,$}}
\newcommand{\lae}{\lower 2pt \hbox{$\, \buildrel {\scriptstyle <}\over {\scriptstyle
\sim}\,$}}
\newcommand{\op}[1]{\ensuremath{\hat{\mathnormal{#1}}}}
\title{Stochastic resetting in interacting particle systems: A review}
\author{Apoorva Nagar$^1$ and Shamik Gupta$^{2}$}
\address{$^1$Indian Institute of Space Science and Technology, Thiruvananthapuram, Kerala, India\\ $^2$Department of Theoretical Physics,  Tata Institute of Fundamental Research, Homi Bhabha Road, Mumbai 400005, India}
\eads{\mailto{apoorva.nagar77@gmail.com}, \mailto{shamikg1@gmail.com}}

\begin{abstract}
We review recent work on systems with multiple interacting-particles having the dynamical feature of stochastic resetting. The interplay of time scales related to inter-particle interactions and resetting leads to a rich behavior, both static and dynamic. The presence of multiple particles also opens up a new possibility for the resetting dynamics itself, namely, that of different particles resetting all together (global resetting) or independently (local resetting). We divide the review on the basis of specifics of reset dynamics (global versus local resetting), and further, on the basis of number (two versus a large number) of interacting particles. We will primarily be dealing with classical systems, and only briefly discuss resetting in quantum systems.
\end{abstract}
\tableofcontents

%%%%%%%%%%%%%%%%%%%%%%%%%%%%%%%%%%%%%%%%%%%%%%%%%%%%%%%%%%%%%%%%%%%%%%%%%%%%%%%%%%%%%%%%%%%%%%%%%%

\section{Introduction}\label{introduction}
Imagine a situation in which a dynamical system evolves starting from a given initial condition, but only for a random period of time, after which the dynamics gets renewed and starts afresh from the given initial condition. The sequence of bare (that is, without any reset events) evolution followed by a reset to the initial condition and the dynamics starting anew is repeated over and over again. Consider different dynamical realizations of the system for the same duration and for the same initial (which is also the reset) condition. Then, different realizations would correspond to varying number of reset events occurring at different time instants. What do we expect to happen? Consider the case where the dynamical system has its natural evolution defined by a deterministic dynamics. Then, in the absence of resets, the dynamical degrees of freedom will have at a given time $t>0$ only a fixed set of values for the given initial condition. On the contrary, when reset is introduced into the dynamics, the following picture is evidently true. At any time $t>0$, the degrees of freedom obtained across different dynamical realizations will have a range of possible values, instead of a fixed set of values as was the case in absence of resets. Thus, for the same initial condition, resetting opens up a pathway to access the available phase space of the dynamical degrees of freedom. One may expect the system to have a stationary state under suitable conditions.

In the above backdrop, let us ask: How different is the situation when the inherent dynamics itself is stochastic? To fix our ideas, let us consider the ubiquitous example of diffusion. Here, a particle starting from a given spatial location will even in the absence of resetting explore the whole of the available phase space once allowed to evolve for long-enough times. Then, introducing repeated resetting at random times to say, the initial location, does just the opposite of what has been anticipated above for deterministic dynamics. Namely, resetting serves to confine the motion of the particle to a region around the initial location, and it is this confinement that eventually leads to a time-independent spatial distribution for the particle. On the contrary, when there is no resetting, the distribution does not ever converge to a time-independent form, and this is owing to the particle exploring larger regions of phase space for longer times.
    
An important implication of introducing the reset dynamics may be appreciated by noting that resetting introduces a time scale in the dynamics, be it stochastic or deterministic, namely, how often are resets happening on the average. This time scale will obviously interplay with the inherent times scales of the bare dynamics, and as we know from our experience, any interplay of time scales in a dynamical system almost invariably results in rich features, both static and dynamic, not only in the stationary state but also while the system is relaxing to the stationary state. 

Now, there are essentially two broad ways in which resetting can be implemented: One, wherein the system in any infinitesimal time interval $[t,t+\mathrm{d}t]$ undergoes a reset with probability $r\mathrm{d}t$ and undergoes evolution according to its intrinsic dynamics with the complementary probability $1-r\mathrm{d}t$, where $r$ is a positive constant. It may then be shown that the time interval $\tau$ between two successive reset events is a random variable distributed as an exponential:
\begin{align}
    \Phi(\tau)=re^{-r\tau};~\tau\in[0,\infty),~~~r>0.
    \label{eq:ptaur}
\end{align}
Hence, this sort of resetting dynamics is known as exponential resetting. The other sort is power-law resetting, wherein the random variable $\tau$ is distributed as a power-law: 
\begin{align}
\Phi(\tau)=\frac{\gamma}{\tau_0(\tau/\tau_0)^{1+\gamma}};~\tau\in[\tau_0,\infty),~\gamma>0,
\label{eq:ptau}
\end{align}
where $\tau_0$ is a microscopic cut-off. All moments of the exponential $\Phi(\tau)$, and in particular, the mean $\langle \tau \rangle$ and the variance $\langle \tau^2\rangle$, are finite. By contrast, for the power-law $\Phi(\tau)$, the mean is finite for $\gamma>1$ and is otherwise infinite, while the variance is finite for $\gamma>2$ and is otherwise infinite. This fact has important consequence on the dynamics of the system undergoing a reset dynamics. When the mean $\langle \tau \rangle$ is finite, it is guaranteed that the variable being subject to the reset dynamics does indeed undergo a reset in a finite time. On the other hand, if the mean is infinite, there could very well be realizations of the dynamics in which the variable does not undergo a reset in a finite time. An infinite variance implies that different realizations of the dynamics may have very different behavior, thereby generating huge fluctuations in the values of the variable undergoing the reset. Besides the aforementioned variation in time in the way resetting is implemented, one may also introduce a variation in space. For example, for particles distributed in space, one may consider all the particles to reset simultaneously, which constitutes the case of global resetting. Alternatively, one may consider local resetting, whereby particles reset individually and independently of each other. 

Let $\mathcal{C}$ denote a generic configuration of a system that is undergoing stochastic resetting. In the absence of resetting, the evolution of the bare system involving update of the configurations in time could be either stochastic or deterministic, and would be subject to a given initial condition. Let us first consider the case of exponential resetting, and ask for the conditional probability
$\mathcal{P}_r(\mathcal{C},t|\mathcal{C}_0,0)$ for the system to be found in configuration $\mathcal{C}$ at time $t>0$, given an initial configuration $\mathcal{C}_0$ at time $t=0$ to which the system resets at exponentially-distributed random times. Noting that the dynamics starts afresh (renews) from $\mathcal{C}_0$ following every reset, the probability $\mathcal{P}_r(\mathcal{C},t|\mathcal{C}_0,0)$ would satisfy a renewal equation of the form~\cite{rev1}
\begin{equation}
\mathcal{P}_r(\mathcal{C},t|\mathcal{C}_0,0) =e^{-rt}\mathcal{P}(\mathcal{C},t|\mathcal{C}_0,0)+r \int_0^t \mathrm{d}\tau~e^{-r\tau}\mathcal{P}(\mathcal{C},t|\mathcal{C}_0,t-\tau).
        \label{eq:renewal-basic0}
\end{equation}
Here, $\mathcal{P}(\mathcal{C},t|\mathcal{C}',t')$ is the conditional probability for the system to be found in configuration $\mathcal{C}$ at time $t>0$, in the absence of resetting and conditioned on the configuration $\mathcal{C}'$ at an earlier time $t'<t$. The first term on the right hand side of Eq.~(\ref{eq:renewal-basic0}) accounts for the event that the system has evolved freely, i.e., without a single resetting, in the time interval $[0,t]$. Here, the factor $e^{-rt}$ indeed gives the probability of no reset for a duration of time equal to $t$. The second term on the right hand side of Eq.~(\ref{eq:renewal-basic0}) accounts for all events in which at time $t$, the time elapsed since the last reset lies between $\tau$ and $\tau+\mathrm{d}\tau;~~\tau \in [0,t]$, with free evolution from the instant of last reset up to time $t$. Indeed, the probability density at time $t$ that a duration of time $\tau$ has elapsed since the last reset (i.e., the last reset happened in the interval  $[t-\tau-\rmd \tau,t-\tau]$) may be written as~\cite{rev1}
\begin{equation}
f(t,t-\tau)=re^{-r\tau}+e^{-r\tau}\delta(t-\tau),
\end{equation} 
which is normalized as $\int_0^t \rmd \tau~f(t,t-\tau)=1$, as it should be.
We can rewrite Eq.~(\ref{eq:renewal-basic0}) as 
\begin{equation}
\mathcal{P}_r(\mathcal{C},t|\mathcal{C}_0,0) =\int_0^t \mathrm{d}\tau~f(t,t-\tau)\mathcal{P}(\mathcal{C},t|\mathcal{C}_0,t-\tau).
\label{eq:renewal-basic1}
\end{equation}
For the case of power-law resetting, a line of reasoning similar to the one above applies, and we have 
\begin{equation}
\mathcal{P}_r(\mathcal{C},t|\mathcal{C}_0,0) =\int_0^t \mathrm{d}\tau~f_\gamma(t,t-\tau)\mathcal{P}(\mathcal{C},t|\mathcal{C}_0,t-\tau),
\label{eq:renewal-basic2}
\end{equation}
where now the quantity $f_\gamma(t,t-\tau)$ has different expressions depending on whether one has $\gamma <1$ or $\gamma >1$.
For $\gamma >1$, we have for large $\tau >
\tau_0$ that
\begin{align}
f_{\gamma>1,\tau>\tau_{0}}(t,t-\tau)=\frac{1}{\tau_{0}}\Big(\frac{\gamma-1}{\gamma}\Big)\Big(\frac{\tau}{\tau_{0}}\Big)^{-\gamma},
\label{fgammagt1}
\end{align}
and hence, 
$\int_0^{\tau_{0}}\rmd\tau~f_{\gamma>1,\tau<\tau_{0}}(t,t-\tau)=1-\int_{\tau_{0}}^{t}\rmd\tau~
f_{\gamma>1,\tau>\tau_{0}}(t,t-\tau)$. On the other hand, when we have $\gamma <1$, one has
\begin{align}
f_{\gamma<1}(t,t-\tau)=\frac{\sin(\pi\gamma)}{\pi}\tau^{-\gamma}(t-\tau)^{\gamma-1},
\label{fgammalt1}
\end{align}
which may be checked to be normalized: $\int_0^t \rmd \tau~f_{\gamma<1}(t,t-\tau)=1$.

A stationary state is said to have been attained when the probability $\mathcal{P}_r(\mathcal{C},t|\mathcal{C}_0,0)$ in the limit $t \to \infty$ reaches a time-independent form. The stationary distribution is defined as $\mathcal{P}_{r,\mathrm{st}}(\mathcal{C}|\mathcal{C}_0)\equiv \mathcal{P}_r(\mathcal{C},t\to \infty|\mathcal{C}_0,0)$, where repeated resetting to the initial condition even at long times makes the latter appear explicitly even in the stationary-state distribution. The stationary state attained in presence of stochastic resetting is a nonequilibrium stationary state (NESS), which is distinctly different from an equilibrium stationary state~\cite{rev1}. The basic roadblock in studying NESSs is the lack of a framework that allows to treat such states on a general footing, akin to the one due to Boltzmann and Gibbs that exists for equilibrium stationary states. Obtaining the stationary-state distribution even for simple nonequilibrium models often proves to be a formidable task, thereby requiring one to resort to numerical simulations and approximation methods.

Resetting dynamics has been thoroughly explored in recent times in a plethora of dynamical scenarios. The exploration began with a study of the reset dynamics of a single Brownian particle ~\cite{r1}, a work that arguably has ushered in a new dawn in the area of statistical mechanics, and, in particular, in the subfield of nonequilibrium statistical mechanics; for a review, see \cite{rev1,rev2}. Over the years, the effects of stochastic resetting have been investigated in a wide spectrum of dynamics, e.g., diffusion~\cite{r1,r2,r3,r4,r5,r6,r7,r8}, random walks~\cite{r9,r10}, random walks on disordered lattices~\cite{Sarkar:2022-1}, L\'{e}vy flights~\cite{r11},  Bernoulli trials~\cite{r12}, discrete-time resets~\cite{r13,Giuggioli:2022}, active motion~\cite{r14} and transport in cells~\cite{r15}, search problems~\cite{r17,r18,r19,r20,r21,r22,r23},  RNA-polymerase dynamics~\cite{r24, r25},  enzymatic reactions~\cite{r26}, dynamics of ecological systems~\cite{r27,r28}, and in discussing Feynman-Kac path integral formalisms~\cite{Roldan:2017}.      

While much of the literature has focused on single-particle systems, which are suitable models for non-interacting particles, most real-life systems involve interactions between particles. Our interest here is to review work in which the effect of resetting has been studied in systems having multiple interacting degrees of freedom. We will primarily be dealing with classical systems. The introduction of multiple interacting particles opens up the possibility of the interplay of inter-particle interactions with the resetting dynamics. Consider the case of resetting in the paradigmatic two-dimensional nearest-neighbour Ising model in contact with a heat bath at a given temperature $T$. Suppose we have $T>T_c$, the critical temperature of phase transition from an ordered or a magnetized phase to a disordered/unmagnetized phase. Let us consider the effects of repeated resetting to an initial ordered configuration. While resetting would like to drive the system to an ordered phase, the bare evolution between two successive resets would tend to do just the opposite, namely, to increase the disorder in the system. Models for predator-prey systems involve two opposing tendencies: the predators performing diffusive motion in search of new prey base would tend to spread out in space, while occasional return to known prey base, which is akin to a resetting move, would tend to bound their spatial distribution.The competition of opposing tendencies from bare evolution and resetting moves is evidently most effective when the underlying time scales are comparable. 

With these introductory general remarks, we now turn to specifics. To this end, we have organized the rest of the review as follows. We first treat classical systems, and broadly divide the systems studied according to the resetting type: in Section~ \ref{sec:global}, we look at global resetting, where all the particles constituting the system reset simultaneously, while in Section~\ref{sec:localresetting}, we discuss local resetting in which the particles reset independently of each other. The section on global resetting is further classified into systems involving two or a larger number interacting particles, covered in Sections~\ref{subsec:twoparticles} and~\ref{subsec:manyparticles}, respectively. Stochastic resets in quantum systems are mentioned briefly in Section~\ref{sec:quantum}. We draw our conclusions in Section~\ref{sec:conclusions}.

\section{Global resetting}\label{globalresetting}
\label{sec:global}
In this section, we look at resetting systems that have a common feature, namely, that the resetting dynamics involves simultaneous relocation of all the constituent particles. The dynamics of relevant quantities in many cases can be expressed as a simple renewal equation. We further classify the systems into those with two or multiple particles.

\subsection{Two interacting particles or degrees of freedom}
\label{subsec:twoparticles}
The inter-particle interaction becomes important even with systems having two interacting particles or dynamical variables. The dynamics in such systems can be written in terms of the evolution of individual variables. We discuss three distinct class of systems involving two interacting degrees of freedom. The first class, discussed in Section~\ref{subsubsec:predator-prey}, involves  predator-prey dynamics in which there is no attractive interaction between the predator and the prey. Indeed, the predator wants to catch the prey as the latter tries to escape being caught by the predator. In the second class, Section~\ref{subsubsec:long-range}, we have two particles that are subject to an effective attractive interaction trying to bring them together, which is countered by resetting moves trying to pull them apart.  While the total number of particles is conserved in these two classes, it is not the case in the third class of systems that we study in Section~\ref{subsubsec:birth-death}, which deals with diffusion with resetting in a system of interacting particles, with the inter-particle interactions inspired by birth-death processes whereby particles may die or be born stochastically.

\subsubsection{Predator-prey models}
\label{subsubsec:predator-prey}
Predator-prey models are used to describe dynamics of biological systems comprising two species, one of which is a predator and the other is a prey. The predators (represented below by subscript $A$) and the preys (subscript $B$) have each their own individual dynamics, in which the influence of the presence of the other species appears thus: The predators need for their sustenance to catch the preys, which however for the same reason of sustenance have to escape being caught by the predators. It may prove beneficial for a predator to once in a while reset its location to locations that have earlier been visited by the prey. In the case in which in one dimension, a diffusing predator adopts the aforementioned chasing strategy for a diffusing prey, it has been shown in Ref.~\cite{Marin:2019} that the mean first-encounter time between the predator and the prey is finite and which decreases with the resetting rate. This is the first work we review in this section. From the point of view of the prey, in the case in which it is being chased by a swarm of predators, a possible scenario is that the prey on encountering one of the predators either perishes or escapes through relocation (akin to a resetting move) to a preferred safe position. In the setting in which both the predators and the prey undergo independent Brownian motion in one dimension, two quantities of interest are: (i) What is the probability that the prey does not perish (the so-called survival probability) up to time $t$? (ii) What is the probability distribution for the total number of encounters till the capture time $t_c$, with the latter itself a random variable? These questions have been addressed in Ref.~\cite{Evans:2022}, which is the second work we review in this section. It has been shown that while the survival probability decays algebraically in time, the probability distribution of the total number of encounters until $t_c$ exhibits an interesting behavior, namely, an anomalous large deviation form. 

While the questions addressed in the two papers mentioned above mainly focus on the dynamics and concern themselves with the speed at which the prey can be caught, the third work that we review in this section looks at the stationary-state population of the predators and the preys. Here, the dynamics of predator-prey system is such that the preys are confined to a single region of space and predators move around according to a power-law (L\'{e}vy) dispersal kernel and undergo resetting at random times to the prey patch. It has been shown that fixing the demographic parameters (the predator mortality and reproduction rates) and the L\'{e}vy exponent, the total population of predators can be maximized for a certain value of the resetting rate.\\

\noindent
\textit{A. A prey and a resetting predator:} In Ref.~\cite{Marin:2019}, the authors have modelled a predator-prey system in one spatial dimension by considering the case of two interacting Brownian particles. The motion of the predator is subordinated to the motion of the prey that moves independently.  Specifically, the prey exhibits ordinary diffusion and its location $x_B(t)$ evolves in time as
\begin{equation}
\frac{\mathrm{d}x_B(t)}{\mathrm{d}t}=\eta_B(t),
\label{eq:prey}
\end{equation}
with $\eta_B(t)$ being a Gaussian, white noise satisfying $\langle \eta_B(t) \rangle=0$ and $\langle \eta_B(t) \eta_B(t')\rangle=2D_B\delta(t-t')$. Here $D_B>0$ is the diffusion coefficient of the prey and angular brackets denote averaging over noise realizations. On the other hand, the motion of the predator consists of free diffusion interspersed with instantaneous resetting to positions previously visited by the prey.  The location $x_A(t)$ of the predator evolves in time as
\begin{equation}
\frac{\mathrm{d}x_A(t)}{\mathrm{d}t}=\eta_A(t)[1-\sigma(t)]+\zeta[t,x_B(\tau);~\tau\le t]\sigma(t),
\label{eq:predator}
\end{equation}
with $\eta_A(t)$ denoting a Gaussian, white noise that is independent of $\eta_B(t)$ and which satisfies $\langle \eta_A(t)\rangle=0$ and $\langle \eta_A(t) \eta_A(t')\rangle=2D_A \delta(t-t')$. Here, $D_A>0$ is the diffusion coefficient of the predator, $\sigma(t)$ is a dichotomous stochastic process taking the value $1$ at Poisson rate $r$, such that in the time interval $[t, t+\mathrm{d}t]$,  the quantity $\sigma(t)$ takes the value $1$ with probability $r\mathrm{d}t$ and the value $0$ with probability $1-r\mathrm{d}t$. The quantity $\zeta[t,x_B(\tau);~\tau\le t]$ denotes the discontinuous stochastic process describing the dynamics of the predator chasing the prey, i.e., at a constant rate $r$, the predator jumps instantaneously (resets) from its current location $x_A(t)$ to a position $x_B(\tau)$ previously visited by the prey at the random time $\tau \le t$. Here, the random variable $\tau$ is taken to be distributed in the interval $[0,t]$ according to a probability density $\widetilde{\phi}(\tau;t)$ given by 
\begin{equation}
\widetilde{\phi}(\tau;t)=\frac{\lambda e^{-\lambda \tau}}{1-e^{-\lambda t}},
\end{equation}
where $\lambda \in (-\infty,\infty)$ is a real parameter that characterizes the range of memory of the predator.  For $\lambda <0$, the predator has a short memory, whereby it resets with a large probability to the most recent positions visited by the prey (active chasing strategy).  On the other hand, for $\lambda >0$, the predator resets preferentially to the initial positions visited by the prey and thus is said to have a long memory (passive chasing strategy). The case $\lambda=0$ corresponds to $\tau$ being chosen uniformly in the interval $[0,t]$. Note that the reset dynamics under consideration is an example of exponential resetting discussed in the introduction.

The main question addressed in the work being reported here is: given the prey and the predator dynamics in Eqs.~(\ref{eq:prey}) and~(\ref{eq:predator}),  when does the predator capture the prey for the first time? The first encounter time is a random variable and the quantity of interest is the mean first-encounter time.  To this end,  $x(t) \equiv |x_A(t)-x_B(t)|$ is defined as the relative distance at time $t$ between the predator and the prey location. The dynamics of $x(t)$ in time is easily obtained from Eqs.~(\ref{eq:prey}) and~(\ref{eq:predator}).  It is evident that for a given value of $x_0\equiv x(t=0)$ of the initial relative distance between the predator and the prey, the first encounter among them corresponds to $x$ attaining the value zero for the first time,  and so this time is nothing but the time of first-passage of $x$ through the value zero. The probability that $x$ does not attain the value zero until time $t$ is called the survival probability. In the absence of resetting, $r=0$, the predator and the prey exhibit independent diffusion, and in this case, the survival probability is given by
\begin{equation}
Q(x_0,t)=\mathrm{Erf}\left(\frac{x_0}{\sqrt{4Dt}}\right),
\end{equation}
where $\mathrm{Erf}(x)$ is the error function, and $D\equiv D_A+D_B$. In the Laplace domain, one then has $\widetilde{Q}(x_0,u)=(1/u)\left(1-\exp(-\sqrt{u/D}~x_0)\right)$, where $u$ is the Laplace variable. Taking the limit $u \to 0$ of $\widetilde{Q}(x_0,u)$ yields the mean first-passage time, i.e., the mean time of first encounter between the predator and the prey. It can be seen that this time is infinite in the absence of resetting.

\begin{figure}[htbp]
\centering
\includegraphics[width=10cm]{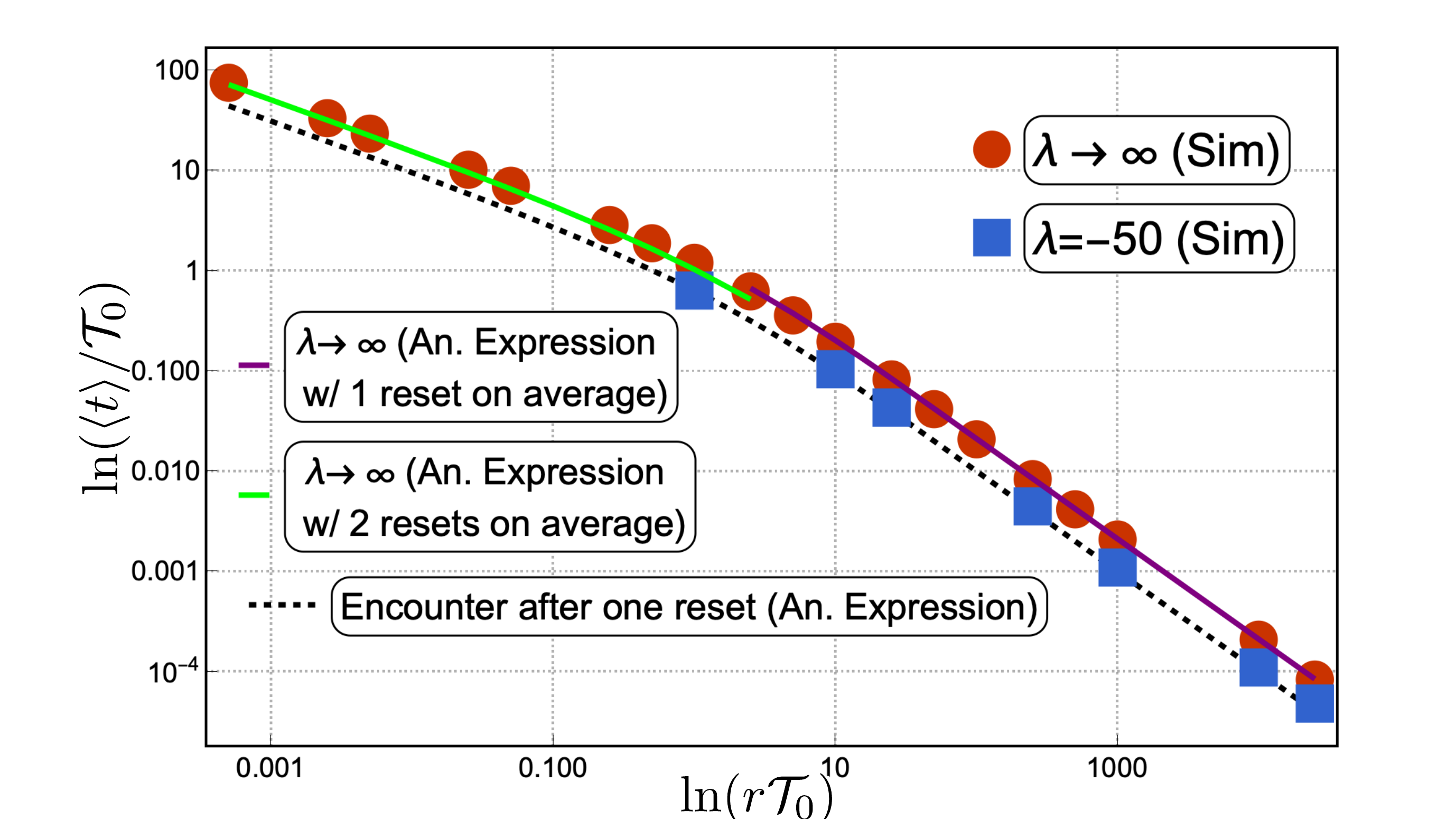}
\caption{For the predator-prey problem studied in Ref.~\cite{Marin:2019} and discussed in the text, Section~\ref{subsubsec:predator-prey}, Subsection~A, the figure shows the mean first-encounter time $\langle t \rangle$ rescaled by $\mathcal{T}_0 \equiv x_0^2/D$ and plotted as a function of $r\mathcal{T}_0$ for different values of the parameter $\lambda$ that characterizes the memory of the predator. The dashed line corresponds to the active chasing strategy of the predator ($\lambda \to -\infty$). The blue squares correspond to data obtained from numerical simulations of the dynamics, Eqs.~(\ref{eq:prey}) and~(\ref{eq:predator}),  with $\lambda = -50$ and with time-step of numerical simulation ranging from $10^{-3}$ to $10^{-6}$ depending on the value of $r$. The red circles correspond to numerical simulations for $\lambda \to \infty$ (passive chasing strategy of the predator).  Each data point in the figure corresponds to $10^7$ simulation runs.  The continuous lines correspond to approximate analytical results obtained following the procedure discussed in the text.  The figure is adapted from Ref.~\cite{Marin:2019}.}
\label{marinMFPT}
\end{figure}

In presence of resetting, $r \ne 0$,  the survival probability may be expressed as a sum, over the number of resets, of the survival probability of a process with exactly $n$ number of resets to a given sequence of reset locations $\{x_i\equiv x(t_i)\}_{1 \le i \le n}$ with $0 < t_1 < t_2 < \ldots t_n < t$.  Specifically,  in the Laplace domain, one obtains
\begin{equation}
\widetilde{Q}_r(x_0,\{x_i\},u)=\widetilde{Q}(x_0,r+u)\left(1+\sum_{n=1}^\infty \prod_{i=1}^n r\widetilde{Q}(x_i,r+u)\right).
\label{eq:SQ}
\end{equation}
 Now,  since each $x_i$ is a random variable, the problem of obtaining the mean first-encounter time from this equation by averaging over distribution of the $x_i$'s becomes analytically intractable.  A headway is achieved by replacing  the quantity $x_i$ in Eq.~(\ref{eq:SQ}) by its typical value $\sqrt{2D_Bi/r}$ (which is obtained in the limit  $\lambda \to \infty$ in which the predator relocates to the initial position of the prey), and by truncating the summation to $n = 1$ and $n = 2$ for $r\mathcal{T}_0 < 1$ and $r\mathcal{T}_0 > 1$, respectively, with $\mathcal{T}_0 \equiv x_0^2/D$.  In Fig.~\ref{marinMFPT} is shown the mean first-encounter time $\langle t \rangle$, rescaled by $\mathcal{T}_0$, when plotted as a function of $r \mathcal{T}_0$ and for different values of $\lambda$. From the figure, it is evident that for a finite reset rate $r$, the mean first-encounter time is finite,  as opposed to the situation when the predator diffuses without resetting for which, as discussed above,  the mean first-encounter time diverges. Thus, resetting proves beneficial for the predator to capture the prey, which constitutes the main result of Ref.~\cite{Marin:2019}. \\
 
\noindent
\textit{B. Resetting prey and diffusing predators:} A model in one dimension with multiple predators and a single prey is studied in Ref.~\cite{Evans:2022}. The model considers a swarm of predators diffusing on a line and labelled sequentially as $n=1,2,3,\ldots$, each with a diffusion constant $D_A>0$, and with only one of them being active, i.e., being able to detect and capture the prey, while the other predators are passive, at any given time instant. The single prey diffuses with a diffusion constant $D_B>0$. An encounter of the prey with the active predator leads to two possibilities: (a) with probability $0 \le p \le 1$, the prey escapes and resets to the origin, while the current predator becomes passive and triggers the next labelled predator to become active, (b) with probability $1-p$, the hunt is successful and the prey dies. Only one of the predators being active at any given time, the system can be described by two degrees of freedom with one representing the prey and the other representing the active predator. 

The position of the active predator ($x_A$) and the prey ($x_B$) evolves according to two independent Brownian motion:
 \begin{eqnarray}
\frac{\mathrm{d} x_A(t)}{\mathrm{d} t} &= & \eta_A(t), \label{langeA.1} \\
\frac{\mathrm{d} x_B(t)}{\mathrm{d} t} &= & \eta_B(t), \label{langeB.1}
\end{eqnarray}
where $\eta_A(t)$ and $\eta_B(t)$ are independent Gaussian white noise terms
with zero mean and correlators given by $\langle \eta_A(t)\eta_A(t')\rangle = 2 D_A \delta(t-t')$, $\langle \eta_B(t)\eta_B(t')\rangle= 2 D_B \delta(t-t')$ and $\langle \eta_A(t)\eta_B(t')\rangle =0$. 

The dynamics can be re-expressed in terms of the relative coordinate $x(t)\equiv x_A(t)-x_B(t)$, which also performs a Brownian motion between two successive
encounters between the predator and the prey, according to 
\begin{equation}
\frac{\mathrm{d} x(t)}{\mathrm{d} t}= \eta(t) \equiv \eta_A(t)-\eta_B(t),
\end{equation}
where the noise $\eta(t)$ is also a Gaussian white noise with 
zero mean and correlator $\langle \eta(t)\eta(t')\rangle= 2 (D_A+D_B)\delta(t-t')$. When the prey encounters the active predator at some time $t_i$, 
the relative coordinate $x(t_i)$ becomes $0$ (the ``relative'' particle getting absorbed) with probability $(1-p)$. With probability $p$, the predator fails to catch the prey as the latter resets its position  to the origin, and a new predator is designated as
active. Since all the predators are performing independent Brownian motion, the resetting move can be seen as the relative co-ordinate performing a sudden jump to $x_A(t_i)$, which is a Gaussian-distributed random variable: 
\begin{equation}
\mathrm{Prob.}\left(x_A(t_i) \in [z, z+ \mathrm{d}z]\right)=\frac{1}{\sqrt{4 \pi D_A t_i}}\,e^{-z^2/{4 D_A t_i}}\mathrm{d}z. 
\label{Gaussian}
\end{equation}
This reset is independent of the position of the previous active predator, and the process then has a renewal property.

One is interested in calculating the probability $Q(t_0,t)$ that starting at time $t_0$ with an active predator and a surviving prey (the relative coordinate $x(t)$ is therefore a Gaussian-distributed random variable, distributed according to the distribution in Eq.~(\ref{Gaussian})), the predator and the prey do not encounter each other up to time $t>t_0$. One has
\begin{equation}
Q(t_0,t)= Q_0(t_0,t)+ p\, \int_{t_0}^t \mathrm{d}t'~F_0(t_0,t')\, Q(t',t).
\label{renewal.1}
\end{equation}
The first term on the right, namely, $Q(t_0,t)$, accounts for the survival probability
when there are no encounters between the predator and the prey for times between $t_0$ and $t$. This term can be calculated as the survival probability for a diffusing particle with diffusion constant $D_A+D_B$, with an absorbing boundary at the origin, and averaged over the initial Gaussian distribution~(\ref{Gaussian}) at time $t_0$. In the second term on the right, $t'$ is the time for the first encounter and the factor $p$ gives the probability for the  prey to survive this encounter. The integrand thus contains the probability density $F_0(t_0,t')$ for the first encounter to happen at time $t'$, multiplied by the probability $Q(t',t)$ that the prey survives from then on up to time $t$. More precisely, the quantity $F_0(t_i,t)$ gives the probability density that after a reset at time $t_i$ (corresponding to the $i$-th encounter), the first encounter between the predator and the prey happens between times $t$ and $t+\mathrm{d}t$, with $t>t_i$. This is related to the probability density $Q_0(t_i,t)$ that the predator and the prey do not encounter each other between the times $t_i$ and $t$ as: $F_0(t_i,t)=- \partial Q_0(t_i,t)/\partial t$. The quantity $Q_0(t_i,t)$ is obtained by following the procedure mentioned above as
\begin{eqnarray}
Q_0(t_i,t)&=& \int_{-\infty}^{\infty}\mathrm{d}z~\mathrm{Erf}\left(\frac{|z|}{\sqrt{4(D_A+D_B)(t-t_i)}}\right)\,
\frac{1}{\sqrt{4\pi D_A t_i}}\, \e^{- z^2/{4 D_A t_i}} \nonumber \label{survi.1} \\[1ex]
&=& \frac{2}{\pi}\, {\tan}^{-1}
\left[\sqrt{ \frac{R_D}{\frac{t}{t_i}-1}}\right],
\label{survi.2}
\end{eqnarray}
where we have $R_D \equiv D_A/(D_A+D_B)$. 

One can now calculate the time dependence of the survival probability using the expression in Eq.~(\ref{renewal.1}) above. One obtains an algebraic decay of the survival probability in time~\cite{Evans:2022}:
\begin{equation}
  Q(t_0,t) \sim (t/t_0)^{-\theta}\quad \mbox{for}\quad  t\gg t_0.
  \end{equation}
The exponent $\theta$ solves the equation
\begin{equation}
\frac{1}{p}= \frac{\sqrt{R_D}}{\sqrt{\pi}}\,
\frac{\Gamma(1/2-\theta)}{\Gamma(1-\theta)}
\,
{}_2F_1\left[1,1/2-\theta,1-\theta; 1-R_D\right],
\label{theta_exp.1}
\end{equation}
where ${}_2F_1\left[a,b,c;z\right]$ is the usual Hypergeometric function. We see that the exponent $\theta(p,R_D)$ depends continuously on the two parameters $p$ and $R_D$.

The authors of Ref.~\cite{Evans:2022} have also calculated the probability that starting at time $t_0$, the prey is successfully caught at a given time $t_c$ after having $N-1$ unsuccessful encounters previously. Here, it is convenient to work with Lamperti time $T_i \equiv \ln(t_i/t_0)$, where the index $i$ indicates the $i$-th encounter between the predator and the prey. The process under consideration starts at time $T_0=0$ and has $N$ encounters at epochs $\{T_1,T_2,\ldots, T_N=T_c\}$, with the final $N$-th encounter at time $t_c$ being successful for the predator. The probability that such a process happens is given by
\begin{equation}
P(N,T_c)= \frac{(1-p)}{p}\, \int_0^{\infty} %
\prod_{i=1}^N \mathrm{d}\tau_i  \left[\prod_{i=1}^N p\, f_0(\tau_i)\right]
\, \delta\left(\tau_1+\tau_2+\ldots +\tau_N-T_c\right).
\label{PNT.1}
\end{equation}
Here, $\tau_i \equiv T_i-T_{i-1}$ is the time interval between the $i$-th and the $(i-1)$-th encounter, while $f_0(\tau)$ is the probability distribution of these interval lengths. The quantity $f_0(\tau)$ is related to the quantity $F_0(t_i,t)$ discussed above as $F_0(t_i,t)\mathrm{d} t =  f_0(T-T_i)\, \mathrm{d} T$, and can be calculated explicitly to give
\begin{equation}
f_0(T)= \frac{\sqrt{R_D}}{\pi} \frac{\e^{T}}{(\e^{T}-1+R_D)\sqrt{\e^T-1}}.
\label{f0T.1}
\end{equation}
The integral in the expression (\ref{PNT.1}) above cannot be evaluated in a closed form, but its moments can be calculated by going to the Laplace space, and for large capture times $t_c$, one gets the expressions
\begin{equation}
\langle N\rangle_{t_c}= A_1 \ln t_c +O(1)\, , \quad \mathrm{and}\quad 
\sigma^2_N\equiv \langle N^2\rangle_{T_c}-\langle N\rangle_{T_c}^2= B\, \ln t_c + O(1).
\label{mean_var.1}
\end{equation}
In the above equation, note that the $p$ and $R_D$ dependence is absorbed in the parameters $A_1$ and $B$. We thus observe a slow logarithmic growth as a function of the capture time $t_c$. This behavior is different from processes where resetting occurs at a constant rate, where, e.g., the mean number of resetting grows linearly. The slow growth can be attributed to the power-law distribution of the intervals between successive resetting for the case at hand. \\

\noindent
\textit{C. Effect of resetting of prey in a Lotka-Volterra-type predator-prey system with localized prey}
In Ref.~\cite{Vasquez:2018}, the authors have considered a predator-prey model, where the preys are confined to a region around the origin, while there is no restriction on the distance to which the predators can move. The paper describes the effect of the predators resetting stochastically with a constant rate $r$ to the origin in such a model. The predators move stochastically following a L\'evy dispersal Kernel characterized by a power-law exponent $\beta$. The authors have analyzed the effect of $\beta$ and $r$ on the abundance of predators in the stationary state. 

The model is defined on a $D$-dimensional regular lattice of square cells of length $R$, with the position of the center of each cell represented by the vector $\textbf{n}$ with integer components. The predators occupy the cells and perform independent random motion involving continuous displacements, with unity being the smallest possible displacement ($1<R$). The displacements are symmetric and independent, with a power-law form for the distribution $\psi(x)$ of step length $x$:
\begin{equation}
\psi(x)=\frac{\beta-1}{2}|x|^{-\beta} \quad \text{for} \quad  |x|>1,
\label{psix}
\end{equation}
and $\psi(x)=0$ for $|x|<1$. The probability distribution of a predator jumping to a cell at a distance $l$ away is given by
\begin{align}         
p(\textbf{\textit{l}})=p_0\delta_{\textbf{\textit{l}},\textbf{0}}+(1-p_0)f(\textbf{\textit{l}}),
\end{align}
where $\textbf{\textit{l}}$ is a vector with integer components, and where the first term on the right hand side represents the possibility that a predator remains in the same cell after moving, and $f(\textbf{\textit{l}})$ is a normalized dispersal distribution with an inverse power-law form with the same exponent $\beta$ as in Eq.~(\ref{psix}):
\begin{equation}
f(\textit{l}=|\textbf{\textit{l}}|)=\frac{|l|^{-\beta}}{2\sum_{m=1}^{\infty} m^{-\beta}}\quad \textrm{  for  }l=\pm 1, \pm 2, \ldots.
\end{equation}
One has $p_0=(2/R)\int_{0}^{R} \mathrm{d}r\int_{0}^{r} \mathrm{d}x\  \psi(x)$, which gives 
\begin{align}
&p_0= 1-\frac{1}{R}+\frac{1-R^{2-\beta}}{(2-\beta)R}~\mathrm{for}~\beta\neq 2,\\ 
&p_0= 1-\frac{1}{R}-\frac{\ln(R)}{R}~\mathrm{for}~\beta=2.
\end{align}
The preys occupy only the cell at the origin ($\textbf{n}=\textbf{0}$), and it is also the only cell where the predators can reproduce (the preys also reproduce in this cell). The predators reset with rate $r$ to the origin cell, independent of their current location in the system. Once a predator has reset to the origin, it continues its dynamics from there (jumping between cells), until the next resetting event takes place. The reset dynamics under consideration is an example of exponential resetting discussed in the introduction.   

Let us denote the predator density in a cell at position ${\bf{n}}$ by $\rho_A(\textbf{n},t)$, while the prey density is represented by $\rho_B(t)$. The predator density satisfies the equation
\begin{eqnarray}
&\frac{\partial{\rho_A(\textbf{n},t)}}{\partial{t}}=-\alpha(1-p_0)\rho_A(\textbf{n},t)+\alpha\sum_{\textit{\textbf{l}},\,|\textbf{\textit{l}}|\neq \textbf{0}}p(\textbf{\textit{l}})\rho_A(\textbf{n}-\textbf{\textit{l}},t)\nonumber \\
&+\vartheta \rho_A(\textbf{0})\rho_B(t)\delta_{\textbf{n,0}}-\mu \rho_A(\textbf{n},t)-r\rho_A(\textbf{n},t)+\left[r\sum_{m}\rho_A(\textbf{m},t)\right]\delta_{\textbf{n},\textbf{0}}, 
\label{lotka}
\end{eqnarray}
where the first term on the right corresponds to the event of predators leaving the cell $\textbf{n}$, with $\alpha$ being the movement rate, and the second term describes the rate at which predators arrive into this cell from another via the L\'evy dynamics described above. The third term accounts for reproduction, which happens only in the origin cell with rate $\vartheta$ and is dependent on the abundance of the preys in the origin cell. The fourth term accounts for predator mortality, which is position independent and happens at a rate $\mu$. The last two terms account for the resetting process, with the fifth term corresponding to predators being removed from their current location at rate $r$, and the sixth term representing the increase in density at the origin due to predators from other locations resetting to the cell at the origin. The prey density satisfies
\begin{equation}
\frac{\mathrm{d}\rho_B}{\mathrm{d}t}=\sigma \rho_B\left(1-\frac{\rho_B+\rho_A(\textbf{0})}{K}\right)-\vartheta' \rho_A(0) \rho_B,
\label{preydensity}
\end{equation}
where $\rho_A(\textbf{0})=\rho_A(\textbf{0},t)$ is the predator density at the origin cell, $\sigma$ is the reproduction rate of the preys, and $\vartheta'$ is the predation rate. The parameter $K$ in the first term represents the prey carrying capacity, which enforces the limitation that the prey number cannot grow indefinitely, since the growth is limited by available resources. 

The authors have considered the stationary-state of the dynamics, in which the prey density has two possible solutions, $\rho_B=0$, corresponding to extinction, and the non-trivial solution $\rho_B=K-\rho_{A,\,\mathrm{st}}(\textbf{0})(1+K\vartheta'/\sigma)$. Here $\rho_{A,\,\mathrm{st}}(\textbf{0})$ is the stationary density of predators at the origin cell, which may be obtained as
\begin{equation}
 \rho_{A,\,\mathrm{st}}(\textbf{0})=\frac{1}{1+K\vartheta'/\sigma}\left\{K-\left[\frac{\vartheta}{\mu(2\pi)^D} \int_{B}\mathrm{d}\textbf{k}~\frac{(\mu+r)}{\alpha(1-p_0)[1-\hat{f}(\textbf{k})]+\mu+r}\right]^{-1}\right\}.
\label{azero}
\end{equation} 
Here, $\hat{f}(\textbf{k})\equiv\sum_{\textit{\textbf{l}}} f(\textit{\textbf{l}})e^{-\mathrm{i}\textbf{k}\cdot\textit{\textbf{l}}}$ is the Fourier transform of $f(\textit{\textbf{l}})$. The central quantity of interest in the study is the total population of predators: $N_p \equiv R\sum_{\textbf{n}}\rho_{A,\mathrm{st}}({\textbf{n}})$, which is shown to equal
\begin{equation}
		N_p=R\frac{\vartheta \rho_{A,\,\mathrm{st}}(\textbf{0})}{\mu}\left[K-\rho_{A,\,\mathrm{st}}(\textbf{0})\left(1+K\vartheta'/\sigma\right)\right].
		\label{totpopul2}
\end{equation}
Let us now discuss the behavior of $N_p$ by considering $D=1$ as a representative case. Let us fix $K=1$, $\sigma=1$, $\vartheta=1$, $\vartheta'=1$, and $R=10$. The dependence of $N_p$ on $r$, on $\beta$, and on the birth/mortality rate ($\vartheta/\mu$ respectively) can be evaluated numerically and some representative figures are shown below. 

\begin{figure}[!htp]
\centering
\begin{subfigure}{0.4\textwidth}
\includegraphics[width=\textwidth]{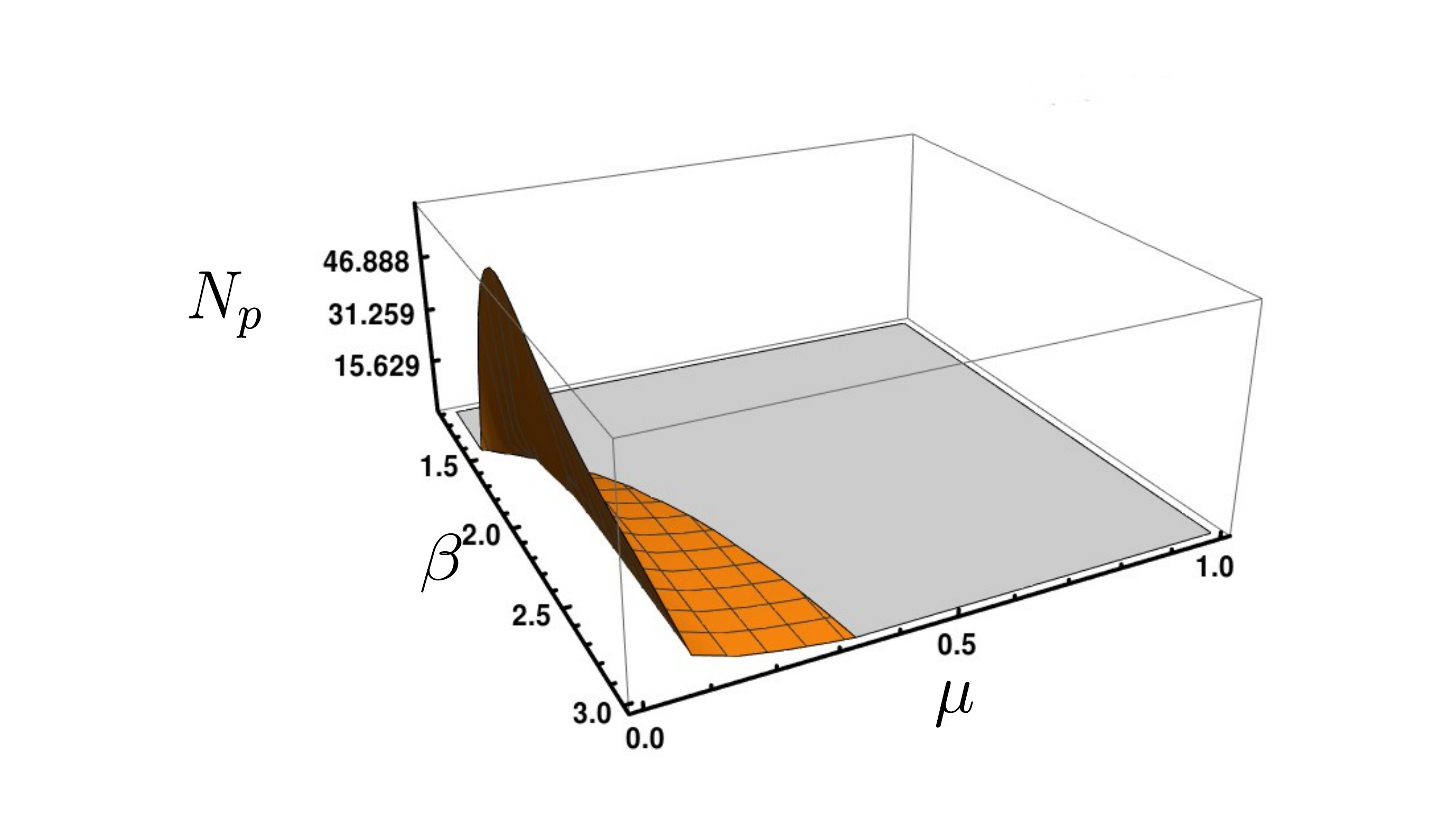}
\caption{$r=0.0$}
\end{subfigure}
\hfill
\begin{subfigure}{0.4\textwidth}
\includegraphics[width=\textwidth]{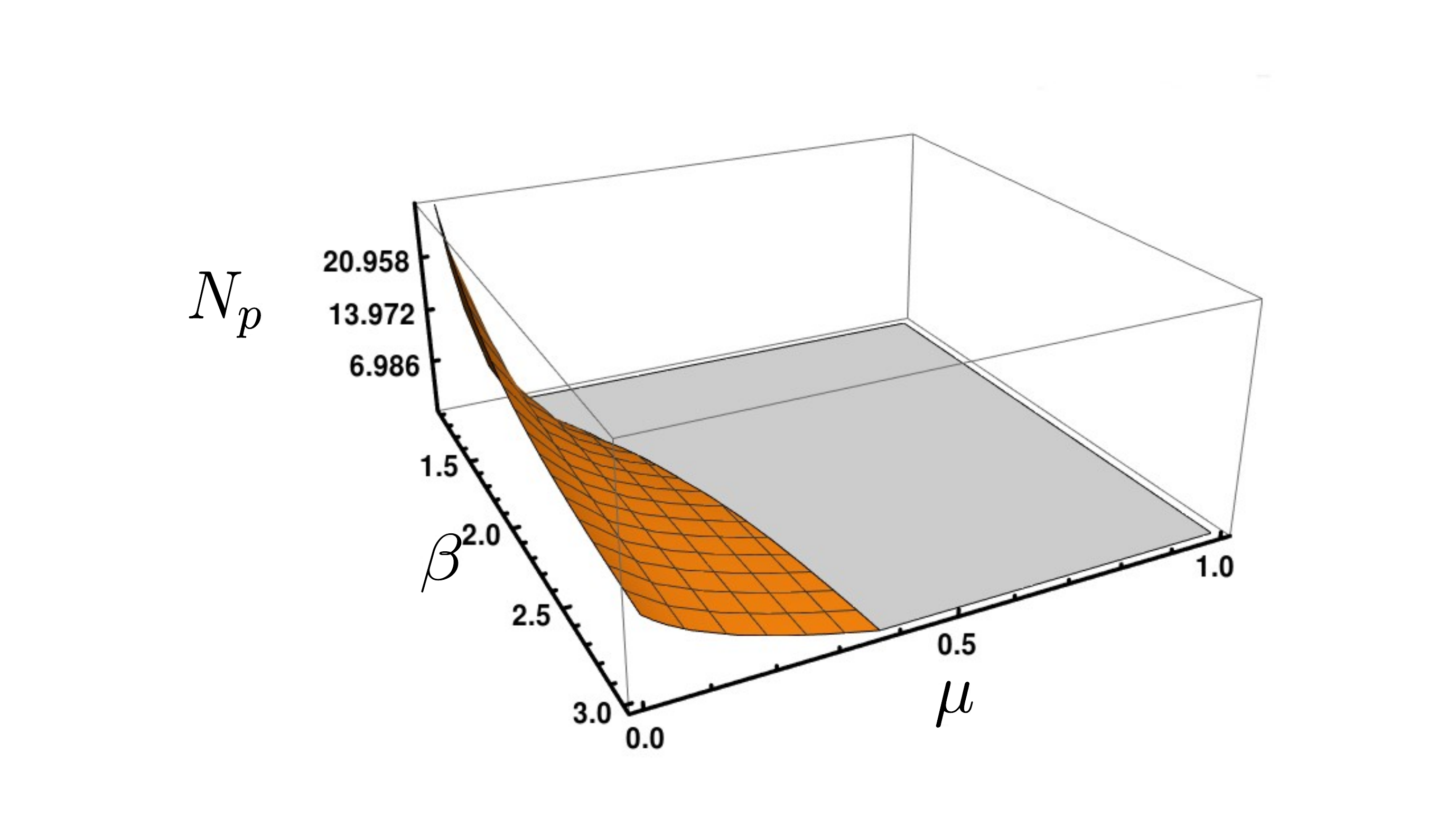}
\caption{$r=0.15$}
\end{subfigure}
\hfill
\begin{subfigure}{0.4\textwidth}
\includegraphics[width=\textwidth]{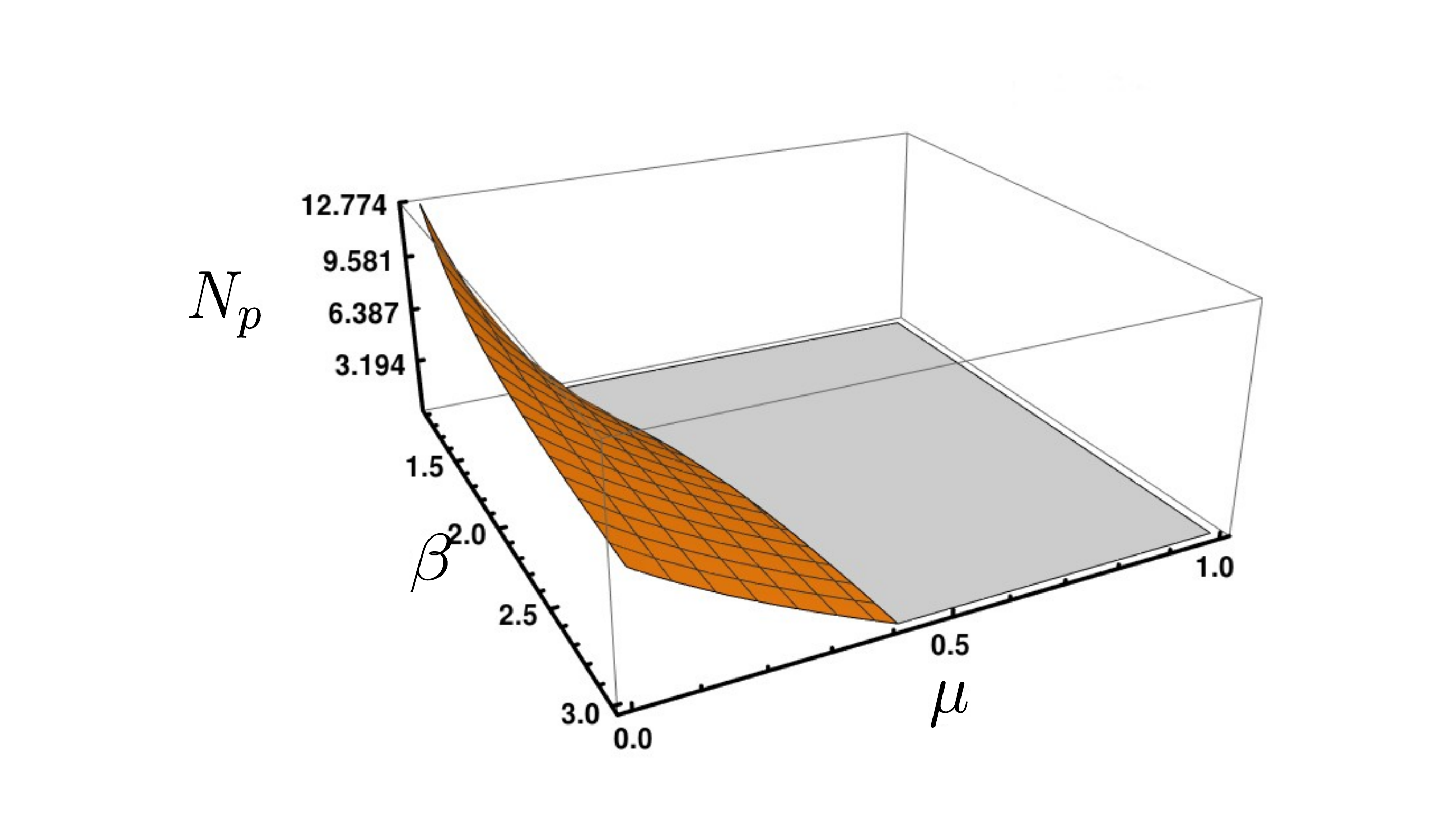}
\caption{$r=0.5$}
\end{subfigure}
\caption{For the predator-prey system studied in Ref.~\cite{Vasquez:2018} and discussed in the text, see Section~\ref{subsubsec:predator-prey}, Subsection~C, the figure shows the total population of predators, denoted by $N_p$, in the ($\mu$, $\beta$)-plane for $\vartheta=0.5$ and for three values of $r$. The figure is adapted from Ref.~\cite{Vasquez:2018}.}
\label{btamu}
\end{figure}

\begin{figure}[htbp]
\centering
\begin{subfigure}{0.4\textwidth}
\includegraphics[width=\textwidth]{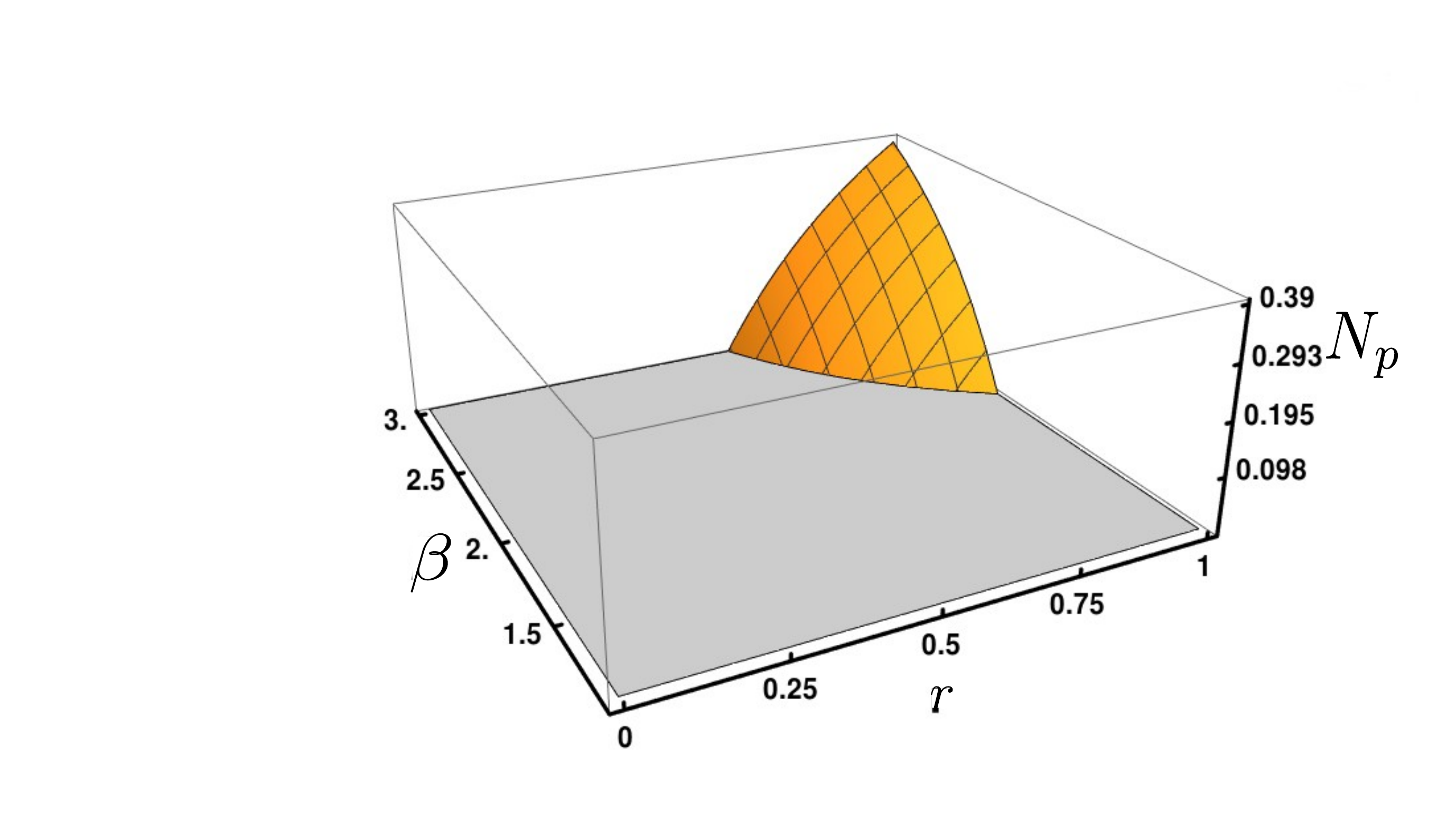}
\caption{$\mu=0.08$}
\end{subfigure}
\hfill
\begin{subfigure}{0.4\textwidth}
\includegraphics[width=\textwidth]{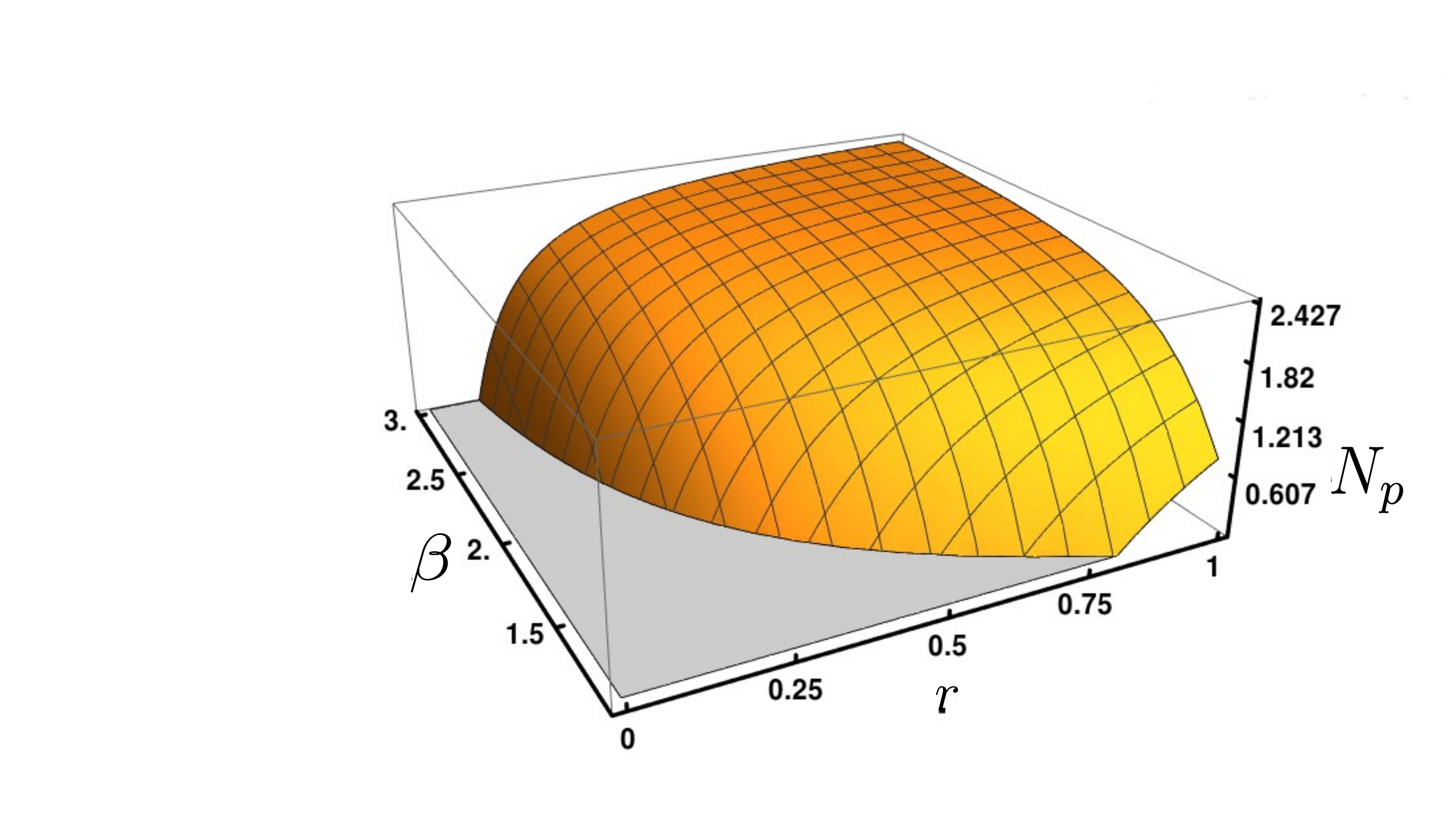}
\caption{$\mu=0.05$}		
\end{subfigure}
\hfill
\begin{subfigure}{0.4\textwidth}
\includegraphics[width=\textwidth]{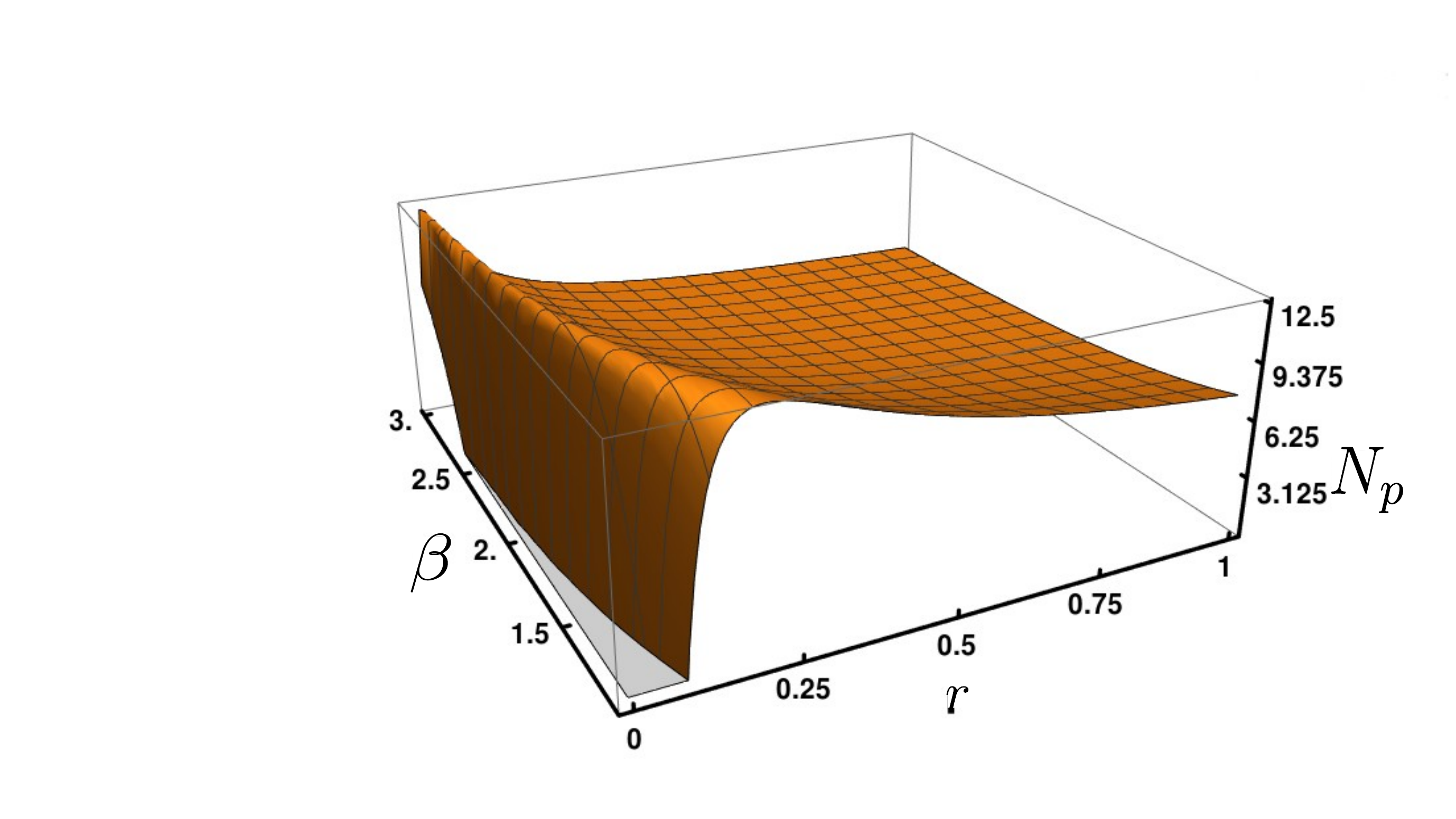}
\caption{$\mu=0.01$}
\end{subfigure}
\caption{For the predator-prey system studied in Ref.~\cite{Vasquez:2018} and discussed in the text, see Section~\ref{subsubsec:predator-prey}, Subsection~C, the figure shows the total population of predators, denoted by $N_p$, in the ($\beta$, $r$)-plane for $\vartheta=0.1$ and three values of $\mu$. The figure is adapted from Ref.~\cite{Vasquez:2018}.}\label{btaerea}
\end{figure}

Figure~\ref{btamu} shows the effect that the introduction of resetting via parameter $r$ has on the total predator population $N_p$. For $r=0$ at fixed $\mu$, we see that for low death rates ($\mu$), the predator population is low for small values of $\beta$; this is due to the large jumps that take the predators very far away from the prey patch and therefore reduce survival probability. Again, at large values of $\beta$, the predator population falls, this time because the predators spend too much time in the prey cell, reducing the prey population and therefore reducing their own numbers. Introduction of resetting provides a counter to the large displacements away from the prey cell, and the dip at low $\beta$ is removed. Increase of $\mu$ has a deleterious effect on predator population, finally leading to extinction. The deleterious effects are less pronounced at the large-$\beta$ end where the predators largely stay close to the prey patch and have a better chance of survival. Figure~\ref{btaerea} displays the variation in $N_p$ with $\beta$ and $r$, for different values of $\mu$. We see that high values of $\mu$, high values of $\beta$ and high values of $r$ provide favourable circumstances for the predators, without which they go extinct. To conclude, in Ref.~\cite{Vasquez:2018}, the authors have shown that the effect of resetting on the studied prey-predator system can be both rich and complex. The upshot is that combined effects of resetting and diffusion can extend the region of species coexistence in ecosystems that face scarcity of resources. 

\subsubsection{Two diffusing particles with an interacting reset mechanism}
\label{subsubsec:long-range}
In Ref.~\cite{Falcao:2017}, resetting has been considered in the framework of two diffusing particles interacting via an attractive interaction in one dimension. The particles are identical and move on a discrete, one-dimensional, infinite lattice with sites separated by unit length. Starting at sites labelled $-l$ and $l$, with $l \in \mathbb{N}$, the positions $x_L(t)$ and $x_R(t)>x_L(t)$ of the left and the right walker, respectively, evolve with time in the following fashion: If at a given time step, the distance between the particles is greater $2$, then in the next time step, a given particle moves towards the other with a probability $1/2+\epsilon$ and away from the other with a smaller probability $1/2-\epsilon$. If the distance is just equal to $2$, then in the next time step, both the particles move to the right or to the left with probability $1/4-\epsilon^2$, or both reset to their initial positions with probability $(1/2+\epsilon)^2$, or the left walker moves to the left and the right walker moves to the right with probability $(1/2-\epsilon)^2$. Thus, when the particles are two sites apart, there is a possibility that the next move lands them on the same site, and if this possibility arises, the particles are reset to their original positions. Here, $\epsilon>0$ is a dynamical parameter, which gives a measure of the effective attraction between the particles. Summarizing, if $x_L-x_R=2$, then the following are the possibilities for the move in the next time step:
\begin{eqnarray}
(x_L,x_R) \rightarrow \left \{  \begin{array}{l} (x_L\!+\!1,x_R\!+\!1), ~~ \textrm{with probability} ~ 1/4-\epsilon^2, \\
                                 (x_L\!-\!1,x_R\!-\!1), ~~\textrm{with probability} ~ 1/4-\epsilon^2, \\
                                 (-l,l), ~~\textrm{with probability} ~ (1/2+\epsilon)^2, \\
                                 (x_L\!-\!1,x_R\!+\!1), ~~\textrm{with probability} ~ (1/2-\epsilon)^2. \\
                                \end{array} \right .\label{dynamicsx2} 
\end{eqnarray}
Note that in the model under consideration, the dynamics of resetting is not imposed from outside (hence, it is unlike the scheme of exponential or power-law resetting discussed in the introduction), but that it emerges out of the chosen internal dynamics of the system. 

The quantity of interest here is the probability density $\mathcal{P}(x_L,x_R,t)$ of finding the particles at given positions $x_L$ and $x_R$ at time $t$. In order to study this quantity, it proves convenient to work with transformed co-ordinates $y \equiv (x_R + x_L)/2$ and  $z \equiv (x_R - x_L)/2$, which represent the center-of-mass and the (half) separation between the particles, respectively.  Following the aforementioned dynamical rules,  one can easily write down the Master equation for the time evolution of the probability density $P(y,z,t)$.  In the continuous space-time limit, this equation reads
\begin{equation}
D \nabla^2 P(y,z,t)+v\frac{\partial P(y,z,t)}{\partial z}+r(t)\delta(y)\delta(z-l)=\frac{\partial P(y,z,t)}{\partial t}, \label{diff}
\end{equation}
where we have $\nabla^2 \equiv \frac{\partial^2}{\partial y^2}+\frac{\partial^2}{\partial z^2}$, $v\equiv \lim_{\Delta x, \Delta t, \epsilon\rightarrow 0}\frac{2\epsilon\Delta x}{\Delta t}$, $D\equiv \lim_{\Delta x, \Delta t \rightarrow 0}\frac{(\Delta x)^2}{4\Delta t}$, and 
\begin{equation}
r(t) \equiv D\int_{-\infty}^\infty \rmd y'~\frac{\partial P(y',0,t)}{\partial z}\label{fdef}.
\end{equation}
In Eq.~(\ref{diff}), the first term on the left hand side represents the diffusive behavior of both $y$ and $z$, the second term is a drift in $z$ due to particles moving towards each other with a bias $v$, while the third term represents resetting to  $z=l$, $y=0$ with rate $r(t)$, with the latter to be determined by the particle dynamics itself.

Equation~(\ref{diff}) has its homogeneous part solved by the Green function $G(y,z,t)$ given by
\begin{eqnarray}
G(y,z,t)&=&\frac{1}{4\pi Dt}\Big[\exp\left(-\frac{\left(z-l+vt\right)^2+y^2}{4Dt}\right )\nonumber \\
&&-\exp\left (-\frac{\left(z+l+vt\right)^2+y^2}{4Dt}\right )\exp\left(\frac{vl}{D}\right)\Big]. \label{hsolution}
\end{eqnarray}
The solution to the full problem that includes the inhomogeneous term arising from resetting can then be written as
\begin{equation}
P(y,z,t)=G(y,z,t)+\int_{0}^{t}\rmd t'~G(y,z,t-t')r(t'). \label{fulltime}
\end{equation}
In the limit $t \to \infty$, the system attains a stationary state. In this limit, one has $G(y,z,t) \to 0$ and also the resetting rate function becoming a constant given by $r_\mathrm{st} \equiv \lim_{t \rightarrow \infty} r(t)$. One then obtains
\begin{equation}
P_\mathrm{st}(y,z)=\frac{r_{st}}{2\pi D}\rme^{-v(z-l)/2D}\left[K_0\left( \frac{v}{2D}\sqrt{(z-l)^2+y^2} \right ) - K_0\left( \frac{v}{2D}\sqrt{(z+l)^2+y^2} \right ) \right], \label{sst}
\end{equation}
where $K_0$ is the zeroth-order modified Bessel function of the second kind, and $r_\mathrm{st}$ is determined by using the conservation of probability $\int \rmd y \rmd z~P_\mathrm{st}(y,z)  =1$, giving $r_\mathrm{st}= v/l$. 

\begin{figure}[!ht]
\begin{center}
\begin{tabular}{cc}
(a) &(b) \\
\includegraphics[height=5cm]{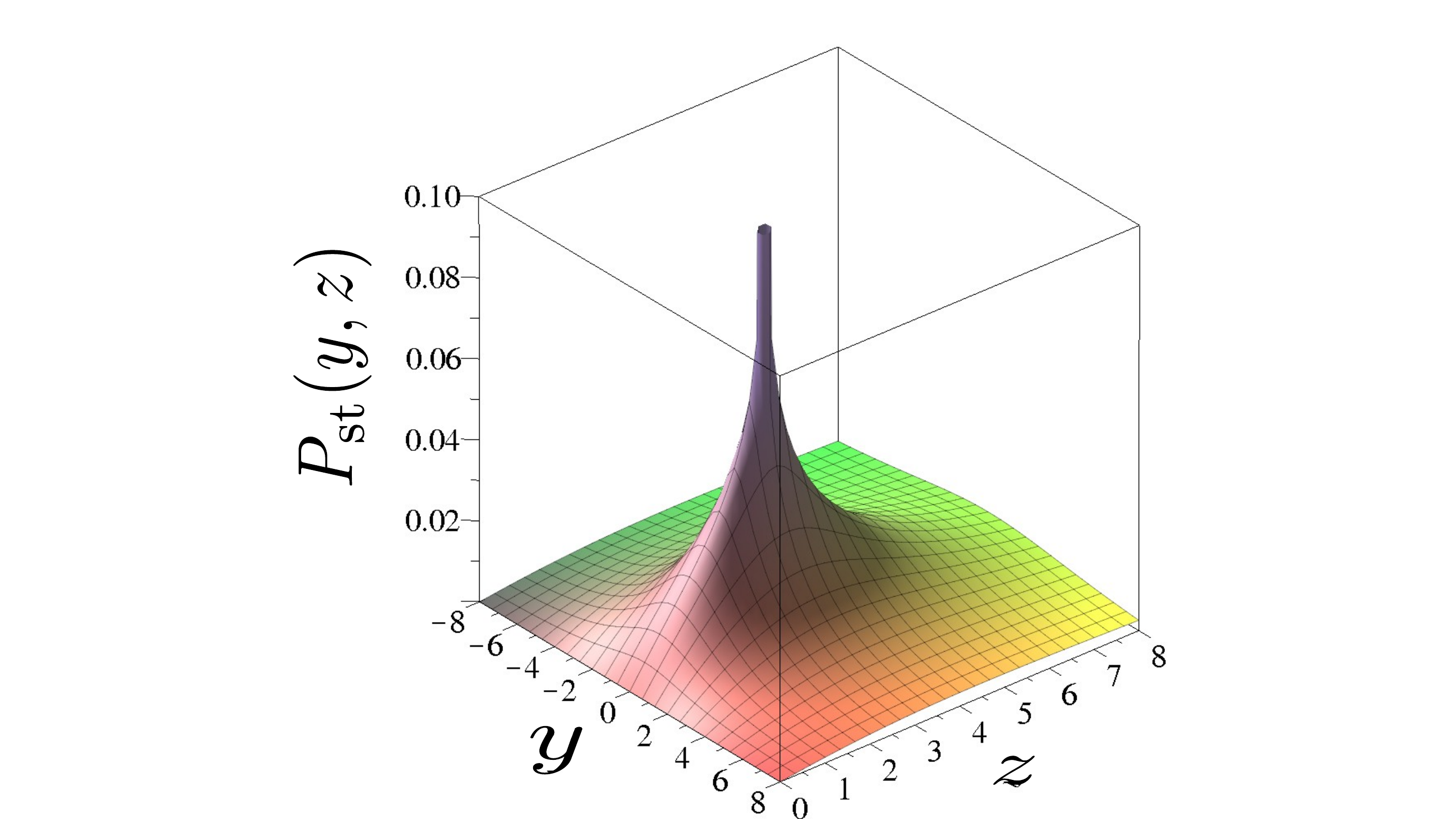} & \includegraphics[height=5cm]{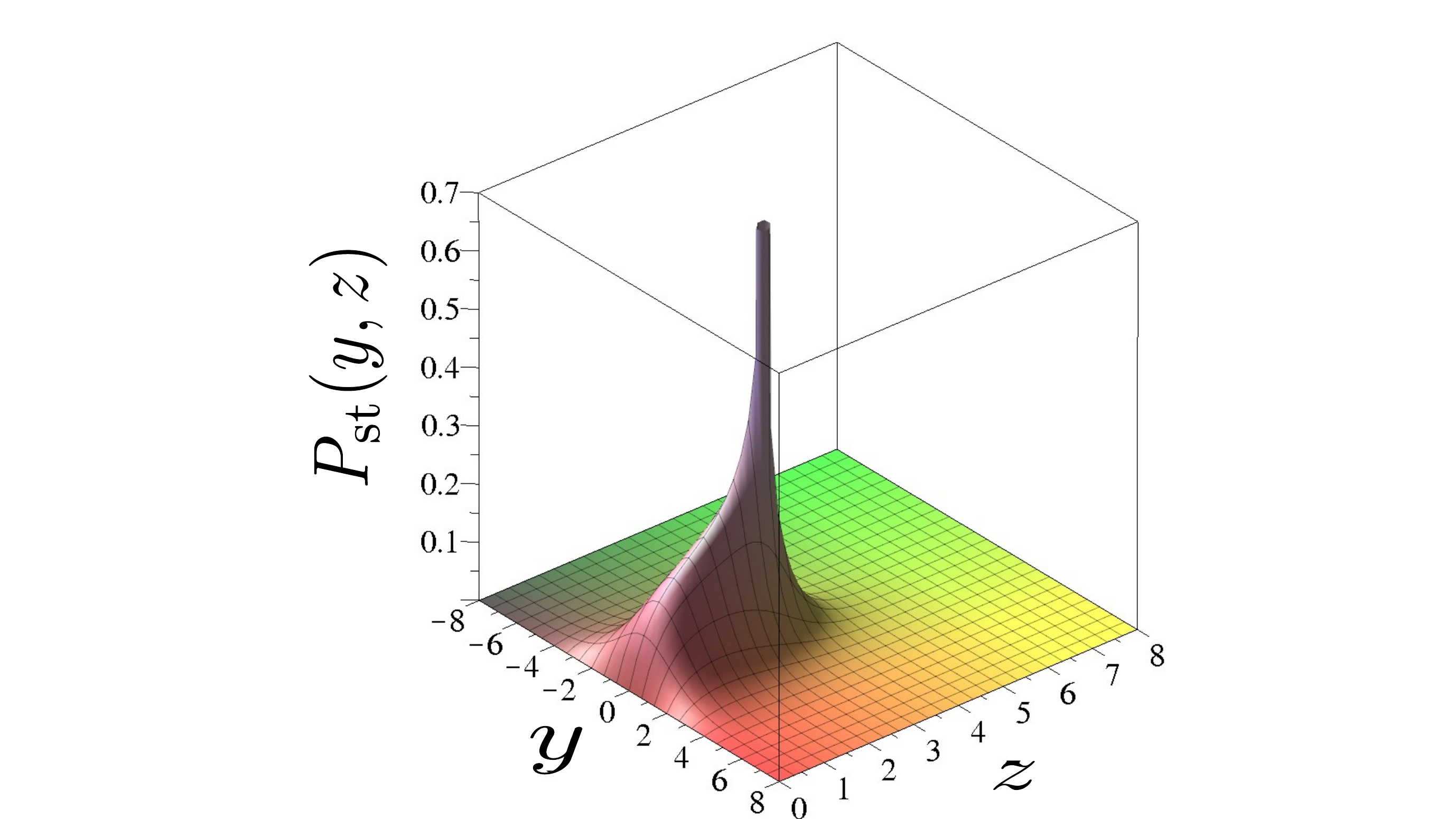}\\
& \\
(c) &(d) \\
\includegraphics[width=8cm]{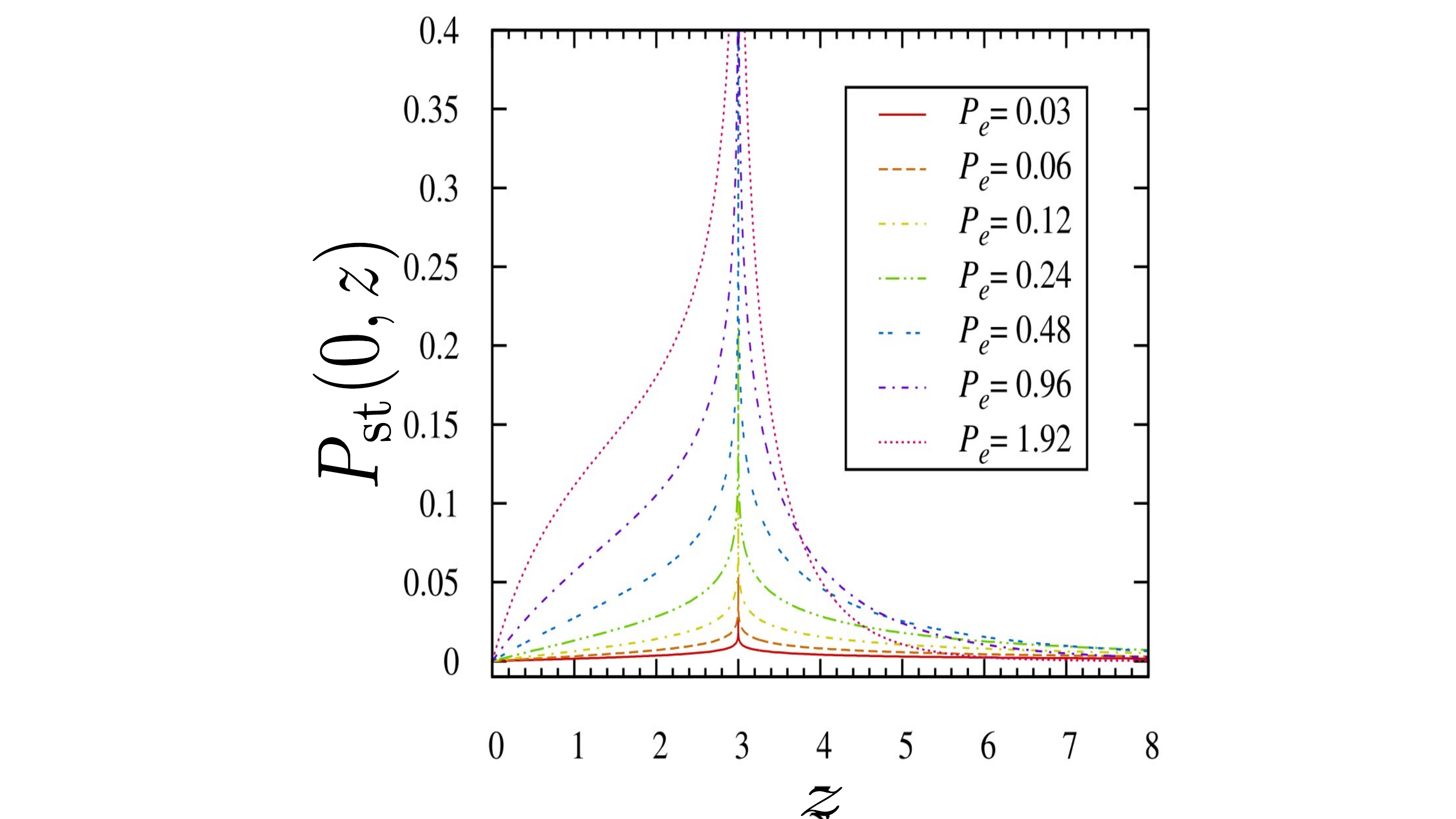} & \includegraphics[width=8cm]{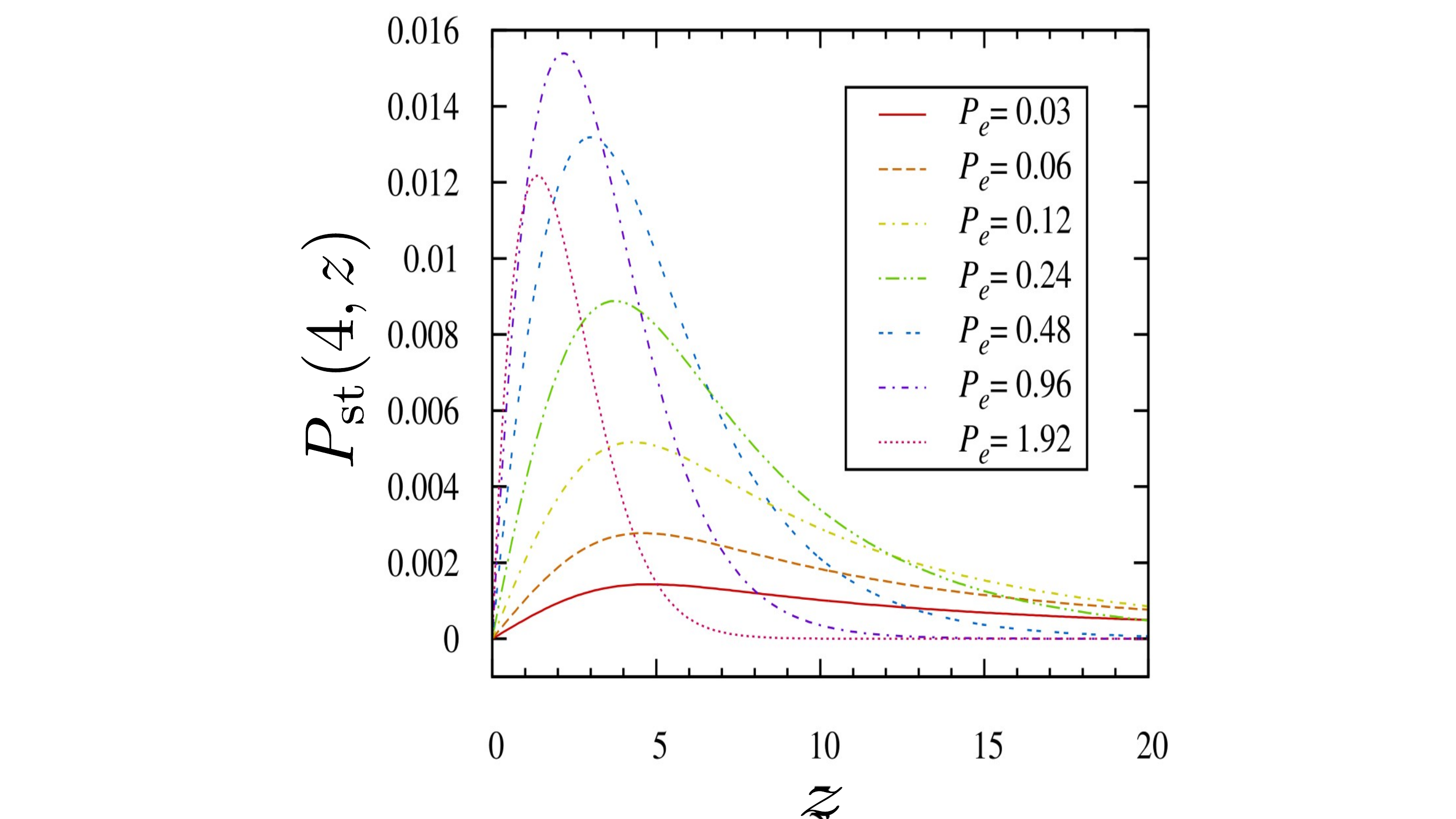}
\end{tabular}
\caption{For the problem of diffusing particles with an interacting reset mechanism studied in Ref.~\cite{Falcao:2017} and discussed in the text, see Section~\ref{subsubsec:long-range}, the figure shows the stationary probability density $P_\mathrm{st}(y,z)$ for different P\'eclet numbers, (a) $P_e=0.3$, (b) $P_e=3$. We also show the cross section of the probability density (c) for $y=0$, (d) for $y=4$ for different P\'eclet numbers. The figure is adapted from Ref.~\cite{Falcao:2017}. \label{fig1}}
\end{center}
\end{figure}

One next introduces the P\'eclet number 
\begin{equation}
P_e \equiv \frac{vl}{2D},
\end{equation}
which characterizes the importance of diffusion and convection in a biased diffusive process. Figure~\ref{fig1}, panels (a) and (b) show the stationary probability distribution for two different P\'eclet numbers, one small (panel (a)) and one large (panel (b)). A large P\'eclet number implies a strong drift, which in turn
promotes resetting, leading to a stronger clustering around the resetting point and a restricted spread along $z$. For small values of P\'eclet number, when diffusion is more important, the stationary-state distribution is almost symmetric around the resetting point, with a larger spread along $z$. It may be seen from Eq.~(\ref{sst}) and Fig.~\ref{fig1} that there is a logarithmic singularity in the distribution at $z=l$, $y=0$, which is the resetting point. Indeed, setting $y=0$ and $z= l +\epsilon$, and using $K_0\left(x\right)\sim -\ln \left(x/2\right)-\gamma_E$ for $0<x\ll1$ (here $\gamma_E$ is the Euler's constant), one gets for $|\epsilon| \ll 1$ that
\begin{eqnarray}
P_\mathrm{st}(0,l+\epsilon) \sim
\frac{r_\mathrm{st}}{2\pi D}\ln \left(\frac{4D}{v|\epsilon|}\right).
\label{stsin}
\end{eqnarray}

One can also evaluate the full time-dependent form for the resetting rate $r(t)$ as a series, which can then be used to look at the behavior of the system as it approaches the stationary state. To obtain this time dependence, one has to substitute Eq.~ (\ref{fulltime}) in Eq.~(\ref{fdef}) to obtain
\begin{equation}
r(t)=k(t)+\int \mathrm{d}t'~k(t-t')r(t'), \label{selff}
\end{equation}
where we have
\begin{equation}
k(t) \equiv \int_{-\infty}^{\infty} \rmd y'~\left .D\frac{\partial G(y',z,t)}{\partial z}\right |_{z=0}=\frac{l}{\sqrt{4D\pi t^3}}\exp\left(-\frac{(tv-l)^2}{4Dt} \right).
\end{equation}
One finally obtains
\begin{equation}
r(t)=\sum_{n=1}^{\infty}\frac{nl}{\sqrt{4\pi D t^3}}\exp\left(-\frac{(tv-nl)^2}{4Dt}\right )=
\sum_{n=1}^{\infty}r_n(t),
\label{fntdef}
\end{equation}
using which one can obtain the full time-dependent solution of Eq.~(\ref{diff}) as
\begin{equation}
P(y,z,t)=  G(y,z,t)+\int_{0}^{t} \rmd t'~G(y,z,t-t')
\sum_{n=1}^{\infty}r_n(t').
\label{fsol}
\end{equation}
For large P\'eclet number (small $D$/large $v$), an approximation for the sum 
(\ref{fntdef}) can be derived to yield
\begin{equation}
r(t)\simeq r_l(t)= \frac{v}{l}\left\{ 1+2R\exp\left(-\frac{4\pi^2tD}{l^2}\right )\cos\left( \frac{2\pi vt}{l}+\arctan \left(\frac{4D\pi}{lv} \right) \right)\right \}, \label{largevf}
\end{equation}
where we have $R \equiv \sqrt{1+(4D\pi/vl)^2}$. Thus, in the large P\'eclet number regime, and in the limit of large $Dt$, one finds that the  effective resetting rate decays as a damped harmonic oscillation towards the value $v/l$.

For small values of P\'eclet number, a good approximation is obtained by replacing the sum  $\sum_{n=1}^{\infty}r_n(t)$  by an integral:
\begin{equation}\fl
r(t)\simeq r_s(t)=\int_{0}^\infty \rmd n~ r_n(t)=\frac{v}{2l}+\frac{v}{2l}\textrm{Erf}\left(\frac{v}{2}\frac{\sqrt{t}}{\sqrt{D}}\right)+\frac{\sqrt{D}}{l\sqrt{\pi t}}\textrm{e}^{-\frac{tv^2}{4D}}. \label{smallvf}
\end{equation}
In this case, $r(t)$ does not have any oscillatory behavior, and $r_s(t)$ approaches the stationary value of $v/l$ monotonically from above.

To summarize, the problem studied in Ref.~\cite{Falcao:2017} is one in which the resetting dynamics is not imposed from outside, but which emerges from the internal dynamics of the system. Resetting in the model acts as a repulsive force acting against the inter-particle attractive interaction. 

\subsubsection{Diffusion with birth-death dynamics and resetting} 
\label{subsubsec:birth-death}
In Refs.~\cite{Silva:2018} and~\cite{Silva:2019} (see also Ref.~\cite{Silva:2022}), the authors have looked at the problem of diffusion with resetting in a system of interacting particles, with the inter-particle interactions inspired by birth-death processes studied in the field of population genetics. The model is defined as follows: there are $n$ indistinguishable particles, each one being of a specific type symbolized by a real number. Thus, the $n$ particles are represented by the set $\{x_1,x_2,\ldots,x_n\}$, with each of the $x_i$'s being a real number. At the initial time $t=0$, all of the types are set to the origin: $x_i=0~\forall~i$. The dynamics involves the type of particle $j$ changing independently of the others with the evolution operator $D \mathrm{d}^2/\mathrm{d}x_j^2$, where $D>0$ is a constant. In other words, $x_j$ performs a Brownian motion with diffusion constant $D$. In addition, at exponentially-distributed random times, with rate $r$, a pair $\{x_j,x_l\}$ chosen randomly undergoes a birth-death process: $x_j + x_l \to 2x_j$, supposing $j<l$ (death of $x_l$, birth of $x_j$). This step is thus akin to a reset move: with a constant rate $r$, the particle at $x_l$ is moved to $x_j$ if $j<l$, that is, there is a sudden relocation (exponential resetting) of the particle at $x_l$. Moreover, with a constant rate $\beta$, one has $x_j \to x'$, with $x'$ distributed according to a given probability distribution $\phi(x')$. One of the main results is that due to resetting, the system of $n$ particles has a stationary state in the limit of large times, in which the joint probability distribution at time $t$, given by $P_{n}(x_1,\ldots,x_n,t)$, has a time-independent form 
\begin{eqnarray}
&P_{n,\,\mathrm{st}}(x_1,\ldots,x_n)= \frac{2}{D(\sqrt{4\pi})^n}\int_{\mathbf{R}^n}\rmd z_1...\rmd z_n\nonumber \\
&\times\left( \frac{\sqrt{(x_1-z_1)^2+...+(x_n-z_n)^2}}{2\sqrt{\frac{n\beta+r {n\choose 2}}{D}}}\right)^{1-\frac{n}{2}}\nonumber \\
  &\times K_{\frac{n}{2}-1} \left(\sqrt{\frac{n\beta+r {n \choose 2}}{D}}\sqrt{(x_1-z_1)^2+..+(x_n-z_n)^2} \right)\nonumber \\
  &\times \left[\beta \sum_{j=1}^{n}\phi(z_j)P_{n-1,\,\mathrm{st}}(z_1,...,z_{j-1},z_{j+1},...,z_n) \right. \nonumber \\
  &+ \left. r \sum_{1 \leq j < l \leq n}\delta(z_l-z_j)P_{n-1,\,\mathrm{st}}(z_1,...,z_{l-1},z_{l+1},...,z_n) \right].
\label{silvaeq2}
\end{eqnarray}  
Here, $K_{\nu}$ is the modified Bessel function of the second kind of order $\nu$.

More detailed results are obtained for the two-particle case, for which an explicit expression for the probability density can be found. Using the following initial conditions: $P_1(x_1,0)=\delta(x_1)$ and $P_2(x_1,x_2,0)=\delta(x_1)\delta(x_2)$, one obtains
\begin{eqnarray}
  &P_{2,\,\mathrm{st}}(x_1,x_2)= \frac{1}{4\pi}\sqrt{\frac{\beta^3}{D^3}}\int_{\mathbf{R}}\rmd u~e^{-\sqrt{\frac{\beta}{D}}|u|}\nonumber \\
  &\times \left\{ \int \int_{\mathbf{R}^2}\rmd z_1 \rmd z_2~[\phi(z_2)\phi(z_1-u)+\phi(z_1)\phi(z_2-u)]\right.\nonumber \\
  &+ K_0 \left(\sqrt{\frac{2\beta+r}{D}}\sqrt{(x_1-z_1)^2+(x_2-z_2)^2} \right)\nonumber \\
  &+ \left. \frac{r}{\beta}\int_{\mathbf{R}}\rmd z_1~\phi(z_1-u)K_0 \left(\sqrt{\frac{2\beta+r}{D}}\sqrt{(x_1-z_1)^2+(x_2-z_1)^2} \right)  \right\}.
\label{silvaeq3}
\end{eqnarray} 
A particularly simple case of interest is $\phi(x)=\delta(x)$, when one obtains
\begin{eqnarray}
  &&P_{2,\,\mathrm{st}}(x_1,x_2)= \frac{1}{4\pi}\sqrt{\frac{\beta^3}{D^3}}\int_{\mathbf{R}}\rmd u~e^{-\sqrt{\frac{\beta}{D}}|u|}\nonumber \\
  &&\times \Big[K_0 \left(\sqrt{\frac{2\beta+r}{D}}\sqrt{(x_1-u)^2+x_2^2} \right)+ K_0 \left(\sqrt{\frac{2\beta+r}{D}}\sqrt{x_1^2+(x_2-u)^2} \right)\nonumber \\
  &&+\frac{r}{\beta}K_0 \left(\sqrt{\frac{2\beta+r}{D}}\sqrt{(x_1-u)^2+(x_2-u)^2} \right) \Big].
\label{silvaeq31}
\end{eqnarray} 
Another important quantity that characterizes the time dependent behavior of the system is the correlation of $X_1(t)$ and $X_2(t)$, where $X_i(t)$ is the position of the $i$-th particle at time $t$. This can be calculated for the case at hand as
\begin{eqnarray}
C(X_1(t),X_2(t)) &\equiv& \frac{\langle X_1(t) X_2(t)\rangle}{\mathrm{Variance}[X_1(t)]\mathrm{Variance}\,[X_2(t)]}\nonumber \\
&=&\frac{r}{2\beta+r}\left[1+\frac{\beta}{\beta+r}\frac{e^{-\beta t}}{1-e^{-\beta t}}(e^{-(\beta+r)t}-1) \right]. 
\end{eqnarray}
One can see that for any given, finite values of $r$ and $\beta$, the system settles to its stationary correlation value exponentially fast in time. For a fixed value of $\beta$, as $r$ is increased from $0$ to $\infty$, the correlation goes to one, which is expected since the limit $r \to \infty$ corresponds to one of the two particles vanishing all the time and re-appearing at the position of the other.
Thus, the sudden moves introduced via the parameters $\beta$ and $r$, which correspond to individual resetting and resetting involving a pair of particles, respectively, lead to a non-trivial stationary state, which can be characterized analytically.

\subsection{Many-particle interacting systems}
\label{subsec:manyparticles}
Here, we look at the effect of resetting in systems having a large number of interacting particles, with the number of particles typically scaling with the system size. The dynamics of the system is described in terms of the evolution of microstates or in terms of mesoscopic variables such as the density of particles. 

We begin this section on multiple-interacting particles with review of two works that dealt with a paradigmatic model of nonequilibrium statistical mechanics, the asymmetric simple exclusion process (ASEP)~\cite{Mallick:2015}. The model involves particles that move on a lattice and interact with each other via hard-core exclusion, thus allowing a maximum of one particle occupying a given site. The symmetric simple exclusion process (SSEP) on a one-dimensional periodic lattice, whereby particles attempt to move to any one of the nearest-neighbour sites with equal probability, has an equilibrium stationary state in the long-time limit. This state has a homogeneous density of particles across the lattice, and the net current across any bond is zero on the average. The review in Section~\ref{subsubsec:symmetricexclusion} looks at the possibilities arising out of resetting such a system into atypical configurations, namely, those in which all the particles are bunched up on one side of the system. The SSEP dynamics between two successive resets tends to drive the system towards a homogeneous-density state, while a reset disrupts this process, eventually leading to an NESS. The authors have characterized the novel stationary-state behavior arising out of the resetting dynamics, by analyzing the density profile and the current in the system.

The totally asymmetric simple exclusion process (TASEP) in one dimension, wherein the particles attempt to move to only one of the nearest-neighbour sites (say the one to the right) shows a very interesting NESS. In particular, in the presence of open boundaries with particles entering at one end and leaving at the other end, the system can exist in three different phases depending on the input and exit rates. This model also has many real-world applications such as in modelling protein synthesis via mRNA-translation. In Section~\ref{subsubsec:asymmetricexclusion}, we review a work investigating the effects of stochastic resetting on a TASEP with open boundaries in one dimension. The repeated resetting to an initial, empty lattice configuration leads to an inhomogeneous density profile, which can be captured efficiently via a combination of an approximate mean-field calculation and numerics.

The symmetric as well as the asymmetric versions of the exclusion process may be mapped to a discrete interface in one dimension. For example, the occupied/empty sites in the exclusion process can be viewed as locally-upward/locally-downward slopes in the equivalent interface model, and particle motion to nearest-neighbour sites may then be linked to the overturning of a local hill/valley of the interface. The behavior of the interface height and its fluctuations, especially when the interface dynamics is interspersed with stochastic resetting at random times, is interesting and non-trivial, and is explored in detail in Section~\ref{subsubsec:interface} for two paradigmatic continuum models of interface dynamics in one dimension, the Edwards-Wilkinson (EW) and the Kardar-Parisi-Zhang (KPZ) model.

The coagulation-diffusion process is a representative nonequilibrium model that has been extensively studied in the literature. Here, one considers hard-core particles occupying the sites of an infinite lattice in one dimension, which move diffusively between the sites such that a particle disappears as soon as it moves to a site that is already occupied by another particle. In Section~\ref{subsubsec:coagulation}, we discuss the effects of stochastic resets to an initial configuration at a constant rate in the one-dimensional coagulation–diffusion process, showing in particular the emergence of an NESS exhibiting a nontrivial behavior of the particle density.  

As opposed to the SSEP and the TASEP, the zero-range process (ZRP) allows any number of particles to occupy a site. On a one-dimensional periodic lattice, the dynamics involves biased hopping of particles to nearest-neighbour sites with a rate that depends on the occupancy at the departure site. At long times, the system reaches an NESS.  For certain classes of the hop rates, on tuning the particle density across a critical threshold, the system exhibits a very interesting nonequilibrium phase transition, from a disordered phase with uniform average density, to a condensed phase in which a finite fraction of particles accumulates on a single site. The ZRP applies to a wide variety of situations, from studying traffic flow, shaken granular gases, polymer dynamics, to addressing the theoretical question of formulating a criterion for phase separation in one-dimensional driven systems. In Section~\ref{subsubsec:zrp}, we consider the effects of resetting in a non-conserving ZRP in which particles are added to the system with a site-dependent creation rate, while they are annihilated with a uniform annihilation rate, and moreover, the system resets to an empty lattice at a constant rate.

The Ising model is a paradigm in the whole of statistical mechanics, be it equilibrium or out of equilibrium. This simple model of spins on a lattice, though originally introduced in the context of magnetism, has found a broad range of applications in systems as varied as phase separation, spin glasses, voting patterns and the activity of neurons. In Section~\ref{subsubsec:Ising}, we review a work that reports on the NESS attained in the Ising model when driven away from thermal equilibrium at temperature $T$ by a stochastic resetting protocol in which the magnetization of the system is at a constant rate reset repeatedly to its fixed initial value. It is observed  that the dynamics in the stationary state exhibits a very rich phase diagram in the $(T,r)$-plane, which can be studied via an exact analysis in one dimension and via scaling arguments and simulations in two dimensions. 

\subsubsection{Stochastic resetting of symmetric simple exclusion process} 
\label{subsubsec:symmetricexclusion}
Here, we review papers \cite{Basu:2019} and \cite{Sadekar:2020}, where the effect of resetting is studied on a SSEP in one dimension. The model consists of a one-dimensional periodic lattice of $L$ sites, each of which may be occupied by at most one particle. There are a total of $N=L/2$ particles in the system. The occupancy at the $j$-th site is given by $n_j=0,1$ with $j=0,1,2,\ldots,L-1$. A configuration $\mathcal{C}$ of the system is specified by the value of $n_j$ for every site: $\mathcal{C}\equiv\{n_j\}$. The system evolves
according to the following stochastic dynamics: Given a configuration
$\mathcal{C}$ at time $t$, in the ensuing infinitesimal time
interval $\mathrm{d}t$, one of the following events may occur: (i) The configuration $\mathcal{C}$ resets to either of two given
configurations $\mathcal{C}_1$ and $\mathcal{C}_2$ with probabilities $r_1 \mathrm{d}t$ and $r_2 {\rmd}t$ respectively, or, (ii) With the complementary probability $1-r \mathrm{d}t$, with $r \equiv r_1+r_2$, a randomly-chosen particle occupying a site attempts
to hop with equal probability to either the left or the right neighbouring site,
with the hop being successful only if the destination site is empty. Note that the reset dynamics under consideration is an example of exponential resetting discussed in the introduction. The configuration $\mathcal{C}_1$  ($\mathcal{C}_2$) corresponds to particles occupying the whole of the left-half (right-half) of the lattice,
with the other half completely empty:
\begin{eqnarray}
\mathcal{C}_1 =\{n_j\};~~n_j=
\left\{
\begin{array}{ll}
               1 & \mbox{for $0 \le j \le L/2-1$}, \\
               0 & \mbox{otherwise},
               \end{array}
        \right. \\ 
        \mathcal{C}_2 =\{n_j\};~~n_j=
\left\{
\begin{array}{ll}
               0 & \mbox{for $0 \le j \le L/2-1$}, \\
               1 & \mbox{otherwise}.
               \end{array}
        \right. 
\end{eqnarray}
For $r_1,r_2 = 0$, we have the usual SSEP without resetting.

Given the above dynamics, one may ask for the conditional probability
$\mathcal{P}(\mathcal{C},t|\mathcal{C}_0,0)$ for the system to be found in configuration $\mathcal{C}$ at time $t>0$, given an initial configuration $\mathcal{C}_0$ at time $t=0$. The probability $\mathcal{P}(\mathcal{C},t|\mathcal{C}_0,0)$ satisfies a renewal equation of a form that may be derived by using Eq.~(\ref{eq:renewal-basic0}), as 
\begin{equation}
\mathcal{P}(\mathcal{C},t|\mathcal{C}_0,0) =e^{-rt}\mathcal{P}_0(\mathcal{C},t|\mathcal{ C}_0,0)+r \int_0^t \mathrm{d}\tau~e^{-r\tau}\left[\tilde{\alpha} \mathcal{P}_0(\mathcal{C},t|\mathcal{C}_1,t-\tau)+(1-\tilde{\alpha})\mathcal{P}_0(\mathcal{C},t|\mathcal{C}_2,t-\tau)\right], 
        \label{eq:renewal-P}
\end{equation}
with $\tilde{\alpha} \equiv r_1/(r_1+r_2)$. Here, $\mathcal{P}_0(\mathcal{C},t''|\mathcal{C}_a,t')$ is the conditional probability for the system to be found in configuration $\mathcal{C}$ at time $t''>t'$, in the absence of resetting and conditioned on the
initial configuration $\mathcal{C}_a$ at time $t'$ ($a=0,1,2$). The first term on the right hand side accounts for the event that the system has evolved freely, i.e., without a single resetting, in the time interval $[0,t]$. The second and third terms on the right hand side account for all events in which at time $t$, the time elapsed since the last reset lies between $\tau$ and $\tau+\mathrm{d}\tau;~~\tau \in [0,t]$, with free evolution from the instant of last reset up to time $t$. The second
(third) term corresponds to the last reset being to configuration $\mathcal{C}_1$ ($\mathcal{C}_2$). 

Multiplying Eq.~(\ref{eq:renewal-P}) by $n_j$ and summing over all configurations, we can calculate the average density $\rho(j,t)$ at site $j$: $\sum_{\{n_j\}} n_j\mathcal{P}(\mathcal{C},t|\mathcal{C}_0,t-\tau)=\rho(j,\tau)$. This density then satisfies the renewal equation:
\begin{eqnarray}
        \rho(j,t)=e^{-rt}\rho_0^{(0)}(j,t)+r\int_0^t {\rm
        d}\tau~e^{-r\tau}\left[\tilde{\alpha}
        \rho_0^{(1)}(j,\tau)+(1-\tilde{\alpha})\rho_0^{(2)}(j,\tau)\right],
        \label{eq:renewal-rho}
\end{eqnarray}
where $\rho_0^{(a)}(j,t)$ is the average density at site $j$ at time
$t>0$, with the initial configuration
$\mathcal{C}_a$ at time $t=0$, and in the absence of resetting. 

In order to solve for $\rho(j,t)$ from Eq.~(\ref{eq:renewal-rho}), we need
to specify the initial configuration $\mathcal{C}_0$. In Ref.~\cite{Sadekar:2020}, the initial configuration has been chosen to be the one prepared by sampling the two configurations $\mathcal{C}_1$
and $\mathcal{C}_2$ with equal probability: $\rho(j,0)=(1/2)[\Theta(L/2-1-j)+\Theta(j-L/2)]$. Here, $\Theta(m)$ is the
Heaviside Theta function: $\Theta(m) = 1$ for $m \ge 0$ and zero
otherwise. The densities $\rho_0^{(0)}(j,t)$, $\rho_0^{(1)}(j,t)$ and $\rho_0^{(2)}(j,t)$ are the result of evolving the system under the usual SSEP dynamics with the initial configurations $\mathcal{C}_0$, $\mathcal{C}_1$ and $\mathcal{C}_2$ respectively. These are given by~\cite{Basu:2019,Sadekar:2020} 
\begin{equation}
        \rho_0^{(a)}(j,t) = \frac{1}{L} \sum_{n=0}^{L-1} e^{-\mathrm{i}
        \frac{2 \pi n j}{L}}e^{-\lambda_n t}\widetilde{\phi}^{(a)}(n),
\end{equation}
with $\lambda_n \equiv 2(1-\cos 2\pi n/L)$, $\widetilde{\phi}^{(0)}(n) = \frac{L}{2}\delta_{n0}$, $\widetilde{\phi}^{(a)}(0)=\frac{L}{2};\, a=1,2,$ and for $n \neq 0$
\begin{eqnarray}
 ~\widetilde{\phi}^{(2)}(n)=-\widetilde{\phi}^{(1)}(n),~\widetilde{\phi}^{(1)}(n) = \left\{ 
\begin{array}{cl}
        1+\mathrm{i}\cot \frac{{\pi}n}{L}  \quad \textrm{for}~ n=1,3,5,...\cr
0    \quad \textrm{for} \;\; \textrm{even}~n \ge 2.                          
\end{array}
\right.
\end{eqnarray}
Using these expressions in Eq.~(\ref{eq:renewal-rho}), one
gets~\cite{Sadekar:2020}
\begin{equation}
        \rho(j,t)=\frac{1}{2} + \frac{1}{L}\left[ \displaystyle
        \sum_{n=1,3}^{L-1}  \left(\frac{r_1 - r_2 }{\lambda_n + r}\right)\left(1+\mathrm{i}\cot\frac{\pi n}{L}\right)
        e^{-\mathrm{i}\frac{2{\pi}nj}{L}} (1-e^{-(\lambda_n + r)t})\right].
        \label{eq:density}
\end{equation}
The stationary profile is obtained by taking the limit $t \to \infty$ in
Eq.~(\ref{eq:density}), and one gets~\cite{Sadekar:2020}
\begin{eqnarray}
\rho(j)=\frac{1}{2}+\frac{1}{L}\left[ \displaystyle \sum_{n=1,3}^{L-1}
        \left(\frac{r_1 - r_2 }{\lambda_n + r}\right)\left(1+{\rm
        i}\cot\frac{\pi n}{L}\right) e^{-{\rm
        i}\frac{2{\pi}nj}{L}} \right]. \
        \label{eq:density-stationary}
\end{eqnarray}
The above exercise of finding the time-dependent and the stationary density profile has been carried out in Ref.~\cite{Basu:2019} for the situation when the initial configuration is $\mathcal{C}_1$, which is also
taken to be the configuration to which the system resets. 

When the two resetting rates are equal, i.e.,
$r_1=r_2$, Eq.~(\ref{eq:density-stationary}) implies that the stationary
density profile is flat for all values of $r=2r_1$. A flat density
profile is also observed in the stationary state of the usual SSEP without resetting. Even though the
density profile is the same in both cases, resetting generically leads to an NESS, as opposed to
the equilibrium stationary state seen in the SSEP dynamics without reset.  

\begin{figure}
\includegraphics[width=8cm]{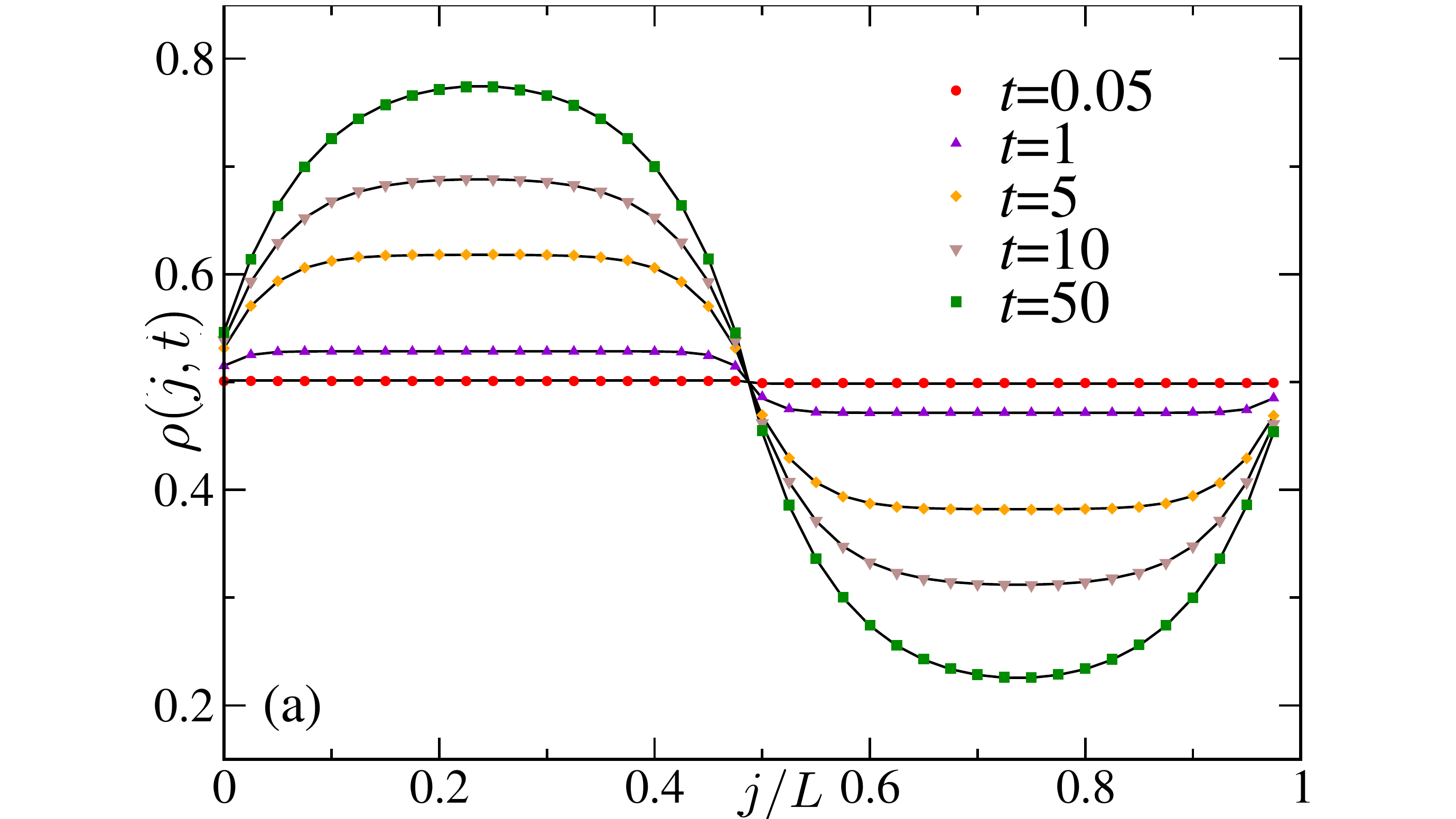}
\includegraphics[width=8cm]{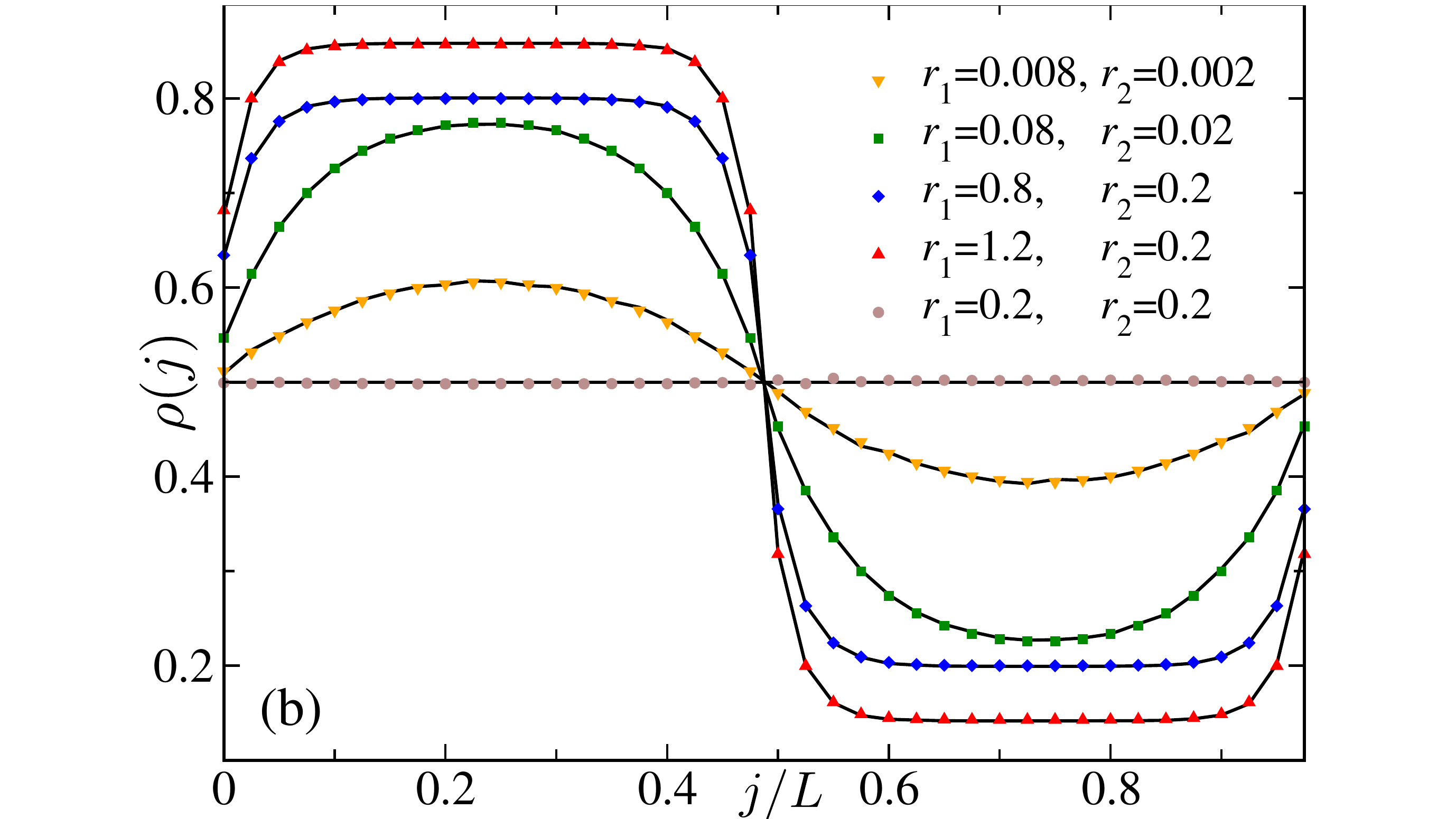}
\caption{For the model of symmetric simple exclusion process subject to stochastic resetting as described in Ref.~\cite{Sadekar:2020} and Section~\ref{subsubsec:symmetricexclusion}, the figure shows numerical (symbols) and analytical (lines) results for the time-dependent density profile (panel (a)) and the stationary density profile (panel (b)). For panel (a), we have $r_1=0.08$ and $r_2=0.02$. The system size is $L=40$. The analytical results correspond to Eqs.~(\ref{eq:density}) and~(\ref{eq:density-stationary}). The initial configuration is the one prepared by sampling the two configurations $\mathcal{C}_1$ and $\mathcal{C}_2$ with equal probability: $\rho(j,0)=(1/2)[\Theta(L/2-1-j)+\Theta(j-L/2)]$, see text. The figure is adapted from Ref.~\cite{Sadekar:2020}.}
 \label{fig:rho}
\end{figure}

Figure \ref{fig:rho}, panels (a) and (b) show numerical and analytical results
for respectively the time-dependent and
the stationary density profile, for $L=40$. The numerical results are obtained by
simulating the dynamics of the model, while the analytical ones are
given by Eqs.~(\ref{eq:density}) and~(\ref{eq:density-stationary}). A very good match between theory and numerics is observed. 

\begin{figure}[t]
\centering
\includegraphics[width=15.5 cm]{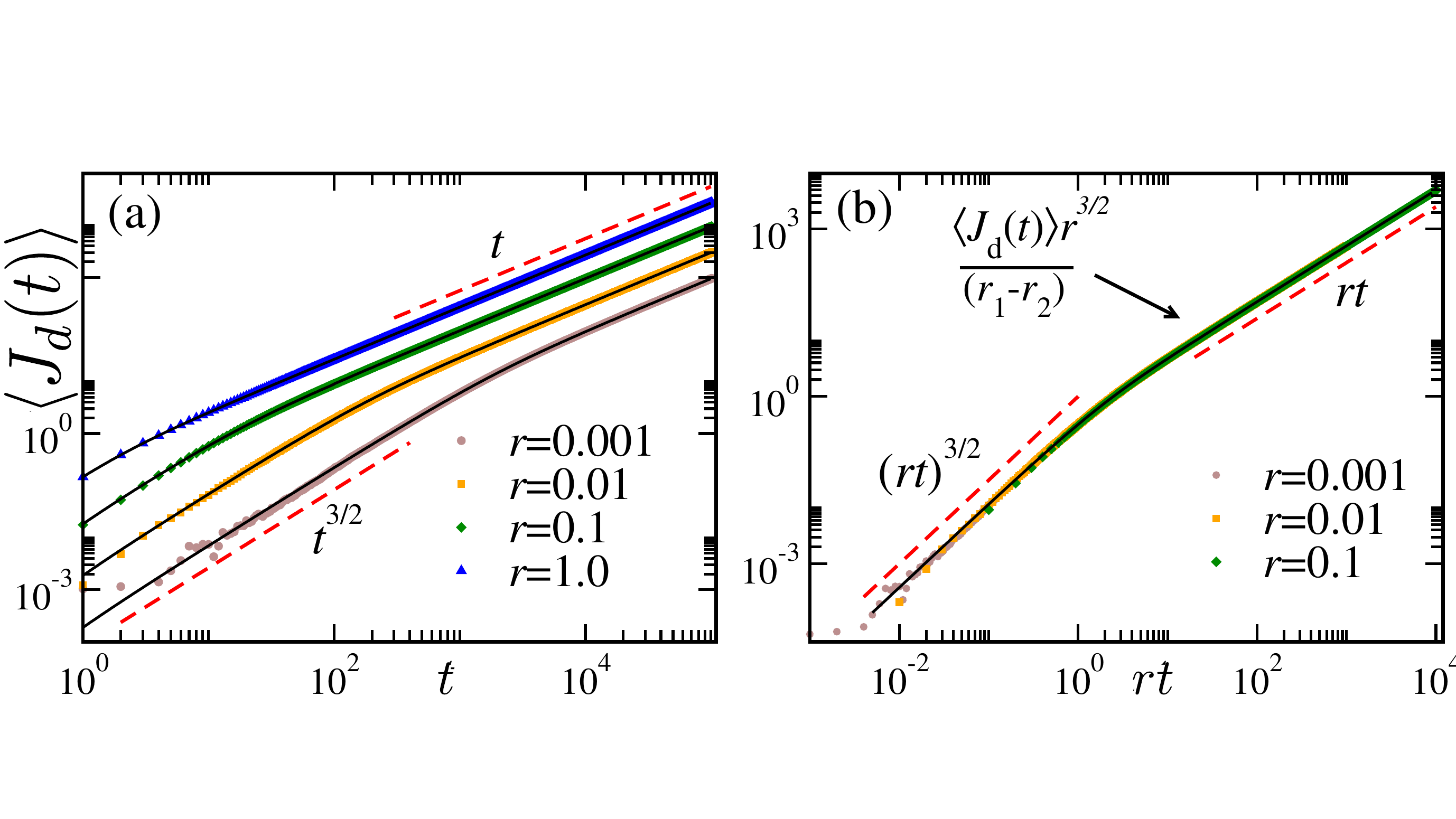}
\caption{For the model of symmetric simple exclusion process subject to stochastic resetting as described in Ref.~\cite{Sadekar:2020} and Section~\ref{subsubsec:symmetricexclusion}, the figure shows the average diffusive current $\langle J_d(t)\rangle$: (a) Plot of $\langle J_d (t)\rangle$ as a function of time $t$ for different values of $r$ and $\tilde{\alpha}=0.8.$ The solid lines correspond to the exact results (see Eqs. (\ref{eq:avJdif_q}) and (\ref{eq:avJdLongTime})) and the symbols correspond to the data obtained from numerical simulations. (b) Scaling collapse of the data in (a) according to Eq.~(\ref{eq:P_JdFirstMoment}). The solid line corresponds to the scaling function implied by Eq.~(\ref{eq:P_JdFirstMoment}). The numerical simulation is done on a lattice of size $L=1000$ and averaged over more than $10^7$ trajectories; we have $\tilde{\alpha}=0.8$. The initial configuration is the one prepared by sampling the two configurations $\mathcal{C}_1$
and $\mathcal{C}_2$ with equal probability: $\rho(j,0)=(1/2)[\Theta(L/2-1-j)+\Theta(j-L/2)]$, see text. The figure is adapted from Ref.~\cite{Sadekar:2020}.}
 \label{fig:diff-cur}
\end{figure}

It is interesting to analyze the current across the central bond $(L/2-1,L/2)$ of the system. The net number of particles crossing this bond in the interval $[t,t+\mathrm{d}t]$ is a stochastic variable, and one takes its average to calculate the current. The average current is denoted by $\langle j_d(t)\rangle$, wherein the angular brackets denote averaging with respect to the probability $\mathcal{P}(\mathcal{C},t|\mathcal{C}_0,0)$.  Since a current will be registered across the bond only when a particle occupying the $(L/2-1)$-th site hops to a vacant $(L/2)$-th site or when a particle occupying the $(L/2)$-th site hops to a vacant $(L/2-1)$-th site, we have  
\begin{eqnarray}
\langle j_d(t) \rangle &= & \left\langle n_{\frac{L}{2}-1}(1-n_{\frac L2})\right\rangle - \left\langle (1- n_{\frac {L}{2}-1})n_{\frac L2}\right\rangle \cr
&= &\rho\left(\frac L2 -1,t\right) - \rho\left(\frac L2,t \right).
\end{eqnarray} 
Using Eq.~(\ref{eq:density}) then gives~\cite{Sadekar:2020}
\begin{eqnarray}
\langle j_d(t) \rangle = \frac {2(r_1-r_2)}{L} \sum_{n=1,3}^{L-1}\bigg[\frac {1-e^{-(r+\lambda_n)t}}{r+\lambda_n}\bigg].
\end{eqnarray}
The average value of the time-integrated current, $J_{d}(t) \equiv \int_0^t \rmd s~ j_d(s)$, can be obtained by using the above equation. In particular, in the continuum limit $L\to \infty$, effecting the transformation $q=2\pi n/L$, one gets~\cite{Sadekar:2020}
\begin{eqnarray}
\langle J_{d}(t)\rangle = (r_1-r_2)\left[\frac{t} {\sqrt{r(4+r)}} - \frac{(2+r)}{r(4+r)} + \int_0^{2 \pi} \frac {\mathrm{d} q}{2 \pi}\frac{e^{-(\lambda_q+r)t}}{(\lambda_q+r)^2}\right].  \label{eq:avJdif_q}
\end{eqnarray}
At long times, the linear term dominates, and we have~\cite{Sadekar:2020}
\begin{eqnarray}
\langle J_{d}(t\to \infty)\rangle = \frac{(r_1-r_2)t} {\sqrt{r(4+r)}}.  \label{eq:avJdlong}
 \label{eq:avJdLongTime}
\end{eqnarray}
This linear behavior, verified in numerics, see Fig.~\ref{fig:diff-cur}, is a simple consequence of the fact that the system possesses a well-defined NESS,  as expounded above. This implies a time-independent average current in the stationary state, i.e., $\langle j_d(t) \rangle$ is a constant at long times. Consequently,  the time-integrated current grows linearly with time at long times.  For short times $t \ll r^{-1}$, it can be shown that the behavior of $\langle J_d)t) \rangle$ is initial-condition dependent: While $\langle J_d(t) \rangle \sim \sqrt{t}$ if the system starts with the initial condition $\mathcal{C}_1$, a super-linear behavior $\langle J_d(t) \rangle \sim t^{\frac{3}{2}}$ is seen when the initial condition is chosen with equal probability to be either $\mathcal{C}_1$ or $\mathcal{C}_2$.

\begin{figure}[t]
\centering
\includegraphics[width=12cm]{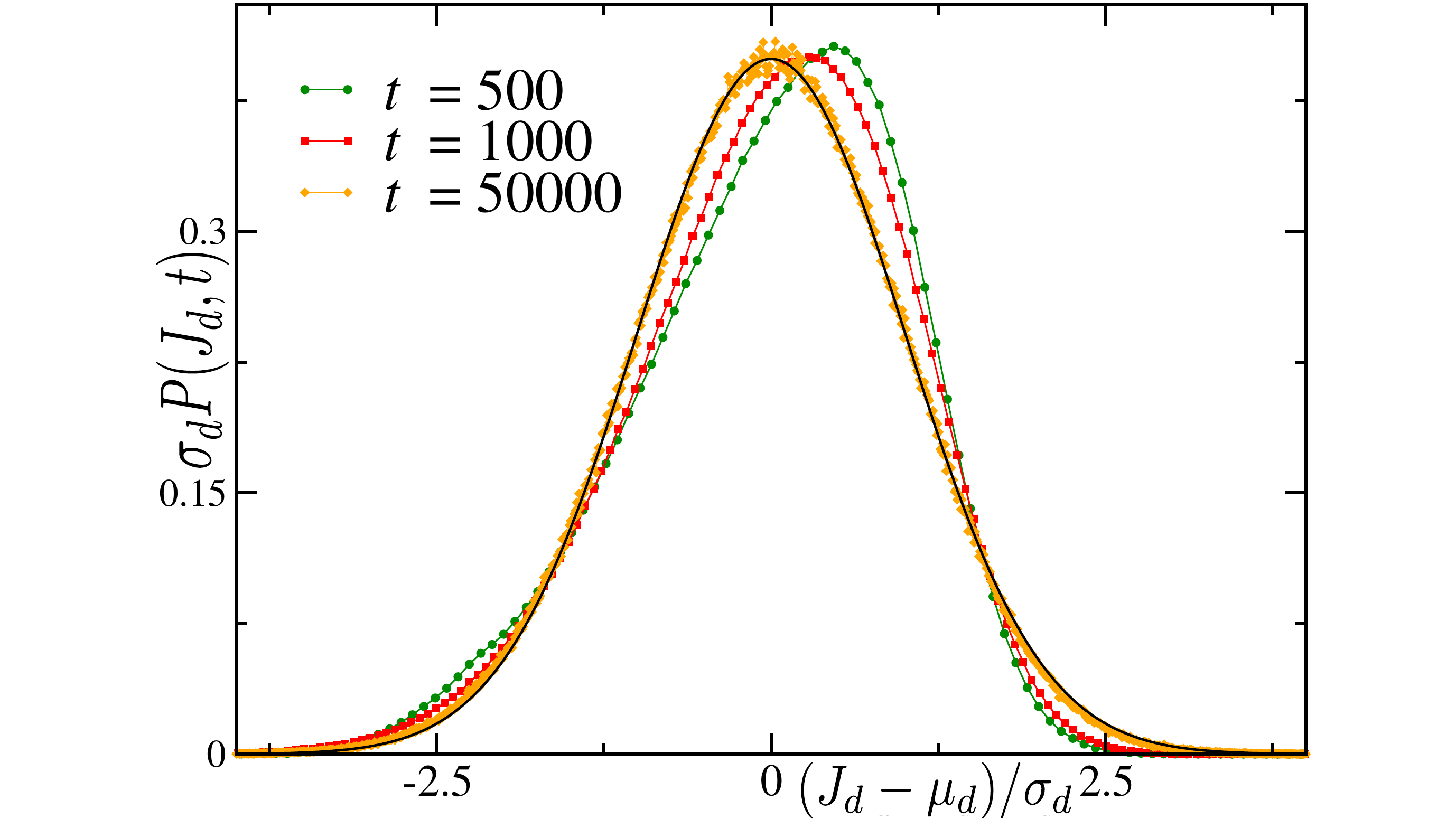}
\caption{For the model of symmetric simple exclusion process subject to stochastic resetting as described in Ref.~\cite{Sadekar:2020} and Section~\ref{subsubsec:symmetricexclusion}, the figure shows the probability distribution of the diffusive current $J_d$. The symbols correspond to the data obtained from numerical simulations and the black solid line refers to the analytical result for large $t$ [see Eq.~(\ref{eq:PJd_Gauss})], with $\mu_d \equiv \langle J_d\rangle$. Here, $r_1=0.008$ and $\tilde{\alpha}=0.8$. The system size is $L=1000$. The initial configuration is the one prepared by sampling the two configurations $\mathcal{C}_1$
and $\mathcal{C}_2$ with equal probability: $\rho(j,0)=(1/2)[\Theta(L/2-1-j)+\Theta(j-L/2)]$, see text. The figure is adapted from Ref.~\cite{Sadekar:2020}.}
\label{fig:P_Jd}
\end{figure}

Fluctuations in the time-integrated current can be characterized using the quantity $\langle J_d^2 \rangle$. For small $r$ and large $t$, Eq.~(\ref{eq:avJdLongTime}) yields 
\begin{eqnarray}
  \langle J_d(t) \rangle \simeq \frac{(r_1-r_2)}{\sqrt{r^3}}\left[\frac{e^{-rt}\sqrt{rt}}{2\sqrt{\pi}}+\frac{1}{4}(2rt-1)\mathrm{Erf}(\sqrt{rt})\right],
\label{eq:P_JdFirstMoment}
\end{eqnarray}
giving $\langle J_d(t)\rangle \simeq (r_1-r_2)t/(2\sqrt{r})$ for $t\gg 1/r$ and $\langle J_d(t)\rangle \simeq 3(r_1-r_2)t^{3/2}/(2\sqrt{\pi})$ for $t \ll 1/r$. This implies as a function of $t$ and at a fixed $r$ a cross-over of the quantity $\langle J_d(t)\rangle r^{3/2}/(r_1-r_2)$ from a behavior $\sim (rt)^{3/2}$ to a linear behavior $\sim rt$, as can be seen compared with the numerics in Fig.~\ref{fig:diff-cur}. One obtains also the expression
\begin{eqnarray}
\langle J_d^2(t) \rangle & =& \frac{(r_1-r_2)^2}{2r^3}+e^{-rt}\left[\frac{b}{2}\sqrt{\frac{t}{\pi}}-\frac{(r_1-r_2)^2}{2r^3}\right]+t\left[\frac{1}{\pi}-\frac{(r_1-r_2)^2}{2r^2}\right]\nonumber \\
&+&\frac{(r_1-r_2)^2t^2}{4r}+\frac{b}{4\sqrt{r}}(1+2rt)\mathrm{Erf}(\sqrt{rt}),
\label{eq:P_JdSecondMoment}
\end{eqnarray}
with $b \equiv 1-1/\sqrt{2}$.
The variance $\sigma^2_{d}(t) =\langle J_d^2(t) \rangle -\langle J_d(t) \rangle^2$ in the long-time limit may then be obtained as
\begin{eqnarray}
\sigma_{d}^2(t) \simeq t \bigg[\frac{1}{\pi} + \frac {\sqrt r}{2}\bigg(1- \frac {1}{\sqrt{2}}-\frac{(r_1-r_2)^2}{4r^2}\bigg)\bigg].
\label{eq.sigma_J_d}
\end{eqnarray}
While the complete time dependence of the distribution $P(J_d,t)$ is not obtained in a closed form, it can be argued using the central limit theorem that the distribution will be a Gaussian in the limit of long times $t \gg 1/r$: 
\begin{equation}
P(J_d,t) \simeq \frac{1}{\sqrt{2 \pi \sigma_d^2(t)}}\exp\left[-\frac{(J_d-\langle J_d(t)\rangle)^2}{2 \sigma_d^2(t)}\right]. \label{eq:PJd_Gauss}
\end{equation}
The authors have checked the above result against numerics, see Fig.~\ref{fig:P_Jd}. 

A particular case explored in more detail is the $r_1=r_2=r/2$ case. Here, the current across the central bond is zero and the density profile is flat, similar to the equilibrium state of the SSEP without  reset. There is a significant difference in the behavior of the equilibrium state without reset and the stationary state obtained with reset which the authors have explored. In the stationary state with resetting, while the average density profile is flat, the typical density profile is highly asymmetric with most particles being situated on the left or on the right side of the lattice. On the other hand, the equilibrium state (no resetting) shows a homogeneous distribution of particles in a typical configuration. The difference shows up in other quantities as well, for example, while the leading behavior of the second moment and the fourth moment of current are $\sqrt{t}$ and $t$, respectively, in the equilibrium case, the same quantities scale as $t$ and $t^2$, respectively, in the NESS attained when resetting is introduced. Thus, one observes that the fluctuations grow much faster in the presence of resetting. 

\subsubsection{Stochastic resetting of the totally asymmetric simple exclusion process}  \label{subsubsec:asymmetricexclusion}

In Ref.~\cite{Karthika2020}, the authors have studied the effect of stochastic resetting to an initial, empty lattice configuration in the totally asymmetric simple exclusion process (TASEP) in one dimension with open boundaries. The system is defined in the following way: hard-core particles occupy sites of a one-dimensional periodic lattice of sites labelled $1,2,\ldots,L$. The dynamics involves these particles hopping with rate unity to the right neighbouring site, provided the destination site is empty. In addition, a particle is input into site $1$ with rate $\alpha>0$, provided the site is empty; also, if the site $L$ is occupied, it is emptied with rate $\beta>0$. The system without resetting exists in the stationary state in three phases depending on the values of $\alpha$ and $\beta$: (i) for $\alpha <\beta$ and $\alpha <1/2$, the system is in a low-density (LD) phase in which the bulk density of particles is less than $1/2$; (ii) for $\beta <\alpha$ and $\beta <1/2$, the system is in a high-density (HD) phase in which the bulk density of particles is greater than $1/2$; (iii) for $\alpha$ and $\beta$ both greater than $1/2$, the system is in the maximal-current (MC) phase, in which the bulk density equals $1/2$. In Ref.~\cite{Karthika2020}, the effect of resetting has been studied in all the three phases. Two different distributions for the inter-reset time intervals have been considered, namely, exponential and power-law. Apart from being an interesting problem from a theoretical perspective, the TASEP dynamics serves as a simple model for the movement of ribosome traffic on the mRNA during protein production. The resetting dynamics then represents the sudden disintegration of the mRNA-ribosome complex, followed by replacement by a new, unloaded mRNA. In the work under consideration, the authors have obtained the reset-averaged density by using approximate expressions for the density as a function of time in the bare TASEP dynamics. In the calculations below,
we will consider a coarse grained description of the system, where the space co-ordinate $x$ varies continuously, with $0 \le x \le L$.

If the resets take place at a constant rate $r$ (exponential resetting), the reset-averaged particle density $\rho(x,t)$ as a function of position and time is obtained by using Eq.~(\ref{eq:renewal-basic0}) as
\begin{equation}
\rho(x,t)=e^{-r t}\rho_0(x,t)+r\int_0^t \rmd \tau~e^{-r \tau}\rho_0(x,\tau),
\label{exptasep}
\end{equation}
where $\rho_0(x,t)$ is the density in the absence of resetting. The initial as well as the reset configuration is an empty lattice. 
For the case of power-law resetting, we have on using Eq.~(\ref{eq:renewal-basic2}) that
 \begin{equation}
\rho(x,t)=\int_{0}^{t} \rmd \tau~f_{\gamma}(t,t-\tau)\rho_0(x,\tau).
\label{powerlawtasep}
\end{equation}

The evolution of the density in the TASEP without resetting is dependent on the entry and exit rates and can be divided into four regimes. The following approximate expressions describe the evolution of the density, with an empty lattice being the configuration at $t=0$:
\begin{enumerate} 
\item For the low-density (LD) phase, where we have $\alpha<\beta$ and $\alpha<1/2$, one has 
\begin{equation}
\rho_0(x,t)=\alpha H\left(t-\frac{x}{v}\right);~~\alpha<\beta,~\alpha<1/2,
\end{equation}
where $H(y)=1$ for $y \ge 0$ and $H(y<0)=0$ is the Heaviside step function. 
\item For the maximal-current (MC) phase, where we have $\alpha>1/2$, $\beta >1/2$, one has
\begin{equation} 
\rho_0(x,t)=\frac{1}{2}\left(1-\frac{x}{t}\right)H(t-x);~~\alpha>1/2,~\beta >1/2.
\end{equation}
\item  In the high-density (HD) phase, where we have $\beta<\alpha$, $\beta<1/2$, the density evolution depends on the value of $\alpha$, since the density starts evolving from the input boundary end and the exit boundary does not come into play till the particles have reached it. Initially, where none of the particles have reached the exit boundary, the system density evolves based only on the input rate $\alpha$ and thus shows an LD-phase-like behavior for $\alpha<1/2$ and an MC-phase-like behavior for $\alpha>1/2$. One thus has two regimes. The first regime is $\beta<\alpha$ with $\beta,\alpha<1/2$. Here, one has 
\begin{eqnarray}
\rho_0(x,t)&=\alpha H\left(t-\frac{x}{v}\right) H\left(\frac{L}{v}+\frac{L-x}{v_r}-t\right)+(1-\beta)\nonumber \\
&\times H\left(t-\left(\frac{L}{v}+\frac{L-x}{v_r}\right)\right);~~\beta<\alpha,~\beta,\alpha<1/2.
\end{eqnarray}
Here. $v \equiv 1-\alpha$ and $v_r \equiv [\alpha(1-\alpha)-\beta(1-\beta)]/[1-\alpha-\beta]$ are the speed of the forward and the backward density front, respectively. The second regime is $\beta<\alpha$, with $\beta<1/2$, $\alpha>1/2$. Here, we have 
\begin{eqnarray}
\rho_0(x,t)&=\frac{1}{2}\left(1-\frac{x}{t}\right) H(t-x)\nonumber \\
&\times H(T-t)+(1-\beta) H(t-T);~~\beta<\alpha,~\beta<1/2, \alpha>1/2, \nonumber \\
\end{eqnarray}
where $T$ is the time at which the reverse moving front of density $1-\beta$ reaches the position $x$ on the lattice and is given by
\begin{equation} 
T=\frac{2x(2\beta-1)+4L(1-\beta)^{2}+\sqrt{\left[-2x(2\beta-1)-4L(1-\beta)^{2}\right]^{2}-4x^{2}(2\beta-1)^{2}}}{2(2\beta-1)^{2}}.
\end{equation} 
\end{enumerate}

\begin{figure}[]
\begin{center}
\includegraphics[scale=0.75]{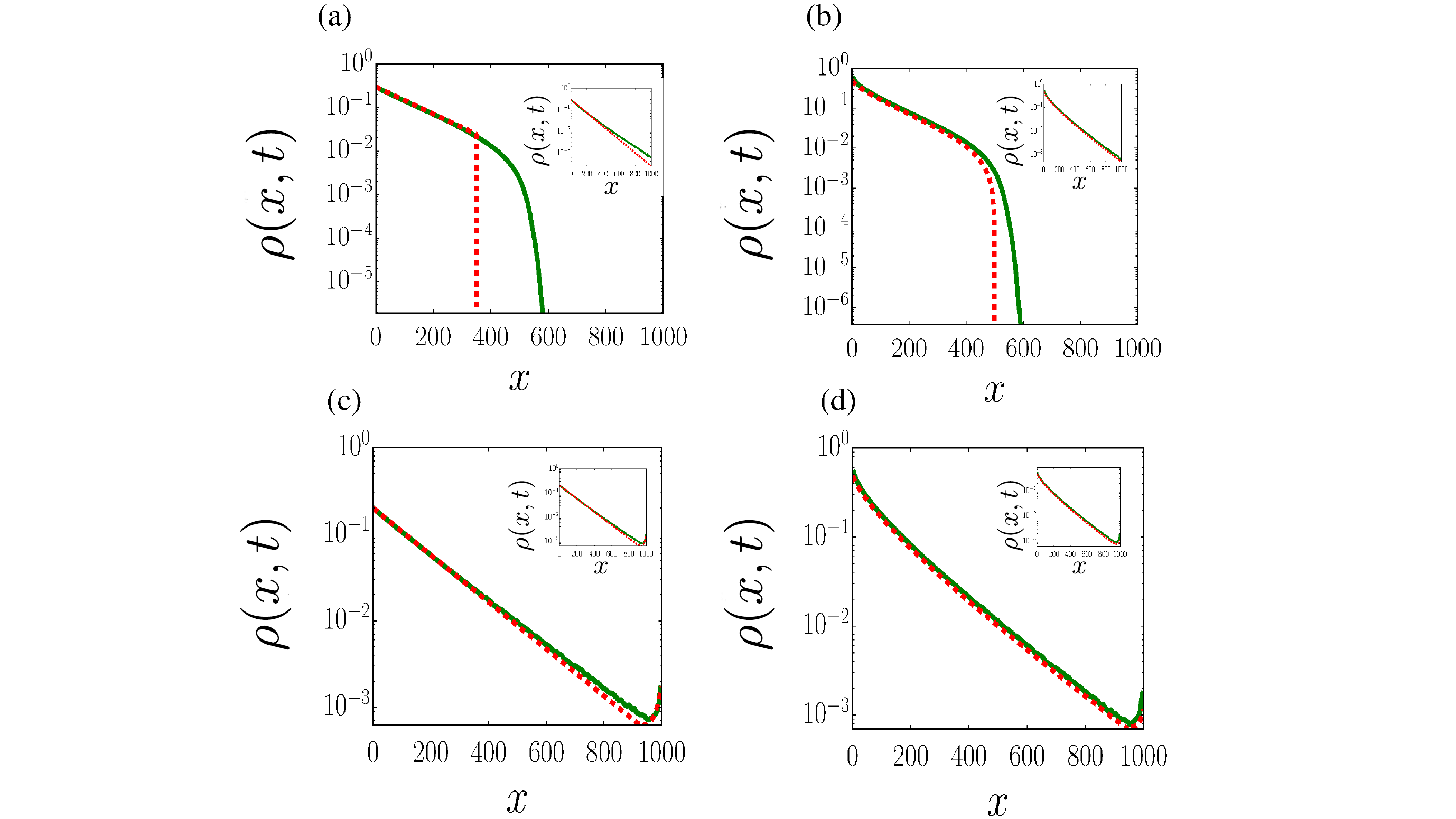}
\caption{The totally asymmetric simple exclusion process subject to exponential resetting, which has been studied in Ref.~\cite{Karthika2020} and discussed in Section~\ref{subsubsec:asymmetricexclusion}. Here, the dotted lines stand for analytical results and the solid lines stand for numerical results. The parameter values are as follows: $r=0.005$ and $L=1000$ for all panels. In panel (a), we have $\alpha=0.3$, $\beta=0.9$, $t=500$ (corresponds to the LD phase of the bare dynamics), in panel (b), we have $\alpha=0.8$, $\beta=0.9$, $t=500$ (corresponds to the MC phase), in panel (c), we have $\alpha=0.2$, $\beta=0.1$, $t=2000$ (corresponds to the HD phase), and in panel (d), we have $\alpha=0.7$, $\beta=0.1$, $t=2000$ (corresponds to the HD phase). The inset in each panel contains the density profile for $t=10,000$, with the rest of the parameter values the same as in the corresponding panel. The figure is adapted from Ref.~\cite{Karthika2020}.}\label{expdecay}
\end{center}
\end{figure}

\begin{figure}[]
\begin{center}
\includegraphics[scale=0.75]{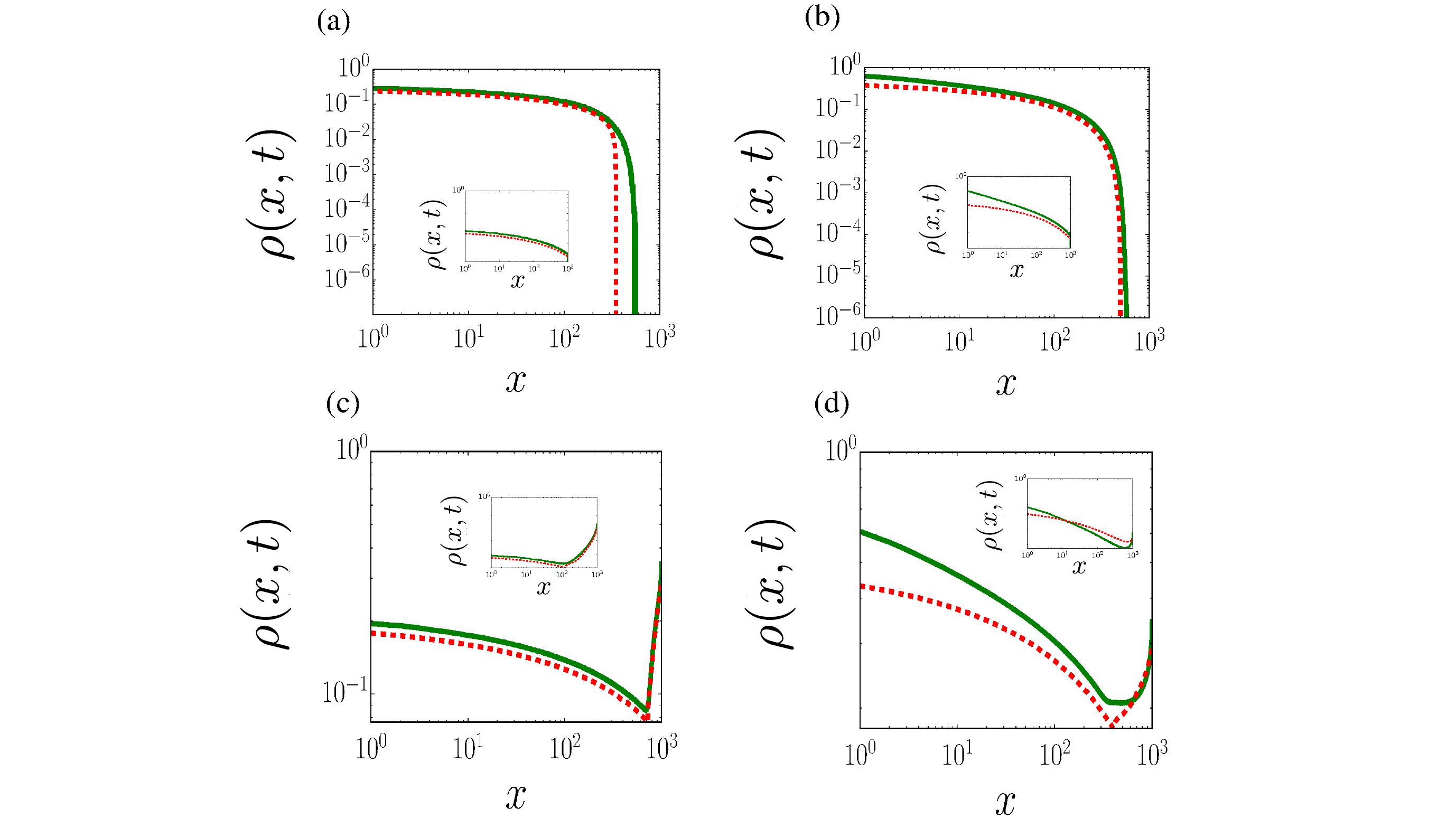}
\caption{The totally asymmetric simple exclusion process subject to power-law resetting, which has been studied in Ref.~\cite{Karthika2020} and discussed in Section~\ref{subsubsec:asymmetricexclusion}. Here, the dotted lines stand for analytical results and the solid lines stand for numerical results. The parameter values are as follows: $\gamma=0.75$ and $L=1000$ for all panels. In panel (a), we have $\alpha=0.3$, $\beta=0.9$, $t=500$ (corresponds to the LD phase of the bare dynamics), in panel (b), we have $\alpha=0.8$, $\beta=0.9$, $t=500$ (corresponds to the MC phase), in panel (c), we have $\alpha=0.2$, $\beta=0.1$, $t=2000$ (corresponds to the HD phase), and in panel (d), we have $\alpha=0.7$, $\beta=0.1$, $t=2000$ (corresponds to the HD phase). The inset in each panel contains the density profile for $t=10,000$, with the rest of the parameter values the same as in the corresponding panel. The figure is adapted from Ref.~\cite{Karthika2020}.}\label{powlaw75}
\end{center}
\end{figure}

\begin{figure}[]
\begin{center}
\includegraphics[scale=0.75]{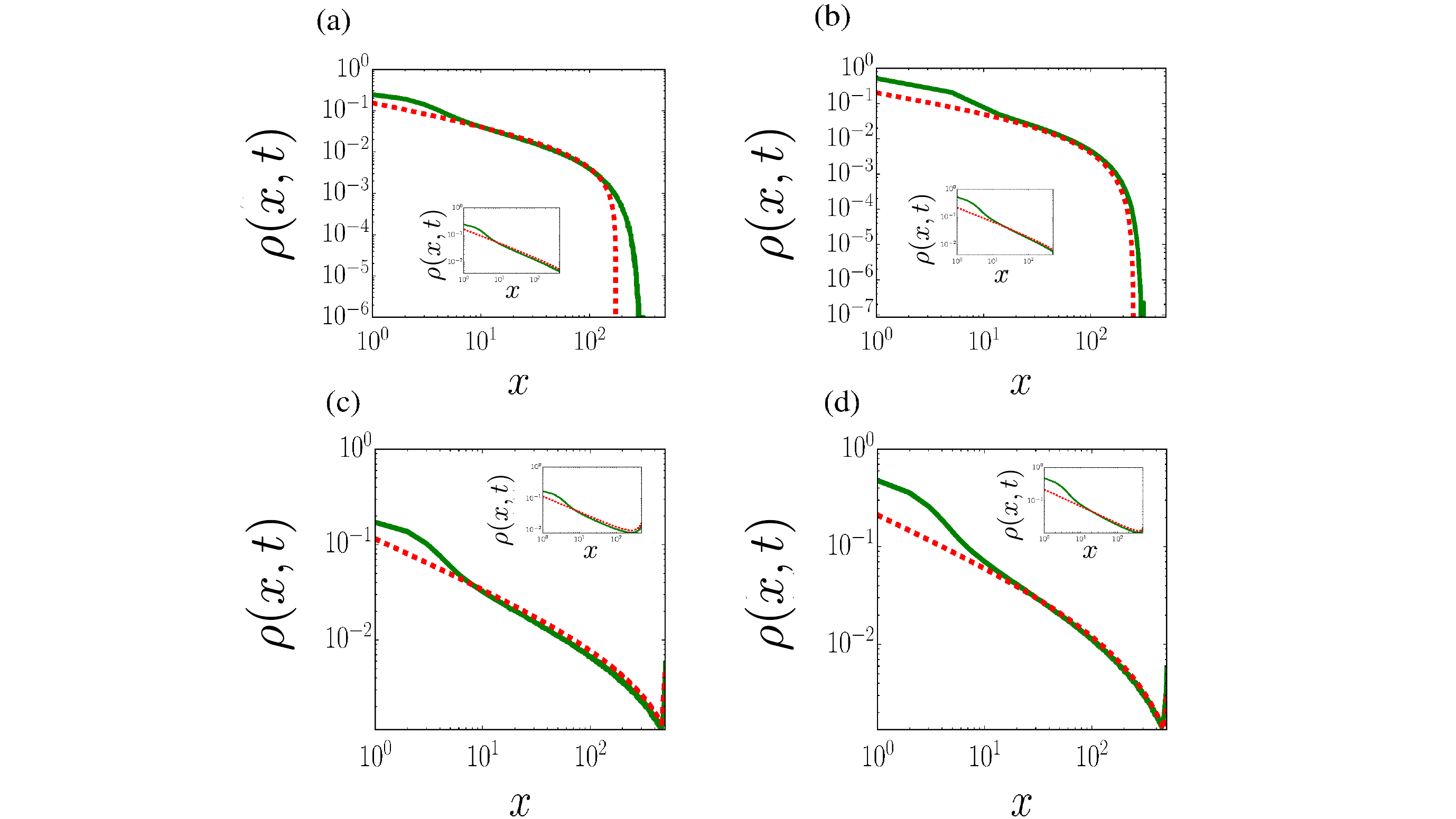}
\caption{The totally asymmetric simple exclusion process subject to power-law resetting, which has been studied in Ref.~\cite{Karthika2020} and discussed in Section~\ref{subsubsec:asymmetricexclusion}. Here, the dotted lines stand for analytical results and the solid lines stand for numerical results. The parameters are as follows: $\gamma=1.5$ and $L=500$ for all panels. In panel (a), we have $\alpha=0.3$, $\beta=0.9$, $t=250$ (corresponds to the LD phase of the bare dynamics), in panel (b), we have $\alpha=0.8, \beta=0.9$, $t=250$ (corresponds to the MC phase), in panel (c), we have $\alpha=0.2$, $\beta=0.1$, $t=1000$ (corresponds to the HD phase), and in panel (d), we have $\alpha=0.7$, $\beta=0.1$, $t=1000$ (corresponds to the HD phase). The inset in each panel contains the density profile for $t=10,000$, with the rest of the parameter values the same as in the corresponding panel. The figure is adapted from Ref.~\cite{Karthika2020}.}\label{powlaw15}
\end{center}
\end{figure}

The integrations in the expressions (\ref{exptasep}) and (\ref{powerlawtasep}) above can be carried out using the approximate expressions for density evolution given in the paragraph above, and thus the reset-averaged density for the TASEP can be calculated for all the four regimes of density evolution. The results from these approximate calculations have been compared to numerical simulations in Figs.~ \ref{expdecay}, \ref{powlaw75}, and \ref{powlaw15}.

The results for the exponential distribution of inter-reset times are shown in Fig.~\ref{expdecay}. In the limit of large times, the system reaches a stationary state, where one observes an exponential decay of the reset-averaged density along the length of the lattice in the LD and the MC phase. In the HD phase, one observes an initial decay and a final rise in the density as a function of $x$. The maximum at the output boundary in the HD case is caused by the interplay of the backward moving density front with the resetting dynamics. In the case of resetting with a power-law distribution, the value of $\gamma$ is very important, and one finds a sudden change in behavior when $\gamma$ crosses one. For $\gamma<1$, the expression for the density depends on the ratio $x/t$, and the system does not reach a stationary state in the limit of large time. The reason for this is that the mean of the inter-reset distribution diverges for $\gamma<1$. The leading behavior of the average density at small $x/t$ is of the form $a-b(\frac{x}{t})^{1-\gamma}$ in the LD and the MC phase with $a$ and $b$ being constants independent of $x,t$ (Fig.~\ref{powlaw75}), while it is more complex with an initial decay and a subsequent increase as a function of $x$ in the HD phase. For $\gamma>1$, the inter-reset time distribution has a finite mean, and so one observes a stationary density profile as $t \rightarrow \infty$. It is seen that the leading behavior of the density far from the boundaries at large times for both the LD and the MC phase is a power-law decay with exponent $1-\gamma$ (Fig.~\ref{powlaw15}). In the HD phase, one observes an initial decay and an eventual rise at the two ends of the lattice, both characterized by a power law with $\gamma -1$ being the exponent.
 
\subsubsection{Resetting of fluctuating interfaces} \label{interface}
\label{subsubsec:interface}
Here,  we review Refs.~\cite{Gupta:2014,Gupta:2016}. Let us consider a well-studied model of fluctuating
interfaces, namely, the Edwards-Wilkinson (EW) interface \cite{EW:1982}. This model describes a surface generated by random
deposition of particles onto a substrate, followed by their diffusion
along the surface~\cite{Barabasi:1995}. To describe the model in one dimension, consider a substrate
of length $L$, and an interface characterized by the height at time $t$ given by $H(x,t)\ge0$ above
position $x \in [0,L]$. The  interface
evolves in time according to the linear equation~\cite{EW:1982} 
\begin{align}
\frac{\partial H}{\partial t}=\nu\frac{\partial^{2}H}{\partial x^{2}}+\eta(x,t),
\label{eq:eom-ew}
\end{align}
where $\nu$ is the diffusivity, while the Gaussian, white noise
$\eta(x,t)$ satisfies
$\langle\eta(x,t)\rangle=0,\langle\eta(x,t)\eta(x',t')\rangle=2D\delta(x-x')\delta(t-t')$.
Here, angular brackets denote averaging over noise, while the constant $D>0$
characterizes the strength of the noise. It is usual to start with a
flat interface: $H(x,0)=0~\forall~x$, and consider periodic boundary conditions:
$H(0,t)=H(L,t) ~\forall~ t$.

Denote by $\overline{H(x,t)}\equiv(1/L)\int_{0}^{L}\rmd x~ H(x,t)$ the
instantaneous spatial average of the height, and by $h(x,t)\equiv
H(x,t)-\overline{H(x,t)}$ the relative height. This gives the width of the
interface at time $t$ as $W(L,t)\equiv\sqrt{\langle
h^{2}(x,t)\rangle}$. The width exhibits the Family-Vicsek scaling \cite{Family:1985}, $W(L,t)\sim
L^{\chi}\mathcal{W}(t/T^\star)$, with the crossover time scale
$T^\star \sim L^z$ being such that over this time, the height fluctuations
spreading laterally correlate the entire interface.
The scaling function $\mathcal{W}(t/T^\star)$ behaves as a constant as $t/T^\star \to \infty$, and
as $(t/T^\star)^{\beta}$ as $t/T^\star\to 0$. Here, $z,\chi,\beta$ are respectively
the dynamic exponent, the roughness exponent, and the growth exponent,
with $z=\chi/\beta$ \cite{Barabasi:1995}. The behavior of $\mathcal{W}(s)$ is such that $W(L,t)$ grows with time as $t^\beta$ for $t\ll T^\star$, and 
saturates to an $L$-dependent value $\sim L^\chi$ for $t \gg T^\star$.
For the EW interface in one dimension, one has $z_{\mathrm{EW}}=2,\chi_{\mathrm{EW}}=1/2,\beta_{\mathrm{EW}}=1/4$.
It then follows that for an interface in the thermodynamic limit $L\to\infty$, the width grows
forever with time, and so there is no stationary state for the
distribution of fluctuations $h$. For the EW interface in this limit, the $h$-distribution at time $t$, while starting from a flat
interface at $t=0$, is given by a Gaussian that has a time-dependent
variance $W_{\mathrm{EW}}^{2}(t)\equiv D\sqrt{2/(\pi\nu)}t^{2\beta_{\mathrm{EW}}}$ \cite{Barabasi:1995}:
\begin{align}
P_{\mathrm{EW}}(h,t|0,0) =\frac{1}{\sqrt{2\pi W_{\mathrm{ EW}}^{2}(t)}}e^{-\frac{h^{2}}{2W_{\mathrm{EW}}^{2}(t)}}.
\label{eq:ew-ht-distr}
\end{align}
For finite $L$, the $h$-distribution at long times $t \gg
T^\star$ is a Gaussian with a time-independent variance $\sim L^{2\chi}$, which corresponds to an
equilibrium stationary state.

Let us now consider the EW interface in the limit $L\to\infty$, and envisage a
dynamical scenario in which the evolution (\ref{eq:eom-ew}) is
repeatedly interrupted with a resetting to the initial flat configuration. Let us consider the case of power-law resetting, whereby two successive resets are separated by random time intervals $\tau$ distributed
according to the power-law~(\ref{eq:ptau}).

Now, in absence of resetting, the
fluctuations grow unbounded in time, and do not have a stationary state. Does resetting lead at long times to
a stationary state with bounded fluctuations? Are 
reset-induced stationary fluctuations different from the Gaussian
fluctuations observed in the stationary state for finite $L$? These questions are best answered by obtaining the quantity $P^{\mathrm{r}}_\mathrm{EW}(h,t|0,0)$, which is the $h$-distribution at
time $t$ and in presence of resetting, while starting from a flat interface at time $t=0$. Using Eq.~(\ref{eq:renewal-basic2}), we have
\begin{align}
P^{\mathrm{r}}_\mathrm{EW}(h,t|0,0)=\int_{0}^{t}\rmd\tau~f_{\gamma}(t,t-\tau)P_\mathrm{EW}(h,\tau|0,0).
\label{eq:pr-time-eqn}
\end{align}
Note that in terms of $h$, the resetting dynamics under consideration corresponds to an
instantaneous jump in its value from $h\ne 0$ to $h=0$, the latter
characterizing the interface at the initial time $t=0$. 

Let us first discuss the case of $\gamma >1$. Using Eq.~(\ref{fgammagt1}) and the
smallness of $\tau_0$ to write
$\int_{0}^{\tau_0}\rmd\tau~f_{\gamma>1,\tau<\tau_0}(t,t-\tau)P_\mathrm{EW}(h,\tau|0,0)\approx
P_\mathrm{EW}(h,\tau_0|0,0)\int_{0}^{\tau_0}\rmd\tau~f_{\gamma>1,\tau<\tau_0}(t,t-\tau)$
give
\begin{align}
&P_{\mathrm{EW}}^{\mathrm{r},\gamma>1}(h,t|0,0)=\frac{e^{-\frac{z}{\sqrt{\tau_{0}}}}\nu^{1/4}}{(8\pi\tau_0)^{1/4}\sqrt{D}}\left[\frac{\gamma-1}{\gamma}+\frac{1}{\gamma}\Big(\frac{t}{\tau_{0}}\Big)^{1-\gamma}\right]\nonumber \\
&+\frac{2\Big(\frac{\gamma-1}{\gamma}\Big)\Big(\frac{z^2}{\tau_{0}}\Big)^{1-\gamma}}{\sqrt{\pi}|h|}\left[\Gamma\left(\beta,\frac{z}{\sqrt{t}}\right)-\Gamma\left(\beta,\frac{z}{\sqrt{\tau_{0}}}\right)\right],\label{eq:ew-agt1-time-depn}
\end{align}
where $z\equiv h^{2}\sqrt{\pi\nu}/(2^{3/2}D)$, $\beta\equiv 2\gamma-3/2$, while $\Gamma(s,x)$ is the upper incomplete gamma function. While $P_{\mathrm{EW}}^{\mathrm{r},\gamma>1}(h,t|0,0)=P_{\mathrm{EW}}^{\mathrm{r},\gamma>1}(-h,t|0,0)$ implies $\langle h \rangle(t)=0~\forall~t$, the square of the width of the
interface $[W_{\mathrm{EW}}^{\mathrm{r}}(t)]^{2}\equiv\int_{-\infty}^{\infty}\rmd h~h^{2}P_{\mathrm{EW}}^{\mathrm{r}}(h,t|0,0)$ is given by
\begin{align}
&[W_{\mathrm{EW}}^{\mathrm{r},\gamma>1}(t)]^{2}=\left[\frac{\gamma-1}{\gamma}+\frac{1}{\gamma}\Big(\frac{t}{\tau_{0}}\Big)^{1-\gamma}\right]D\sqrt{\frac{2\tau_{0}}{\pi\nu}}\nonumber
\\
&+\frac{2^{3/2}D\sqrt{\tau_{0}}}{\sqrt{\pi\nu}}\Big(\frac{\gamma-1}{\gamma(2\gamma-3)}\Big)\Big[1-\Big(\frac{t}{\tau_{0}}\Big)^{3/2-\gamma}\Big].
\label{eq:EW-width}
\end{align}

In the limit $t\to\infty$, one obtains a non-trivial stationary state:
\begin{align}
&P_{\mathrm{EW,st}}^{\mathrm{r},\gamma>1}(h|0)  =\frac{1}{\tau_{0}^{\beta_{\mathrm{EW}}}}\mathcal{G}_{\mathrm{EW}}\left(\frac{h}{\tau_{0}^{\beta_{\mathrm{EW}}}}\right); \nonumber \\
&\mathcal{G}_{\mathrm{EW}}(s)
=\left(\frac{\gamma-1}{\gamma}\right)\frac{\nu^{1/4}}{2^{3/4}\pi^{1/4}\sqrt{D}}e^{-\frac{s^{2}\sqrt{\pi\nu}}{2^{3/2}D}}\nonumber \\
&+\Big(\frac{\gamma-1}{\gamma}\Big)\frac{(\pi\nu/D^2)^{1-\gamma}}{2^{2-3\gamma}\sqrt{\pi}}\frac{1}{|s|^{4\gamma-3}}\gamma\Big(2\gamma-\frac{3}{2},\frac{s^{2}\sqrt{\pi\nu}}{2^{3/2}D}\Big),
\label{eq:ss-EW}
\end{align}
where $\gamma(a,x)$ is the lower incomplete gamma function. The stationary state is strongly non-Gaussian with power-law tails $\sim
|h|^{3-4\gamma}$, unlike the Gaussian stationary state for finite $L$.
Also, here one obtains in the distribution a cusp around the resetting
point $h=0$, implying the stationary state to be an NESS; this may be contrasted with the equilibrium
stationary state obtained for finite $L$ in the absence of resetting.

From Eq. (\ref{eq:EW-width}), it follows that for $\gamma>3/2$, the
width at long times relaxes to a stationary value: 
\begin{align}
[W_{\mathrm{EW,st}}^{\mathrm{r},\gamma>3/2}]^{2}=\Big(\frac{(\gamma-1)(2\gamma-1)}{\gamma(2\gamma-3)}\Big)D\sqrt{\frac{2\tau_{0}}{\pi\nu}},
\end{align}
while for $1<\gamma<3/2,$ the width grows indefinitely with time, with the long-time behavior given by
\begin{align}
[W_{\mathrm{EW}}^{\mathrm{r},1<\gamma<3/2}(t)]^{2}\approx\frac{2^{3/2}D\sqrt{\tau_{0}}}{\sqrt{\pi\nu}}\Big(\frac{\gamma-1}{\gamma(3-2\gamma)}\Big)\Big(\frac{t}{\tau_{0}}\Big)^{3/2-\gamma}.
\end{align}
The $h$-distribution relaxes to a stationary state for all
$\gamma>1$, yet it has fat enough tails for $1<\gamma < 3/2$ that the
interface width diverges with time, while a time-independent finite value
results for $\gamma>3/2$. Consequently, the interface width exhibits a
crossover at $\gamma^{(\mathrm{w})}=3/2$. 

Let us now discuss the situation $\gamma<1$.
For large $t \gg \tau_0$, using Eq.~(\ref{fgammalt1}) in Eq. (\ref{eq:pr-time-eqn}) gives
\begin{align}
&P_{\mathrm{EW}}^{\mathrm{r},\gamma<1}(h,t|0,0)=\frac{\nu^{1/4}\Gamma\left(\gamma\right)\sin(\pi
\gamma)}{2^{3/4}\pi^{7/4}\sqrt{D}t^{1/4}} G_{1,3}^{3,0}\left(\begin{array}{c}
\frac{3}{4}\\
\frac{1}{2},0,\frac{3}{4}-\gamma
\end{array}|\frac{z^2}{4t}\right),
\label{eq:ew-a1t1}
\end{align}
where $G_{p,q}^{\, m,n}\Big(\begin{array}{c}a_{1},\ldots,a_{p}\\
b_{1},\ldots,b_{q}
\end{array}\vert\, z\Big)$ is the Meijer G-function. Equation (\ref{eq:ew-a1t1}) suggests the following scaling form of the distribution for
different times:
\begin{align}
P_{\mathrm{EW}}^{\mathrm{r},\gamma<1}(h,t|0,0)=\frac{1}{t^{\beta_{\mathrm{EW}}}}g_{\mathrm{r},\mathrm{EW}}\Big(\frac{h}{t^{\beta_{\mathrm{EW}}}}\Big),
\label{eq:ew-alt1-scaling}
\end{align}
where the scaling function is 
\begin{align}
g_\mathrm{r,\mathrm{EW}}(s)=\frac{\nu^{1/4}\Gamma\left(\gamma\right)\sin(\pi \gamma)G_{1,3}^{3,0}\left(\begin{array}{c}
\frac{3}{4}\\
\frac{1}{2},0,\frac{3}{4}-\gamma
\end{array}|\frac{s^{4}\nu\pi}{32D^{2}}\right)}{2^{3/4}\pi^{7/4}\sqrt{D}}.
\label{eq:ew-scaling-function-alt1}
\end{align}
Equation (\ref{eq:ew-alt1-scaling}) implies collapse of the data
for $P^{\mathrm{r},\gamma<1}_\mathrm{EW}(h,t|x_0,0)$ at different times on plotting
$t^{\beta_\mathrm{EW}}P_{\mathrm{EW}}^{\mathrm{r},\gamma<1}(h,t|0,0)$ versus $h/t^{\beta_\mathrm{EW}}$. 
The mean $\langle h \rangle$ is zero due to the fact that we have $P_\mathrm{EW}^{\mathrm{r},\gamma<1}(h,t|0,0)$
being even under $h \to -h$, while the width grows with time as 
\begin{align}
[W_{\mathrm{EW}}^{\mathrm{r},\gamma<1}(t)]^{2}=\mathcal{C}t^{2\beta_\mathrm{EW}},
\end{align}
with $\mathcal{C}$ a finite constant:
\begin{align}
\mathcal{C} \equiv \frac{16 \Gamma(\gamma)\sin(\pi \gamma)D\sqrt{t}}{\pi^{5/2}\sqrt{\nu}}\int_0^\infty \rmd y~y^2 G_{1,3}^{3,0}\left(\begin{array}{c}
\frac{3}{4}\\
\frac{1}{2},0,\frac{3}{4}-\gamma
\end{array}|y^{4}\right).
\end{align}

In the limit $t \to \infty$,  $P_{\mathrm{EW}}^{\mathrm{r},\gamma<1}(h,t|0,0)$ does not relax to a stationary state. One may contrast this feature with the
case for $\gamma>1$ discussed above, where the distribution of fluctuations does relax to a well-defined stationary state
(\ref{eq:ss-EW}) on introducing resetting. 
An analysis of the distribution
$P^{\mathrm{r},\gamma<1}_\mathrm{EW}(h,t|0,0)$ reveals that on crossing $\gamma=3/4$, the behavior of the distribution as $h\to 0$ crosses over from being with a cusp for
$\gamma<3/4$ to being divergent for $3/4< \gamma<1$. This crossover behavior is explained by
analyzing Eq. (\ref{eq:pr-time-eqn}) in the limit $h \to 0$:
\begin{align}
P^{\mathrm{r},\gamma<1}_\mathrm{EW}(h \to 0,t|0,0) \sim
\int_{0}^{t}\rmd\tau~\tau^{-\gamma-\beta_\mathrm{EW}}(t-\tau)^{\gamma-1},
\label{eq:pr-time-eqn-smallh}
\end{align}
where we have used Eq. (\ref{fgammalt1}) and the fact that $P_\mathrm{EW}(h\to 0,\tau>0|0,0)$ = a finite constant. The
integral on the right hand side is finite for $\gamma+\beta_\mathrm{EW}<1$, whereby it contributes a cusp, and is divergent for $\gamma+\beta_\mathrm{EW} \ge 1$. A crossover in behavior is then expected at
$\gamma^{(\mathrm{d})}=1-\beta_\mathrm{EW}=3/4$. For a general interface with growth
exponent $\beta$, this then predicts a similar crossover from cusp to
divergence in the $h$-distribution close to the
resetting location at $\gamma^{(\mathrm{d})} \equiv 1-\beta$. 

\begin{table}[!h]
\centering
\begin{tabular}{|c|c|c|}
\hline 
Inter-reset & & \tabularnewline
time distribution & & \tabularnewline
$\sim\tau^{-(1+\gamma)}$ & $\gamma>1$ & $0<\gamma<1$ \tabularnewline
\hline 
\hline 
Long-time & {\em Stationary}  & {\em Time-dependent} \tabularnewline
distribution & \multicolumn{1}{c|}{$\sim|h|^{-(\gamma+\beta-1)/\beta}$} & $(1/t^{\beta})g_{\mathrm{r}}(|h|/t^{\beta})$\tabularnewline
of fluctuations            &  ({\em Power-law tails}) & ({\em Scaling form}) \tabularnewline
& & \underline{Around resetting point:} \tabularnewline
& & $\gamma<\gamma^{(\mathrm{d})}$: {\em Cusp} \tabularnewline
& & $\gamma > \gamma^{(\mathrm{d})}$: {\em Divergence} \tabularnewline
& & Cross-over \tabularnewline
& & at $\gamma^{(\mathrm{d})}\equiv1-\beta$\tabularnewline
\hline 
& %
$\gamma<\gamma^{(\mathrm{w})}$: {\em Diverging}& \tabularnewline
Interface width & $\gamma > \gamma^{(\mathrm{w})}$: {\em Stationary} &{\em Diverging} \tabularnewline
& Cross-over at $\gamma^{(\mathrm{w})}\equiv1+2\beta$ &\tabularnewline
\hline 
\end{tabular}
\caption{Summary of long-time behavior of a one-dimensional fluctuating interface subject to
stochastic resetting at power-law times, see Section~\ref{subsubsec:interface}. Here, $\beta$ is the growth
exponent characteristic of the universality class for the interface
dynamics in the absence of resetting.}
\label{table1}
\end{table}

\begin{figure}[!h]
\centering
\includegraphics[width=10cm]{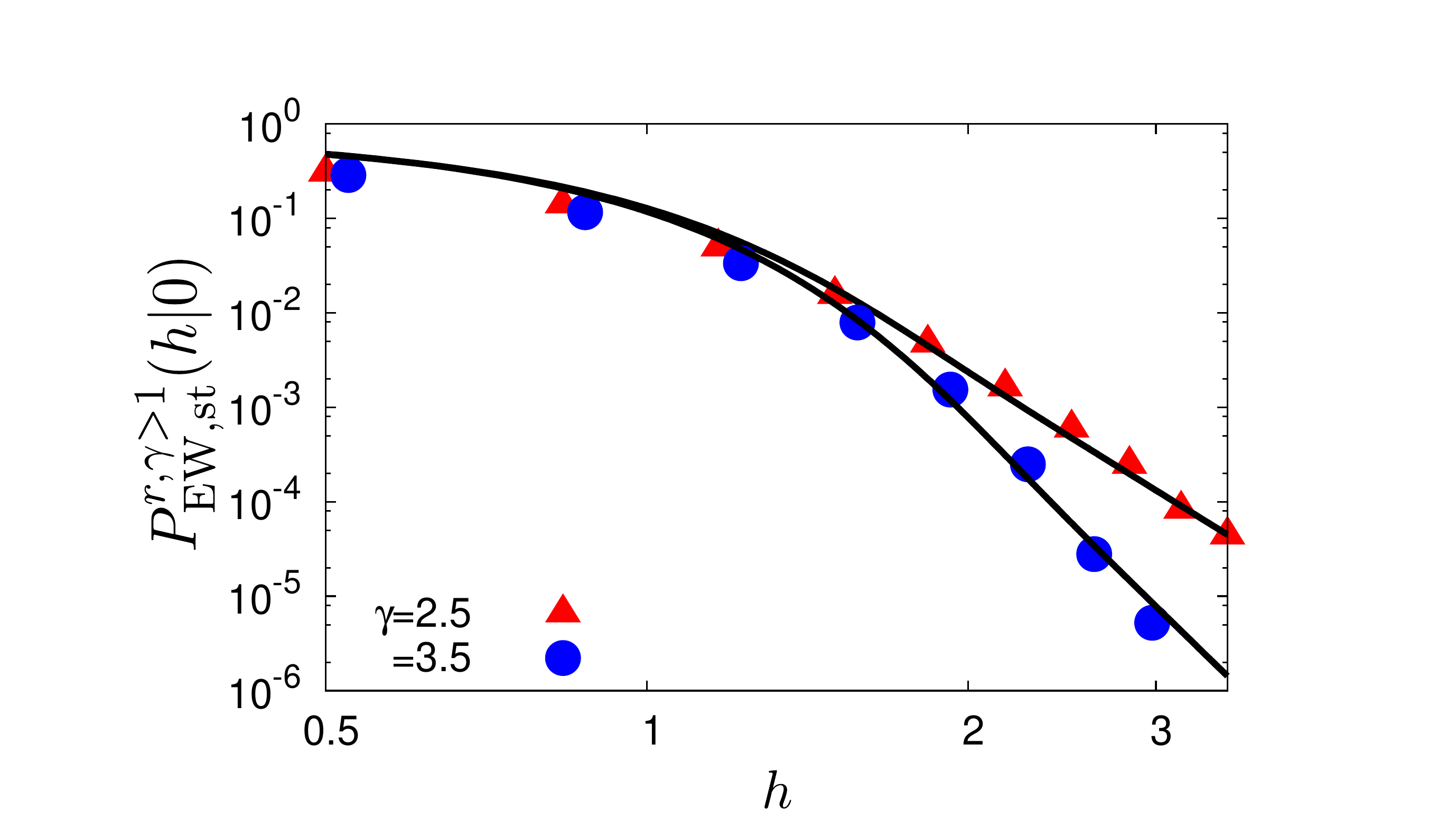}
\caption{The figure shows the stationary-state height distribution for the one-dimensional Edwards-Wilkinson interface subject
to resetting at power-law times, with $\gamma>1$, see Ref.~\cite{Gupta:2016} and Section~\ref{subsubsec:interface}. Here, the points refer
to numerical simulations of a discrete interface of size $L=2^{14}$, while lines refer to the exact result (\ref{eq:ss-EW}). The different parameter
values are: $\nu=1.0$, $D=1.0$, $\tau_0=0.1$. The figure is adapted from Ref.~\cite{Gupta:2016}.}
\label{fig:ew-ss-agt1}
\end{figure}

\begin{figure}[!h]
\centering
\includegraphics[width=15cm]{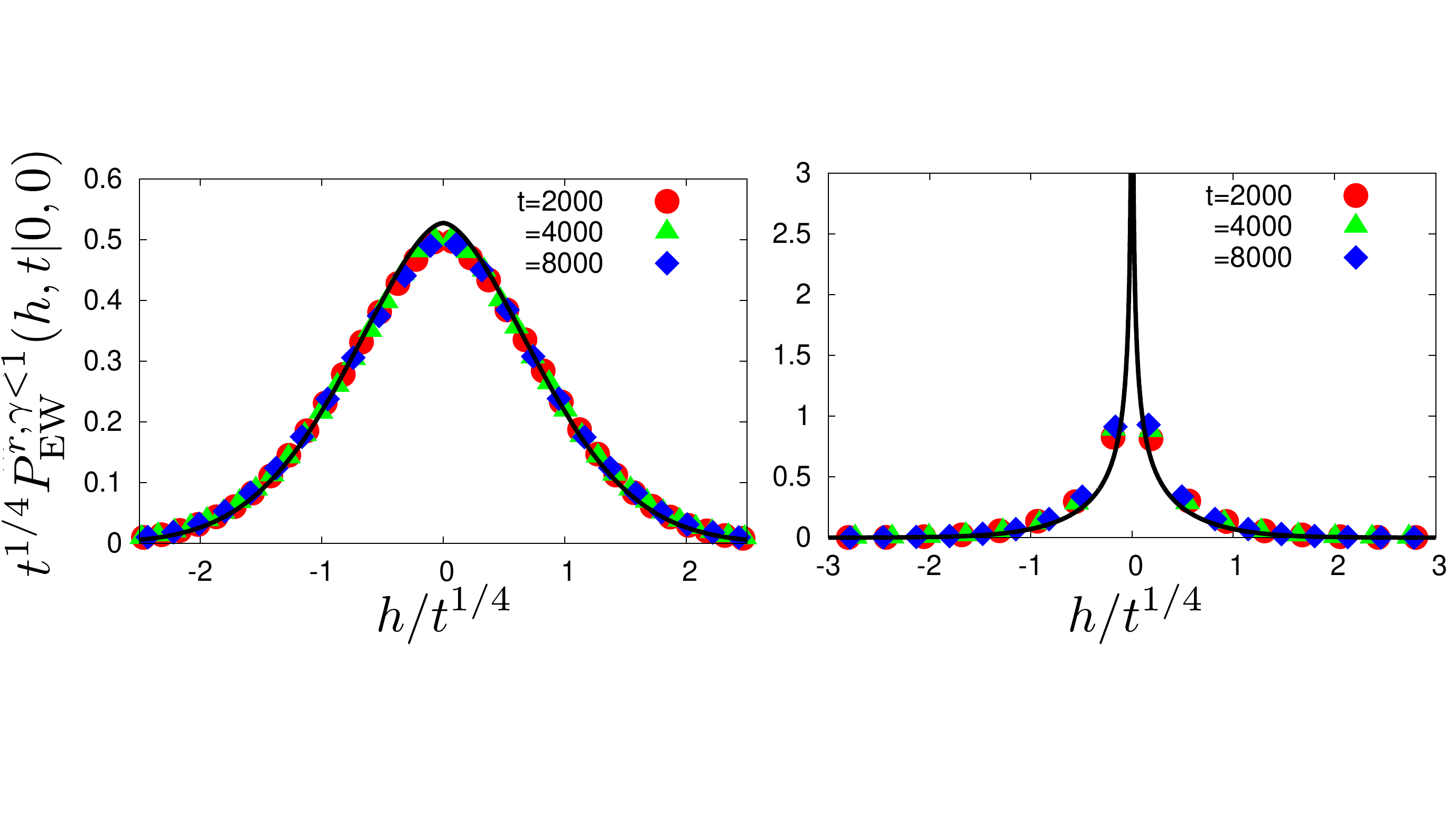}
\caption{The figure shows the time-dependent height distribution for the one-dimensional Edwards-Wilkinson interface subject
to resetting at power-law times, with $\gamma<1$, see Ref.~\cite{Gupta:2016} and Section~\ref{subsubsec:interface}. The data points are obtained
from numerical simulations of a discrete interface of size $L=2^{14}$.  The different parameter
values are: $\nu=1.0$, $D=1.0$, $\tau_0=0.1$, $\gamma=0.25$ (left panel) and $\gamma=0.85$ (right panel). The displayed collapse of the data
for different times follows the scaling form (\ref{eq:ew-alt1-scaling}),
with $\beta_\mathrm{EW}=1/4$. In the figure, the lines denote the exact scaling function (\ref{eq:ew-scaling-function-alt1}). The figure is adapted from Ref.~\cite{Gupta:2016}.}
\label{fig:ew-time-alt1}
\end{figure}

On the basis of the above discussions, we summarize the principal findings in Table~\ref{table1}.
The results for the EW interface may be confirmed by performing numerical simulations on a discrete one-dimensional periodic interface $\{H_{i}(t)\}_{i=1,2,\ldots,L}$ that evolves
at times $t_{n}=n\Delta t,$ with $n$ an integer and time step $\Delta t\ll1$.
Starting with a flat interface, $H_{i}(0)=0~\forall~i$, its evolution
according to the EW dynamics is interrupted by a reset to the initial
configuration, with two successive resets separated by time intervals $\tau$ sampled from the distribution
(\ref{eq:ptau}). In Fig.~ \ref{fig:ew-ss-agt1}, we show for two
representative values of $\gamma>1$ a comparison of the simulation
results with the stationary-state distribution (\ref{eq:ss-EW}), displaying a very good
agreement between theory (lines) and simulation (points). In Fig.~\ref{fig:ew-time-alt1}, we show for two
representative values of $\gamma<1$ a collapse of the simulation data
for different times in accordance with the scaling form (\ref{eq:ew-alt1-scaling}), with the lines
showing the exact scaling function (\ref{eq:ew-scaling-function-alt1}).
It may be noted that on crossing $\gamma=3/4$, the behavior of
$P^{\mathrm{r},\gamma<1}_{\mathrm{EW}}(h,t|0,0) $ as $h\to 0$ changes, from being with a cusp to being divergent,
as implied by the obtained analytical results. In Ref.~\cite{Gupta:2016}, the predictions summarized in Table~\ref{table1} were confirmed also for an interface that has a nonlinear evolution in time. The interface in question is
the one-dimensional Kardar-Parisi-Zhang (KPZ) interface, with time evolution given by 
\begin{align}
\frac{\partial H}{\partial t}=\nu\frac{\partial^{2}H}{\partial
x^{2}}+\frac{\Lambda}{2}\Big(\frac{\partial H}{\partial
x}\Big)^{2}+\eta(x,t),
\label{eq:eom-kpz}
\end{align}
and with the exponents $z,\chi,\beta$ having the values $z_{\mathrm{KPZ}}=3/2,\chi_{\mathrm{KPZ}}=1/2,\beta_{\mathrm{KPZ}}=1/3$. In the absence of analytical results, the predictions in Table~\ref{table1} were tested and confirmed by performing extensive  numerical simulations.  

Resetting of fluctuating interfaces has also been considered for the case in which the distribution $\Phi(\tau)$ is an exponential, as in Eq.~(\ref{eq:ptaur})~\cite{Gupta:2014}. In this case, $P_{\mathrm{}}^\mathrm{r}(h,t|0,0)$ relaxes to a stationary form $P^\mathrm{r}_\mathrm{EW, st}(h|0)$ that has the scaling form 
\begin{align}
P^\mathrm{r}_\mathrm {EW, st}(h|0) \sim \sqrt{\kappa} r^{1/4}G^\mathrm{EW}(h\sqrt{\kappa}r^{1/4}) \;,
\label{ph-scaling}
\end{align}
with $\kappa \equiv \sqrt{\pi \nu}/(D2^{3/2})$ and the function $G^\mathrm{EW}(x)$ given by
\begin{align}
G^\mathrm{EW}(x)=\frac{1}{\sqrt{\pi}}\int_0^\infty \frac{\mathrm{d}y}{y^{1/4}}\exp\Big(-y-\frac{x^2}{\sqrt{y}}\Big) \;,
\label{GEW-1}
\end{align}
which is symmetric in $x$, $G^\mathrm{EW}(-x) = G^\mathrm{EW}(x)$, yielding
zero mean, and variance $\int x^2 \mathrm{d}x~G^\mathrm{EW}(x)= \sqrt{\pi/4}$. From the scaling form in (\ref{ph-scaling}),
one easily obtains the scaling of the stationary width with $r$ as $W_\mathrm{EW, st}^\mathrm{r} \sim r^{-1/4}$. The function $G^\mathrm{EW}(x)$ behaves asymptotically as
\begin{eqnarray}\label{asympt_GEW}
G^\mathrm{EW}(x) \sim 
\begin{cases}
&\frac{1}{\sqrt{\pi}}\Big[\Gamma \left(\frac{3}{4}\right)-x^2 \Gamma
\left(\frac{1}{4}\right)+\frac{8}{3} \sqrt{\pi } |x|^3\Big] \;,\; x \to 0 \,, \\
& c|x|\exp[-3/2^{2/3} \, |x|^{4/3}] \;, \; x \to \pm \infty \;,
\end{cases}
\end{eqnarray}
where $\Gamma(x)$ is the Gamma function and $c$ is a constant that can be computed. Interestingly, due to the
$|x|^3$ term in (\ref{asympt_GEW}), $G^\mathrm{EW}(x)$ is non-analytic close to $x=0$. In the limit $x \to \pm \infty$, the stretched
exponential behavior (\ref{asympt_GEW}) is significantly different from a Gaussian tail.  For a general interface,  $P^\mathrm{r}_\mathrm{st}(h|0)$ is universal in the limit
$r \to 0$ and $h \to \infty$, keeping $hr^\beta$ fixed, and has the scaling form  
\begin{align}
P^\mathrm{r}_\mathrm{st}(h|0)\sim 
r^{\beta}G(hr^\beta) \;, \; G(x) = \int_0^\infty \frac{\mathrm{d}y}{y^\beta} e^{-y} g\left(\frac{x}{y^\beta} \right) \;,
\label{ph-scaling-general}
\end{align}
where $ g\left(\frac{x}{y^\beta} \right)$ is a scaling function. This expression implies in particular the behavior of the stationary width 
$W^\mathrm{r}_\mathrm{st} \sim r^{-\beta}$ as $r \to 0$. These predictions have been tested and confirmed for the EW and the KPZ interface in Ref.~\cite{Gupta:2014}

\subsubsection{Coagulation-diffusion process with stochastic reset}
\label{subsubsec:coagulation}
The coagulation-diffusion process describes particles moving diffusively between the sites of an infinite lattice in one dimension, with one particle allowed per site and is such that a particle disappears as soon as it moves to a site that is already occupied by another particle~\cite{ben-Avraham}. The dynamical feature of disappearance is usually denoted by the chemical reaction $A+A \rightarrow A$, where $A$ denotes the particle. Here, we review Ref.~\cite{Durang:2014} that concerns the possibility of input (deposition) of particles at random sites on the lattice with rate $\lambda$. An important  quantity of interest is the density of particles in the system as a function of time, which can be calculated by using the method of empty intervals. The method involves studying the dynamical quantity $E_n(t)$ that gives the probability at time $t$ that a randomly-chosen segment of $n$ consecutive sites contains no particles.

In Ref.~\cite{Durang:2014}, the effect of resetting has been explored in the aforementioned coagulation-diffusion model defined on a chain of $\mathcal{N}$ sites. The stochastic reset takes the system to one of a set of configurations defined by the probabilities $F_n$ for having $n$ consecutive empty sites. The dynamics of resetting involves the following: in order to update a configuration, a particle is chosen randomly. This particle moves to the left with probability $\mathcal{P}_{g}=\frac{D}{2D+r/\mathcal{N}}$, or to the right with probability
$\mathcal{P}_{d}=\mathcal{P}_{g}$, else the entire system is reset to one of a set of configurations specified by $F_n$ with probability $\mathcal{P}_r = \frac{r/\mathcal{N}}{2D+r/\mathcal{N}}$. Here $r \ge 0$ is the reset rate (the reset dynamics is thus a case of exponential resetting). To define $F_n$, we may consider configurations where each site is occupied with probability $p$ independently of other sites, so then the probability of finding $n$ consecutive empty sites is simply  $F_n = (1-p)^n$. A sweep of the lattice consists of $\mathcal{N}$ such updates. 

In the absence of reset, the density on an infinite lattice approaches zero in the long-time limit as $\rho(t) \sim t^{-1/2}$. The effect of reset can be studied for the model at hand by using the following equation for the probability $E_n(t)$ that $n$ consecutive sites are empty:
\begin{equation}
\hspace{-0.3truecm}\partial_t E_n(t) = \frac{2D}{a^2} \left( E_{n-1}(t) + E_{n+1}(t) - 2 E_n(t) \right) -r E_n(t) + r F_n, \;\; E_0(t) =1, \;\; E_{\infty}(t)=0.
\label{coagulation1}
\end{equation}
Here, $a$ is the lattice constant, while $D>0$ is the diffusion constant. 
This equation can be solved by using the generating function $G(z,t)\equiv \sum_{m=-\infty}^{\infty} z^m \, E_{m}(t)$. The average density $\rho(t)$ can be expressed as $\rho(t)=1-E_1(t)$, and one obtains the following expression in terms of the distribution $F_n$: 
\begin{eqnarray}
\rho(t) &=& \e^{-(4D+r)t}\left(I_0(4Dt)+I_1(4Dt) - \sum_{m=1}^{\infty} \frac{mE_m(0)I_m(4Dt)}{2Dt} \right) \nonumber \\
 & & + r \int_0^t \rmd t'~\e^{-(4D+r)t'} \left( I_0(4Dt')+I_1(4Dt')
 - \sum_{m=1}^{\infty} \frac{mF_m I_m(4Dt')}{2Dt'} \right), \nonumber \\
\end{eqnarray}
where $I_m$ is the modified Bessel function. From here, one can derive the stationary state density for the particular case of $F_n=(1-p)^n$ as
\begin{eqnarray}
&&\rho_\mathrm{st}=\frac{\sqrt{r(r+8D)} -r}{4D} \nonumber \\
& & -\frac{r}{2D}\sum_{m=1}^{\infty}
\left(\frac{1-p}{2}\right)^m \left(\frac{4D}{r+4D}\right)^m
{}_2F_1\left(\frac{m}{2},\frac{m+1}{2};m+1;\left(\frac{4D}{r+4D}\right)^2\right). \nonumber \\
\end{eqnarray}
Thus, one observes that resetting leads to a finite density in the stationary state, unlike the system without resetting where the density vanishes as $t \rightarrow \infty$. The other important difference is in the time dependence of the approach to stationary state; while one has $\rho(t) \sim t^{-1/2}$ without resetting, the stationary state density is approached exponentially fast once resetting is introduced.

\begin{figure}[tb]
\centering
 \includegraphics[width=10 cm]{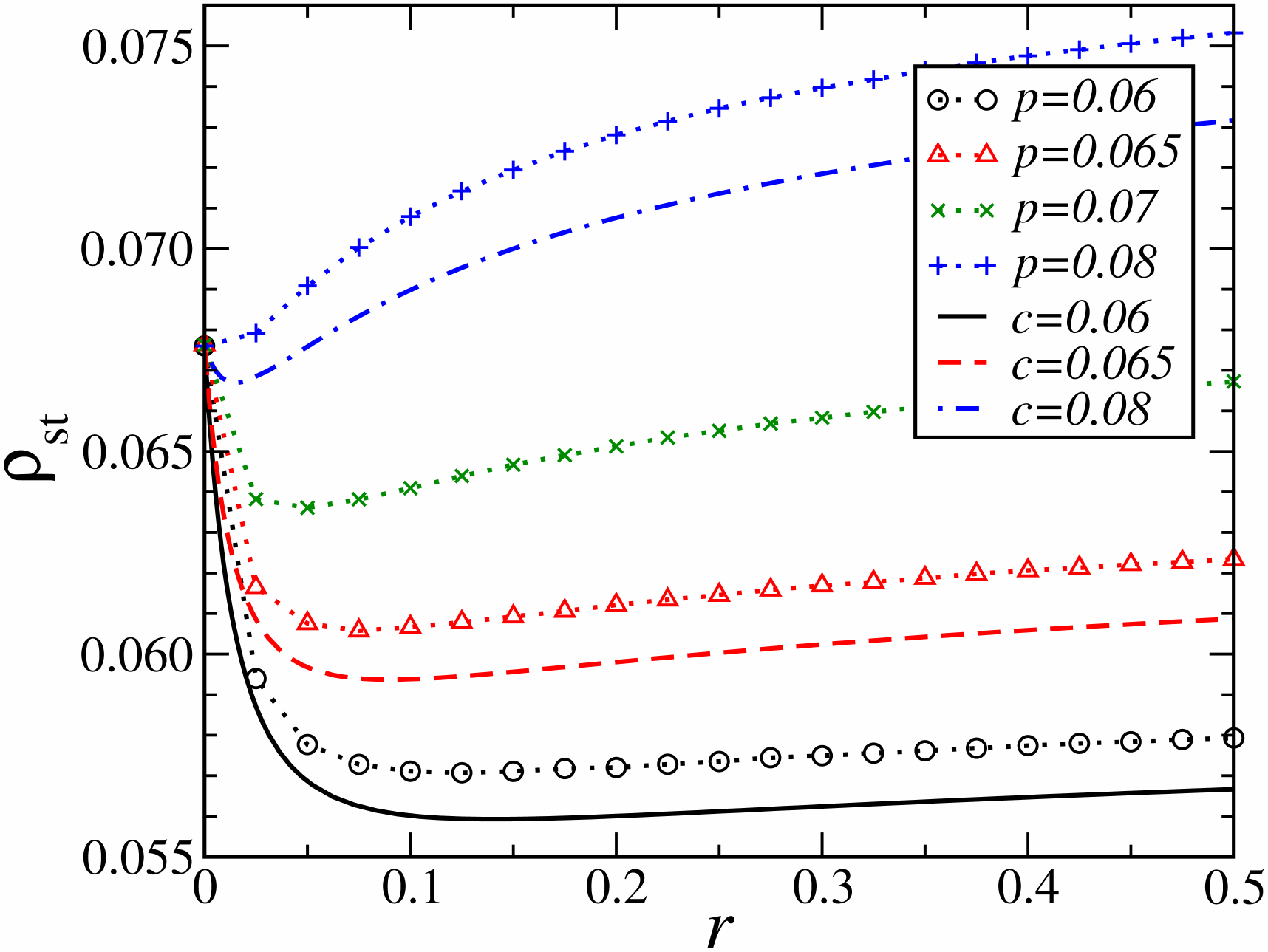}
\caption{For the coagulation-diffusion model studied in Ref.~\cite{Durang:2014} and discussed in Section~\ref{subsubsec:coagulation}, the figure shows the plot of the stationary probability $\rho_\mathrm{st}$ as a
function of the reset parameter $r$ for $\lambda =0.0008$. The lines without symbols give the analytical
solution (\ref{eqDurang1}) in the continuum limit. The figure is adapted from Ref.~\cite{Durang:2014}.}
\label{figDurang}
\end{figure}

Further, the authors have studied the effect of introducing deposition (input) of particles at random sites  with a rate $\lambda$ in the model above. In this case, progress can be made by studying the evolution of the quantity $E_n(t)$ in the continuum limit; the corresponding time evolution is given by
\begin{equation} \label{3.1}
\frac{\partial E(t,x)}{\partial t} = 2D \frac{\partial^2 E(t,x)}{\partial x^2} - \lambda x E(t,x) - r E(t,x) + r F(x)
\end{equation}
with the boundary and initial conditions as $E(t,0) = 1$, $E(t,\infty) = 0$, and $E(0,x) = E_0(x)$,  respectively. The density in the stationary state can be derived by solving the above equation, to get:
\begin{equation}
\rho_\mathrm{st}=cP\left( \frac{c}{\beta}, \frac{\beta r}{\lambda}\right),
\label{density-coagulation}
\end{equation}
where  $c\equiv -F'(0)$ is the average particle-density in the reset configuration $F(x)$ and $\beta\equiv (\lambda/(2D))^{1/3}$. The function $P(u,y)$ is given by
\begin{equation}
P(u,y)=-\frac{1}{u}\frac{Ai'(y)}{Ai(y)} - \pi y \left( Bi'(y)
- Ai'(y) \frac{Bi(y)}{Ai(y)}\right) \int_0^\infty \rmd Y~ F(uY/c) Ai(Y+y), \label{eqDurang1}
\end{equation}
where $Ai$ and $Bi$ are Airy functions. The variation of the density~(\ref{density-coagulation}) as a function of the reset parameter $r$ is shown in Fig.~\ref{figDurang}. While the behavior near $r=0$ is close to the system without reset, the value at very large $r$ is expected to be close to the reset configuration density. The behavior in between these two extremes shows a complex behavior with non-monotonicity seen in certain parameter ranges.   

\subsubsection{ZRP on fully connected lattice with resetting}
\label{subsubsec:zrp}
The zero-range process (ZRP) \cite{Evans:2005} is a model of particles hopping on a lattice, where the rate of departure of a particle is determined by the number of particles present at the departure site. Owing to its simplicity of definition, analytic tractability and non-trivial stationary state possibilities, this is a very well studied model in nonequilibrium statistical mechanics. In particular, the model has been studied in the context of population dynamics, where the particle number is not conserved, with creation and annihilation of particles representing birth and death processes, respectively.

In Ref.~\cite{Grange:2020}, the author has considered a ZRP on a fully connected lattice, where any number of particles can occupy a site and can hop stochastically to any other site uniformly chosen from the available sites with rate $\beta$. A particle at a site may get annihilated at a rate $\delta$. Moreover, at a constant rate $r>0$ (exponential resetting), the system resets to an empty configuration. Every site has an associated number $l \in ]0,1]$, which determines the rate of creation of particles at that site. The $l$ values form a discrete set of $L$ regularly-spaced values in the interval $]0,1]$. In the limit of large $L$, these discrete values form a continuum defined on $[0,1]$. A probability density $q(l)$ determines the fraction of the total number of sites with a particular value of $l$. It is assumed that there is only one site with the maximum allowed value of $l$, i.e., $q(1)=0$. One is interested in finding out $p_l(n,t)$, the probability of having $n$ particles among all sites with a given $l$. Using this distribution, one can calculate the average occupation number for a given $l$ as well as the average density, as $\overline{n_l}(t) \equiv \sum_{n\geq 0 } n p_l( n,t)$, $\rho(t) \equiv \int_0^1 \rmd l~\overline{n_l}(t)$, respectively.

Without resetting, we have only the hopping, creation and annihilation dynamics and the dynamics of $p_l(n,t)$ can be expressed by the following master equation: 
\begin{eqnarray}\label{evolPDE}
\hspace{-2cm}
\frac{dp_l(n,t)}{dt} = \;& \theta( n ) \left\{  \left( \beta \rho(t) q\left( l \right) +  l  n \right)p_l(n-1,t)-( \delta  +  \beta ) n p_l(n,t)    \right\}\nonumber \\
&+(  \beta+ \delta)(n+1)p_l(n+1,t)  -  \left( \beta \rho(t) q\left(  l \right) +   l ( n+1)\right)p_l( n,t), \;\;\;\forall n \geq 0. \nonumber \\
    \end{eqnarray}
Using the above equation, the average $\overline{n_l}(t)$ can be obtained as
\begin{equation}
\overline{n_l}(t) = e^{-(1+\zeta - l )t} \overline{n_l}(0) + \frac{l- l e^{-(1+\zeta - l)t)}}{1+\zeta -l}+\beta q(l) \int_0^t \rmd s~e^{(l-1 - \zeta)(t-s)}\rho(s),
\end{equation}
where $n_l(0)$ is the initial condition, and $\zeta \equiv \beta+\delta-1$. The average density $\rho(t)$ is then given by
\begin{equation}
\rho(t) =\int_0^1 \rmd l~e^{-(1+\zeta - l )t} \overline{n_l}(0) - \int_0^1 \rmd l~ (l-1-\zeta)  C[\rho,1,l,t],
\end{equation}
with 
\begin{equation}
C[\rho, J,l,t] \equiv \int_0^t \rmd s~\frac{-l-\beta \rho(s) q(l)}{l(1-J)-  (1+\zeta - l J )e^{ ( l-1-\zeta) (s-t)}}.
\end{equation}

The density in presence of resetting can be calculated by using a renewal theory approach (see Eq.~(\ref{eq:renewal-basic0})). and is given by
\begin{equation}\label{meanReset}
\overline{n}_{l,r}(t)= e^{-rt} \overline{n}_{l}(t) + r \int_0^t \rmd \tau~e^{-r\tau} \overline{n}_{l}(\tau),
 \end{equation}
with the first term representing the probability that there is no reset up to time $t$, while the second term evaluates the evolution up to time $t$ since the previous reset. In the stationary state, attained as $t \to \infty$, one has
\begin{equation}\label{steadyReset}
  \overline{n}_{l,r}(\infty) =  \frac{l + r \overline{n_l}(0)}{1+\zeta +r -l} + 
     \frac{ \beta\rho_r(\infty)q(l)}{ 1+\zeta +r - l},
\end{equation}
where $\rho_r(\infty)$ is given by
\begin{equation}\label{steadyReset1}
  \rho_r(\infty)  =    \int_0^1 \rmd l \frac{l +  r \overline{n_l}(0)}{1+\zeta +r -l} \left( 1 - \beta \int_0^1 \rmd l \frac{q(l)}{1+\zeta + r -l}\right)^{-1}.  
 \end{equation}
One observes that while the system without resetting shows an exponentially fast decay of the initial condition, resetting makes the conditions at $t=0$ important and appear in the stationary state as a skewed version. 

\subsubsection{Resetting of the Ising model}
\label{subsubsec:Ising}
Here, we review Ref.~\cite{Magoni:2020} that studies the Ising model subject to stochastic resetting. To set the stage, consider the 
 Ising model with ferromagnetic nearest-neighbour interactions and defined on a $d$-dimensional lattice with $N$ sites and periodic boundary conditions. The Ising Hamiltonian is given by 
 \begin{align}
 H = - J \sum_{\langle i,j \rangle} s_i s_j;~s_i=\pm 1.
 \end{align}
 Here, the parameter $J>0$ describes the strength of ferromagnetic interaction between nearest-neighbour pair $\langle ij\rangle$ of Ising spins $s_i=\pm 1$ that may point either up ($s_i=1$) or down ($s_i=-1$). Consider an initial configuration in which the $s_i$'s are independently chosen to be $\pm 1$ with probability $(1 \pm m_0)/2$. Correspondingly, the magnetization, given by $(1/N)\sum_i s_i$ for a given configuration $\{s_i\}$, has, when averaged over configurations, the value $m \equiv (1/N)\sum_i \langle s_i \rangle=m_0$. One may study the time evolution of a given configuration $\{s_i\}_{1\le i \le N}$ within the protocol proposed by Glauber. The Glauber dynamics consists of flipping a single spin with rate \cite{Glauber1963}
\begin{eqnarray}\label{rate}
w(s_i \to - s_i) = \frac{1}{1+e^{\beta \Delta E}},
\end{eqnarray} 
where $\beta = 1/(k_BT) \ge 0$ is the inverse temperature and $\Delta E = 2 J s_i \sum_{j \in \mathrm{n. n.}} s_j$ is the change of energy in flipping the 
$i$-th spin. Here, $\mathrm{n.n.}$ stands for nearest neighbours. Under the Glauber dynamics, it is guaranteed that the long-time distribution of configurations is the canonical equilibrium distribution at temperature $T$: $P_\mathrm{st}(\{s_i\}) \propto \exp(-\beta H)$. 
In this equilibrium state, the equilibrium value $m_\mathrm{eq}$ of the average magnetization shows a continuous phase transition at a critical temperature $T_c$ between a ferromagnetic phase ($m_\mathrm{eq} \ne 0$) observed for $T<T_c$ and a paramagnetic phase ($m_\mathrm{eq}=0$) observed for $T\ge T_c$.

To study the Ising model in presence of resetting, one envisages a situation in which the system goes back randomly in time to a given initial configuration with a constant rate $r>0$ (exponential resetting). The initial configuration is taken to be the one mentioned in the preceding paragraph, i.e., one in which each spin $s_i$ is chosen independently to be either $+1$ or $-1$ with probability $(1+m_0)/2$ and $(1-m_0)/2$, respectively.  In between two successive resetting events, the system evolves following the aforementioned Glauber dynamics. On noting that at a fixed time $t$ of observation, what matters is the time $\tau$ elapsed since the last resetting before time $t$, and that resetting happening stochastically implies that $\tau$ is a random variable, the average magnetization $m(t)$ at time $t$ also becomes a random variable (There are two sources of noise in the system: (i) The first source of noise is due to contact with a heat bath at temperature $T$, with the Glauber dynamics modelling the corresponding dynamics and inducing the canonical equilibrium state in absence of resetting; (ii) The second source of noise is due to the resetting dynamics.). It may then be shown straightforwardly by using Eq.~(\ref{eq:renewal-basic0}) that the probability distribution $P_\mathrm{r}(m,t)$ of the average magnetization  follows the renewal equation~\cite{Magoni:2020}
\begin{equation}\label{renewal1}
P_\mathrm{r}(m,t) = e^{-r t}P_0(m,t)+ r \int_0^t \mathrm{d}\tau~ e^{- r \tau} P_0(m,\tau).
\end{equation}  
Here, $P_0(m,\tau) = \delta(m-m(\tau))$ is the magnetization distribution in absence of resetting ($r=0$), when it evolves deterministically according to the Glauber dynamics, as $m(\tau)$.  In the limit of long times, the second term in Eq.~(\ref{renewal1}) may be neglected, and one obtains the stationary magnetization distribution $P_{r,\,\mathrm{st}}(m)$ satisfying \begin{equation}
P_{r,\mathrm{st}}(m) = r \int_0^{\infty} \mathrm{d}\tau~ e^{-r\tau} \delta(m-m(\tau)).
\label{trick}
\end{equation} 
In order to proceed, one is required to know the Glauber time-dependence $m(\tau)$ of the average magnetization. 

In $d=1$, one has a simple expression for the Glauber rate: $w(s_i \to - s_i)  = (1/2) (1 - \tanh{(2 \beta J)} s_i(s_{i-1} + s_{i+1})/2)$. It may be shown that the one-dimensional Glauber dynamics is exactly solvable, in the sense that one has an exact expression for the time evolution of the average magnetization $m$: 
\begin{equation}
m(t) = m_0 \, e^{-r^*(T) t}.
\label{eq:decay}
\end{equation}
 where we have $r^*(T) \equiv 1-\tanh{(2 \beta J)}$. This exact solvability stems from the fact that in one dimension, the evolution equation for the $n$-point correlation functions involves only $n$-point functions~\cite{Glauber1963}, unlike in higher dimensions. 
Using Eq.~(\ref{eq:decay}) in Eq.~(\ref{trick}),
one obtains the exact stationary-state distribution of the average magnetization in presence of resetting:
\begin{eqnarray}\label{PDF_1d}
P_{r,\,\mathrm{st}}(m) = \frac{r}{m_0 r^*(T)} \left( \frac{m}{m_0}\right)^{\frac{r}{r^*(T)}-1};~~ m \in [0,m_0].
\end{eqnarray}
In one dimension, it is well known that $T_c=0$, and consequently, we need to discuss the behavior of $P_{r,\,\mathrm{st}}$ only for $T\ge T_c$. The result in Eq.~(\ref{PDF_1d}) implies that near $m=0$, we have $P_{r,\,\mathrm{st}}(m) \sim m^{\zeta}$, with the exponent 
$\zeta = r/r^*(T)-1$ varying continuously with temperature. Consequently, $P_{r,\,\mathrm{st}}(m)$ either diverges (for $r<r^*(T)$) or vanishes (for $r>r^*(T)$) as $m \to 0$. In the former case, the most probable or the typical value of the magnetization is $m_\mathrm{typ} =0$, which corresponds to a paramagnetic phase. On the contrary, for $r>r^*(T)$, one has $m_\mathrm{ typ} >0$, which characterizes a magnetized phase. We thus see that resetting can induce magnetization in the system even in a regime in which the bare model without resetting does not support such a phase in the stationary state. This phase is referred to as the ``pseudo-ferro'' phase.

Let us now move to the case of $d=2$. Here, one has a finite critical temperature: $T_c \approx 2.269$, with the choice $J=k_B=1$. 
Unlike in $d=1$, the Glauber dynamics is not exactly solvable in $d=2$. Consequently, in Ref.~\cite{Magoni:2020}, well-established phenomenological behavior of $m(t)$ under the Glauber dynamics and at long times has been invoked and used in the renewal equation (\ref{trick}) to make predictions for the stationary distribution $P_{r,\,\mathrm{st}}(m)$, which were then verified in numerical simulations. We discuss below the cases $T>T_c$, $T<T_c$, and $T=T_c$. 

(a) $T>T_c$: In this case, the average magnetization under the  Glauber dynamics decays at late times as $m(t) \sim a_1 e^{-\lambda_1 t}$, where the amplitude $a_1$ and the leading decay rate $\lambda_1$ are both functions of temperature and can be estimated  from Monte Carlo simulations of the Glauber dynamics. This exponential decay of $m(t)$ holds only for long times $t \gg 1/\Delta \lambda$, where
$\Delta \lambda$ is the first gap in the relaxation spectrum. Using Eq.~ (\ref{trick}) then yields for $r \ll \Delta \lambda$ that
\begin{equation}
P_{r,\,\mathrm{st}}(m) \approx \frac{r}{\lambda_1 a_1^{\frac{r}{\lambda_1}}} m^{\frac{r}{\lambda_1} -1}, \quad \quad m \in (0,a_1].
\label{eq:parstat}
\end{equation}
The above expression is a good approximation at high temperature $T \gg T_c$, when  $\Delta \lambda$ is large, and $a_1 \approx m_0$. The behavior of $P_{r,\,\mathrm{st}}(m)$ is thus qualitatively the same as in the $d=1$ case. In $d=2$, for $T \gg T_c$, we then have $r^*(T) = \lambda_1$. Thus, as in the $d=1$ case, one observes a crossover from the usual para phase for $r < r^*(T)$ to the ``pseudo-ferro'' phase for $r>r^*(T)$ across the crossover line $r^*(T)$ in the $(T,r)$ plane, for $T \gg T_c$.

(b) $T < T_c$: Here, under the bare Glauber dynamics,  the average magnetization has a nonzero equilibrium value $m_\mathrm{eq}$, with a stretched exponential relaxation at late times:
\begin{equation}
m(t) \approx m_\mathrm{eq} \pm a e^{-bt^c},
\label{evferr}
\end{equation}
where the parameters $a, b$ and $0<c<1$ are determined from simulations. In Eq.~(\ref{evferr}), we use $+$ and $-$ signs to stand for the cases $m_0 > m_\mathrm{eq}$ and $m_0 < m_\mathrm{eq}$,  respectively. In presence of resetting, the stationary distribution $P_{r,\,\mathrm{st}}(m)$, obtained using Eq.~(\ref{trick}),  is non-trivial, and for whose detailed form, we refer the reader to Ref.~\cite{Magoni:2020}. For example, for $m_0 > m_\mathrm{eq}$, the distribution has a support  over the interval $[m_\mathrm{eq}, m_0]$; near the lower edge ($m \to m_\mathrm{eq}^+$), one has $P_{r,\,\mathrm{st}}(m)$ vanishing faster than a power law: $P_{r,\,\mathrm{st}}(m) \sim \exp(- B [-\ln(m-m_{eq})]^{1/c})$, with $0<c<1$ and $B \equiv r\,b^{-1/c}$. The main point to note is that for $T<T_c$, one has $m_\mathrm{typ} \ne 0$, and so resetting is able to sustain in this temperature range the magnetized phase of the bare model.

\begin{figure}
\centering
\includegraphics[width=12cm]{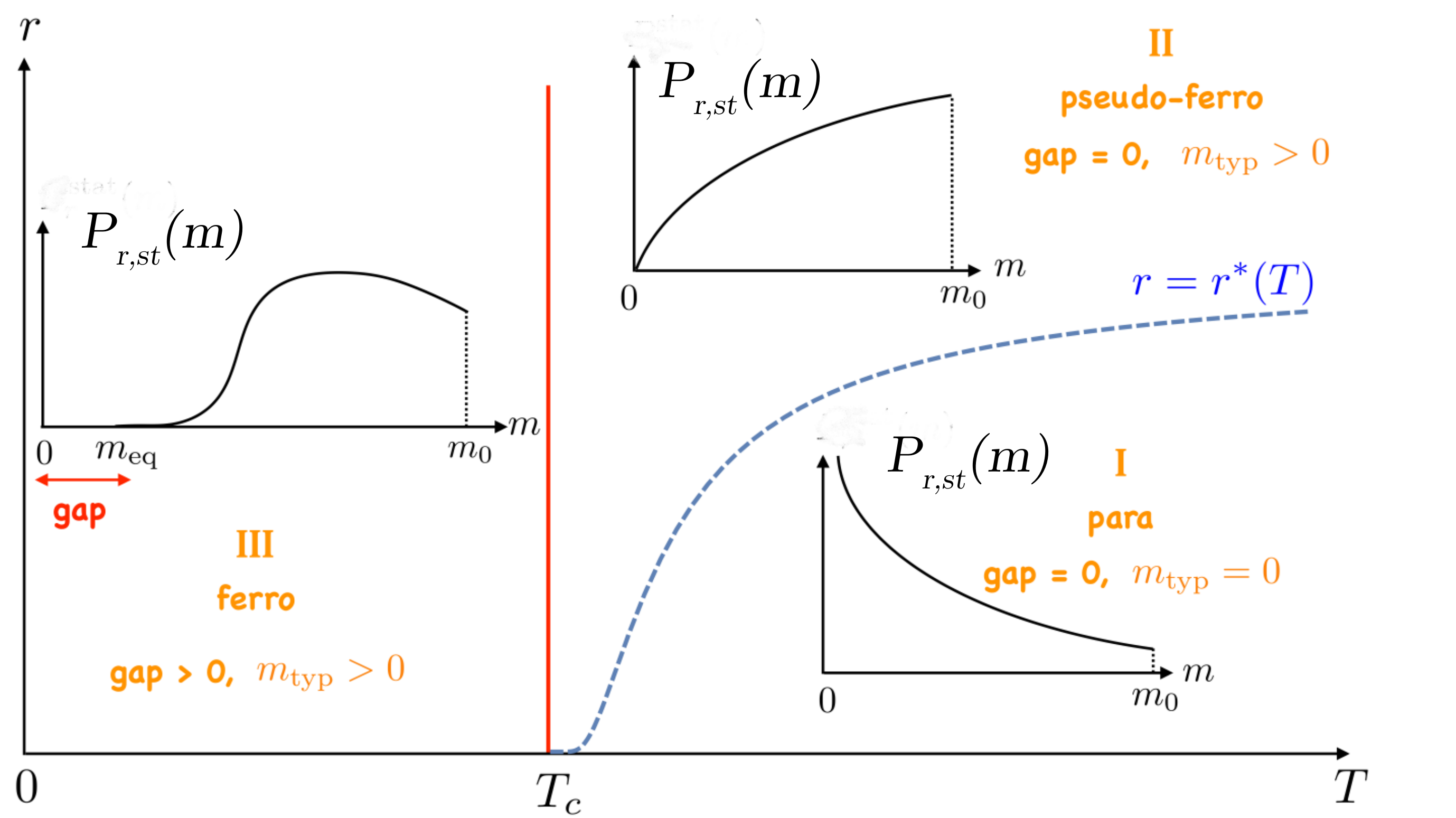}
\caption{For the two-dimensional nearest-neighbour Ising model subject to stochastic resetting at constant rate $r$ studied in Ref.~\cite{Magoni:2020} discussed in Section~\ref{subsubsec:Ising}, the figure shows the schematic phase diagram in the $(T,r)$-plane, with $T$ being the temperature. For $T>T_c$, there is a crossover line $r^*(T)$ separating the para phase for $r<r^*(T)$ and the ``pseudo-ferro'' phase for $r>r^*(T)$. In the para phase, the stationary distribution $P_{r,\,\mathrm{st}}(m)$ has a divergent peak at $m=m_\mathrm{typ}=0$ (and no gap at $m=0$), while in the pseudo-ferro phase, one has $P_{r,\,\mathrm{st}}(m)$ vanishing at $m=0$ with a peak at $m_\mathrm{typ} >0$ (and no gap at $m=0$). For $T<T_c$, one observes a nonzero gap opening up at $m=0$, with the distribution peaking at $m_\mathrm{typ} > 0$. The figure is adapted from Ref.~\cite{Magoni:2020}.}
\label{fig:ising-reset-phdiagram}
\end{figure}

(c) $T=T_c$: Exactly at the critical point, the magnetization $m(t)$ under the bare Glauber dynamics has a non-monotonic decay with time, whose analytical form is known only  at short and long times. This renders difficult the exact determination of the reset-induced stationary distribution $P_{r,\,\mathrm{st}}(m)$ by using Eq.~ (\ref{trick}) for all $m$. However, when $r$ and $m$ are both small, one may use in Eq.~ (\ref{trick}) the late time form of $m(t) \approx b_c \, t^{-\varphi}$, where the exponent $\varphi \equiv \beta/(\nu z)$. Here, $\nu$ and $\beta$ are respectively the correlation length and the order parameter critical exponent associated with the equilibrium magnetization phase transition of the Ising model at $T=T_c$, while $z$ is the dynamical critical exponent associated with the Ising-Glauber dynamics at $T = T_c$. It is then found that there is a scaling regime as $m \to 0$, $r \to 0$ but with $m \, r^{-\varphi}$ fixed, where the distribution $P_{r,\,\mathrm{st}}(m)$ has the scaling form 
\begin{eqnarray}\label{scaling_Tc}
P_{r,\,\mathrm{st}}(m) \approx r^{-\varphi} G(m \, r^{-\varphi}).
\end{eqnarray}
Here, the scaling function $G(y) = Ay^{-1 - 1/\varphi} e^{-(b_c/y)^{1/\varphi}}$, with $A \equiv b_c^{1/\varphi}/\varphi$, vanishes extremely fast as $y \to 0$. The distribution $P_{r,\,\mathrm{st}}(m)$ has  a support $m \in [0, \Gamma]$, where the upper cut-off $\Gamma$ depends on system parameters. While there is no strict gap at $m=0$ (as in the ferro phase), $P_{r,\,\mathrm{st}}(m)$ vanishes extremely fast as $m \to 0$. The schematic phase diagram in the $(T,r)$-plane, showing the rich behavior of the stationary distribution $P_{r,\,\mathrm{st}}(m)$, is shown in Fig.~\ref{fig:ising-reset-phdiagram}.

A related work that we now review very briefly is that of stochastic resetting (exponential resetting) of the so-called Kuramoto model, studied in Ref.~\cite{Sarkar:2022}. The model involves a system of $N \gg 1$ limit-cycle oscillators of distributed natural frequencies that are globally coupled through the sine of their phase differences. The dynamics is inherently deterministic, unlike the Ising model for which the bare dynamics in absence of resetting, which may be modelled by the Glauber dynamics, is inherently stochastic. A remarkable feature of the Kuramoto model is that provided the strength $K$ of global coupling is beyond a threshold value $K_c$, the system at long times evolves spontaneously to a stationary state that is globally synchronized. In such a state, a macroscopic number of oscillator phases evolve in time while maintaining a constant phase difference among them. In the limit $N \to \infty$ and for a unimodal and symmetric distribution for the natural frequencies, there exist two kinds of qualitatively different phases depending on the value of the coupling $K$: For $K \le K_c$, the stationary state is unsynchronized, while for $K>K_c$, the system is synchronized. When the system is unsynchronized, the oscillator phases evolve independently of each other in time, with the phase difference changing continually in time. Thus, a synchronized (respectively, an unsynchronized) phase exemplifies an ordered (respectively, a disordered) phase. Moreover, the synchronization order parameter undergoes a continuous phase transition (more precisely, a bifurcation), remaining zero for $K\le K_c$, and increasing continuously as a function of $K$ as the latter is increased beyond $K_c$.  The question then is: What happens when the Kuramoto model is subject to random and repeated interruptions of its dynamics with a reset at a constant $r$ to a synchronized configuration?  Reference~\cite{Sarkar:2022} unveiled how such a protocol of stochastic resetting dramatically modifies the phase diagram of the bare model, allowing, in particular, for the emergence of a synchronized phase even in parameter regimes for which the bare model does not support such a phase. The phase diagram bears a strong resemblance with the one for the two-dimensional Ising model in the presence of stochastic resetting discussed in the preceding paragraph. Indeed, stochastic resetting induces in the Kuramoto model a stationary ordered phase (the pseudo-synchronized phase) for values of the coupling strength $K$ satisfying $K < K_c$, for which the bare model does not exhibit any ordered phase, as discussed above. There are other similarities as well: in the Ising case, a threshold resetting rate separates the pseudo-ferro phase from a disordered phase, the paramagnetic phase, Fig.~\ref{fig:ising-reset-phdiagram}, just as in the case of the Kuramoto model, it separates the pseudo-synchronized phase from the unsynchronized phase. 

\section{Local resetting}
\label{sec:localresetting}
In this section, we look at systems in which the particles reset individually and independently. Such dynamics has been given the name of local resetting in Ref.~\cite{Miron:2022}. We consider three different problems. The first one deals with models of aggregation undergoing local resetting, see Section~\ref{subsec:binary}. Various physical processes like polymerization, coalescence of aerosols and emulsification involve the coagulation of smaller clusters to form larger ones \cite{Wattis:2006}. One of the most important  models for such processes was introduced in the seminal work by Smoluchowski in the form of the coagulation equation named after him. The density distribution of the aggregates by size is the quantity of central interest in such systems. While the aggregation dynamics promotes clusters of larger size, the introduction of a resetting dynamics that destroys clusters can counteract this tendency. The second (Section~\ref{subsec:ssep-local}) and the third (Section~\ref{subsec:tasep-local}) problem deal respectively with the SSEP and the TASEP considered previously under the theme of global resetting, the difference being that now, the dynamics involves particles resetting individually and independently of each other.

\subsection{Binary aggregation with resetting}
\label{subsec:binary}
Here, we summarize the work~\cite{Grange:2021}, in which the author has studied a model of binary aggregation with constant kernel that is subject to exponential resetting, whereby aggregates of any size explode into monomers at independent stochastic times. The density of aggregates has an NESS, in which the density is a function of the size of the aggregate rescaled by a function of the resetting rate. It has been shown that the stationary-state density of aggregates of a given size is maximized on setting the resetting rate to the quotient of the aggregation rate by the size of the aggregate (minus one). We now turn to details.

Reference~\cite{Grange:2021} describes the effect of resetting on a model of aggregation, wherein one has a process of binary aggregation, which involves a cluster of size $i$ merging with a cluster of size $j$ to form a cluster of size $i+j$ at a constant rate $K>0$ that is independent of the size and the shape of the clusters. Resetting here involves the stochastic breaking of a cluster to smaller units: an aggregate of size $k$ breaks down into $k$ monomers at a constant rate $r$ (exponential resetting). The aggregates are assumed to be well mixed in a solvent, and the monomers resulting from resetting are immediately available for aggregation. At any time instant, the system will have clusters of different sizes, and every cluster has the possibility of undergoing resetting independently of the others. Thus, we have a case of local as opposed to global resetting. The quantity of interest is the concentration of aggregates of a given size $k$ as a function of time, denoted by $\rho_k(t)$, with $k\ge 1$. It is assumed that this density is well defined at all times. The density satisfies the time evolution
\begin{equation}\label{masterck}
\frac{\rmd\rho_k}{\rmd{{t}}} = \sum_{i+j = k}  \rho_i \rho_j - \rho_k \sum_i  \rho_i - r  \rho_k + r \delta_{k1}\sum_{i}    i  \rho_i.
\end{equation}
The first two terms on the right hand side arise out of the binary aggregation process, while the last two terms arise due to resetting. Further progress can be made by introducing the generating function $\tilde{\rho}(t,z) \equiv \sum_{k\geq 1 } \rho_k(t) z^k$, which satisfies the equation
\begin{equation}\label{evolC}
\frac{\partial \tilde{\rho}(t,z)}{\partial t} =  [\tilde{\rho}(t,z)]^2 - (2N(t) + r)\tilde{\rho}(t,z) + rMz,
\end{equation} 
where $N(t) \equiv \sum_{k\geq 1} \rho_k(t) =\tilde{\rho} ( t, 1 )$ is the total density of aggregates and $M \equiv \sum_{k \geq 1} k \rho_k(t)$ is the total mass density, which is a constant on account of the fact that both aggregation and resetting conserve mass. One may set this constant to unity, $M=1$, which is equivalent to picking a unit of volume. The initial condition is taken to be a monomer-only configuration, $\rho_k(0) = \delta_{k1}$.

The equation for time evolution of $N(t)$ can be found using Eq.~(\ref{evolC}) as
\begin{equation}\label{evolN}
\frac{ d N}{d t} =  -N^2 - rN + r,
\end{equation}
which can be solved to obtain
\begin{equation}\label{N(t)}
  N(t) = \frac{N_+(  N(0 ) - N_-) {{-}} N_- (N(0) - N_+ )e^{-\sqrt{r(r+4)}t}}{  N(0 ) - N_-  - ( N(0) - N_+  ) e^{-\sqrt{r(r+4)}t}  }.
\end{equation}
Here, $N(0)$ is the initial density of aggregates, and $N_+$ and $N_-$ are given by 
\begin{equation}
 N_+\equiv \frac{r}{2}\left(-1 + \sqrt{1 + \frac{4}{r}}\right),\;\;\;\;\;\;\; N_- \equiv \frac{r}{2}\left(-1 - \sqrt{1 + \frac{4}{r}}\right).
\end{equation}
At long times, the system reaches a stationary state in which one has a time-independent density of states given by $N(\infty) = N_+$.

For the initial condition $\rho_k(0) = \delta_{k1}$, the full solution for the generating function can be found as
\begin{equation}\label{solGener}
  \tilde{\rho}  (t,z) = N( t ) + \frac{1}{2}\left( r  - \sqrt{r^2 + 4r(1-z)}\right) + \frac{ \sqrt{r^2 + 4r(1-z)} e^{-  \sqrt{r^2 + 4r(1-z)} t}}{ -1 + \frac{2 \sqrt{r^2 + 4r(1-z)}}{ 2(z-1) - r +  \sqrt{r^2 + 4r(1-z)}}   + e^{ -  \sqrt{r^2 + 4r(1-z)}   t}}.
\end{equation}
In the stationary state, one then has
\begin{equation}\label{generIn}
{\tilde{\rho}}_\mathrm{st}(z)\equiv\tilde{\rho} (\infty,z)=\frac{r}{2} \sqrt{1 + \frac{4}{r}}-  \frac{1}{2} \sqrt{r^2 + 4r(1-z)},
\end{equation}   
which can be inverted to obtain the density of aggregates of various sizes in the stationary state as
\begin{equation}\label{ckstat}
\begin{split}
&\tilde{\rho}_\mathrm{st}(z) = \sum_{k\geq 1}  \rho_k(\infty) z^k,\\
&\rho_k(\infty) \equiv \sqrt{r(r+4)}\frac{\Gamma\left( k - \frac{1}{2}\right)}{2\sqrt{\pi} \Gamma( k + 1 )}\left( 1 + \frac{r}{4}\right)^{-k}.
\end{split}
\end{equation}
One can show that for large aggregate sizes, the quantity $\rho_k(\infty)$  assumes a scaling form:
\begin{equation}
\begin{split}
\rho_k(\infty) &\underset{k\to\infty}{\sim}  \frac{\sqrt{r(r+4)}}{2\sqrt{\pi}}\left( \log\left( 1 + \frac{r}{4}\right) \right)^{\frac{3}{2}}g\left( \frac{k}{\sigma(r)} \right);\\
g(x)& \equiv x^{-\frac{3}{2}}e^{-x},\;\;\;\;\;{\mathrm{and}}\;\;\;\sigma(r) \equiv \left(\log\left( 1 + \frac{r}{4}\right) \right)^{-1}.
\end{split}
\end{equation}

The author has also calculated the value of the resetting rate that maximizes the stationary-state density for a given cluster size $k$ (except for $k=1$). The optimal resetting rate $r_k^\ast$ is defined as satisfying $\rmd \rho_k(\infty)/\rmd r|_{r_k^\ast}=0$, and is obtained as $r_k^\ast = 2/(k-1);~k>1$. The maximum value for the density of aggregates of size $k$ in the stationary state is given by
    \begin{equation}\label{toExpand}
       {{\rho_k^\ast(  \infty ) = \frac{\Gamma\left( k - \frac{1}{2}\right)}{\sqrt{\pi} \Gamma( k + 1 )} \sqrt{\frac{2}{k-1}}\left(1 + \frac{1}{2(k-1)} \right)^{-k}. }}
       \end{equation}
The optimal resetting rate is a decreasing function of $k$, as rare resetting events favour the formation of large aggregates. 

\subsection{Symmetric simple exclusion process with local resetting}
\label{subsec:ssep-local}
Here, we review Ref.~\cite{Miron:2022} that considers the paradigmatic symmetric simple  exclusion process undergoing local resetting.
The model involves a one-dimensional periodic lattice of $L$ sites that are
labelled by the index $l=0,...,L-1,$ and which are occupied by $N$ hard-core particles so that every site is either empty or occupied by at  most one particle. The average particle density is $\overline{\rho}\equiv N/L$. The configurations evolve in continuous time according to the following rules: Each particle makes an attempt to hop to its left
or right neighbouring site with rate $1$ and to reset its position
to site $l=0$ with a constant rate $r>0$ (exponential resetting) independently of the other particles. In either case, the attempt is successful if and 
only if the destination site is empty. To study the dynamics, one defines $n_{l}\left(t\right)$ as the occupation variable of site $l$ at time $t$, taking values $0$ and $1$ depending respectively on whether the site is empty or occupied. The $n_{l}\left(t\right)$'s evolve in time as
\begin{equation}
n_{l}\left(t+\mathrm{d}t\right)-n_{l}\left(t\right)=\Gamma_{l}\left(t\right),\label{eq:markov chain}
\end{equation}
where $\Gamma_{l}\left(t\right)$ reads as
\begin{equation}
\Gamma_{l}\left(t\right)=\begin{cases}
n_{l\pm 1} & \text{with probability }\left(1-n_{l}\right)\mathrm{d}t,\\
-n_{l} & \text{with probability}\left(1-n_{l\pm 1}\right)\mathrm{d}t,\\
\sigma_l & \text{with probability}\left(1-n_{0}\right)R_{l}\mathrm{d}t,
\end{cases}
\label{eq:markov Gamma}
\end{equation}
where we have
\begin{equation}
\begin{array}{c}
\sigma_l=-1\text{ and }R_{l}=rn_{l}\,\,\,\,\,\,\,\,\,\,\,\,\,\,\,\,\,\,\,\,\,\,\,\text{ for }l\ne0,\\
\sigma_l=+1\text{ and }R_{l}=r\sum_{m=1}^{L-1}n_{m}\,\,\text{  for }l=0.
\end{array}\label{eq:Gamma_sigma_R}
\end{equation}
The expression for $R_{0}$ stems from particle conservation implied by the dynamics, i.e.,  $\sum_{l=0}^{L-1}\left[n_{l}\left(t+\mathrm{d}t\right)-n_{l}\left(t\right)\right]=0$. 

To solve for the stationary-state averaged $n_l(t)$'s (averaged over stochastic evolution), one invokes the mean-field approximation. To this end, one averages Eqs. (\ref{eq:markov chain}) and (\ref{eq:markov Gamma}),
replaces the mean occupation $\left\langle n_{l}\right\rangle$
by the density field $\rho_{l}\in\left[0,1\right]$, and approximates  $\left\langle n_{l}n_{k}\right\rangle$ as  $\left\langle n_{l}n_{k}\right\rangle \approx\left\langle n_{l}\right\rangle \left\langle n_{k}\right\rangle =\rho_{l}\rho_{k}$. We also have  $\left\langle R_{l=0}\right\rangle=r\sum_{m=1}^{L-1}\rho_{m}$ and $\left\langle R_{l\ne 0}\right\rangle =r\rho_{l}$.
As $\mathrm{d}t\to 0$, the mean-field-approximated time-evolution equation  for $\rho_{l}\left(t\right)$ is obtained as
\begin{equation}
\frac{\partial \rho_{l}(t)}{\partial t}=\rho_{l+1}-2\rho_{l}+\rho_{l-1}+\sigma_{l}\left(1-\rho_{0}\right)\langle R_{l}\rangle.
\label{eq:density eqn}
\end{equation}
The stationary density profile is obtained by setting the left hand side to zero and analyzing separately the case
$l\ne0$ (termed the ``bulk'' equation) and the case $l=0$ (termed the ``boundary'' equation). The former gives $
\rho_{l}=c_{1}A_{-}^{l}+c_{2}A_{+}^{l} \label{eq:rho_bulk_sol_1}$, where $c_{1,2}$ are constants, and we have 
$A_{\pm}=1+a/2~\left(1\pm\sqrt{1+(4/a)}\right)$, with $a \equiv r(1-\rho_0)$. Using the symmetry $\rho_{l}=\rho_{L-l}$
gives $c_{2}=c_{1}A_{+}^{-L}$. The particle conservation condition $N=\overline{\rho}L=\rho_{0}+\sum_{l=1}^{L-1}\rho_{l}$ 
is then used to determine $c_{1}$, yielding~\cite{Miron:2022} 
\begin{equation}
\rho_{l}=\frac{\left(1-A_{-}\right)\left(L\overline{\rho}+a/r-1\right)}{2\left(A_{-}-A_{-}^{L}\right)}\left(A_{-}^{l}+A_{+}^{l-L}\right);~~l \ne 0.\label{eq:rho_bulk_sol_2}
\end{equation}
To determine $\rho_{0}$, consider Eq.~(\ref{eq:density eqn}) for $l=0$ and in the stationary state, when this equation reads as
\begin{equation}
\rho_{1}=\rho_{0}-\frac{r}{2}\left(1-\rho_{0}\right)\left(N-\rho _0\right),\label{eq:rho_1_eqn}
\end{equation}
where $N-\rho_0=\sum_{m=1}^{L-1}\rho_{m}$ and the $l\to L-l$ symmetry has been used to get $\rho_{1}+\rho_{L-1}\to 2\rho_{1}$. To proceed, consider two different physical scenarios: the case of a constant particle density $\bar{\rho}$ and the case of a constant particle number $N$.

\begin{figure}
\begin{centering}
\includegraphics[scale=0.6]{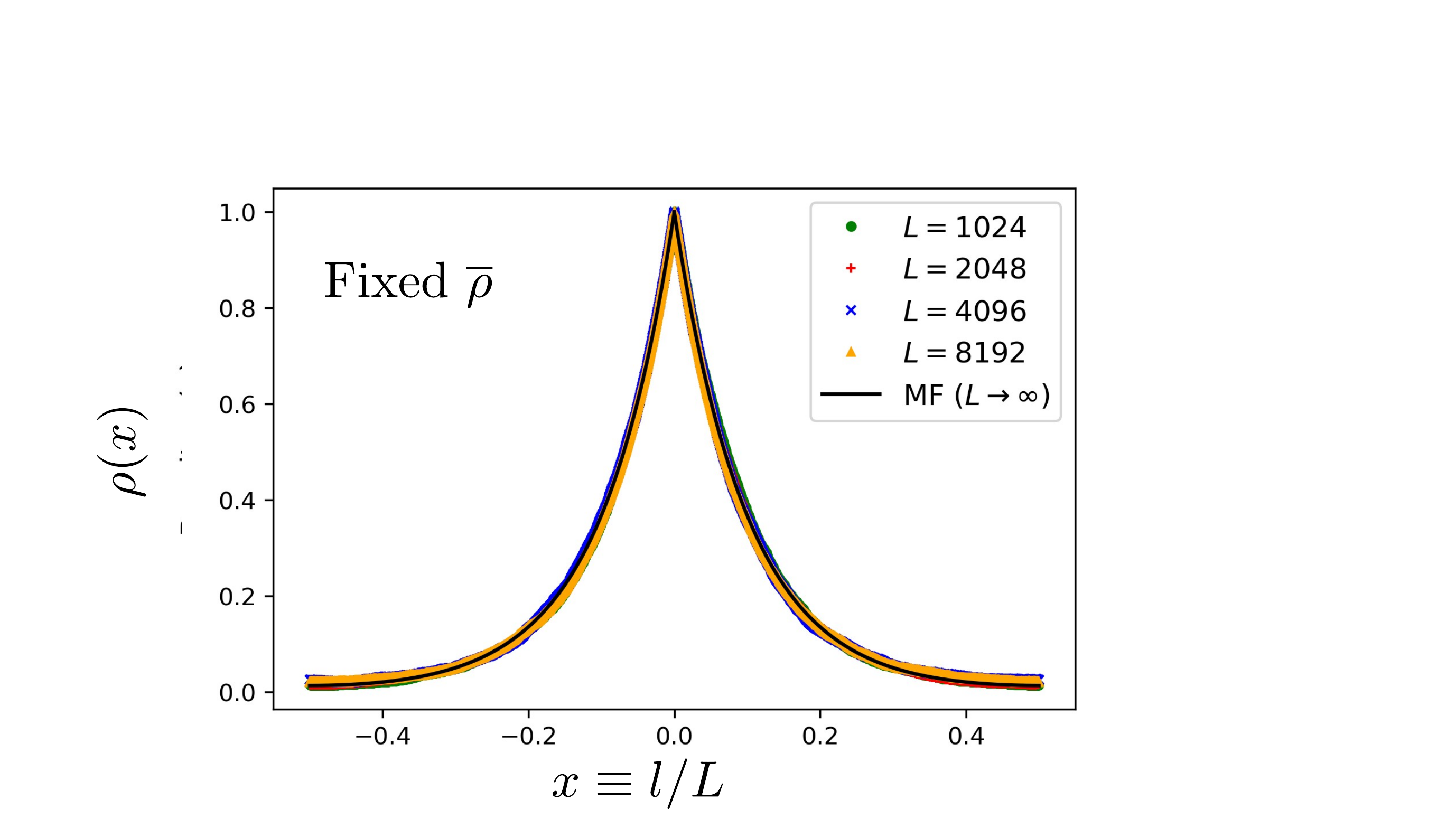}
\par\end{centering}
\caption{For the problem of symmetric simple exclusion process subject to local resetting studied in Ref.~\cite{Miron:2022} and discussed in Section~\ref{subsec:ssep-local}, the figure shows for $r=1$ and $\overline{\rho}=0.2$ the data collapse of the density profile 
versus the scaling variable $x \equiv l/L$. The solid black
curve denotes the mean-field (MF) theoretical prediction in Eq.~(\ref{eq:rho_bulk_sol_3}) for $L\to\infty$. Here, the site indices $l$ have been shifted to $l=-L/2+1,...,0,...,L/2$. Correspondingly, we have $x\in \left[-0.5,0.5\right)$. The figure is adapted from Ref.~\cite{Miron:2022}. \label{Miron-Reuveni-fig1}}
\end{figure}

\textit{Fixed density $\overline{\rho}$:} Here, the number
of particles, $N$, increases linearly with system-size $L$.
Then, for large $L$, only a small time gap exists between the time instant the site $l=0$ is vacated by a particle hopping to a vacant neighbouring site, and the time it is reoccupied due to a resetting event~\cite{Miron:2022}. In the limit of $L\to \infty$, the reoccupation becomes immediate, so that the density near the origin becomes unity. Considering the ansatz~\cite{Miron:2022} 
\begin{equation}
\begin{cases}
\rho_{0}\cong 1-(\alpha/L)^{\mu},\\
\rho_{1}\cong 1-(\beta/L)^{\nu},
\end{cases}\label{eq:ansatz}
\end{equation} 
and substituting into Eq.~(\ref{eq:rho_1_eqn}) yields $\mu=1+\nu$. One then determines $\nu$ by substituting Eq.~(\ref{eq:ansatz}) into Eq.~(\ref{eq:rho_bulk_sol_2}) for $\rho_l$ evaluated at site $l=1$, yielding  $\nu=1$ and hence, one gets $\mu=2$. With this, one obtains for $l\ne 0$ that~\cite{Miron:2022}
\begin{equation}
\rho_{l}\cong 2^{-1}\alpha\overline{\rho}\left(e^{\alpha}-1\right)^{-1}\left(e^{\alpha\left(L-l\right)/L}+e^{\alpha l/L}\right).
\label{eq:rho_bulk_sol_3}
\end{equation}
One then obtains the parameter $\alpha$ by requiring that the ansatz for $\rho_0$ in Eq. ~(\ref{eq:ansatz}) is consistent with the density profile $\rho_l$ given above at the site $l=0$. This gives~\cite{Miron:2022} 
\begin{equation}
\overline{\rho}\alpha\coth\left[\alpha/2\right]=2,\label{eq:alpha_equation}
\end{equation}
which is a transcendental equation for $\alpha$ that must be numerically solved for a given value of $\overline{\rho}$. Finally, one obtains~\cite{Miron:2022} 
\begin{equation}
\rho_{l}\cong \cosh{\left[\frac{\alpha}{2}\left(\frac{L-2 l}{L}\right)\right]}/\cosh{\left[\frac{\alpha}{2}\right]}.\label{eq:rho_bulk_sol_4}
\end{equation}
In the asymptotic limits of $\alpha \to 0$ and $\alpha \to \infty$, analysis yields~\cite{Miron:2022}
\begin{equation}
\overline{\rho}\cong \begin{cases}
1-\alpha^{2}/12; & \alpha\ll1,\\
2/\alpha; & \alpha \gg1.
\end{cases}\label{eq:rho_alpha_relation}
\end{equation}
One thus finds that for a fixed mean density $\overline{\rho}$, $\rho_l$ is a scaling function of $l/L$, at large $L$. This implies that the density profile spans the entire system, as shown in Fig.\ref{Miron-Reuveni-fig1}. 

\begin{figure}[t]
\begin{centering}
\includegraphics[scale=0.6]{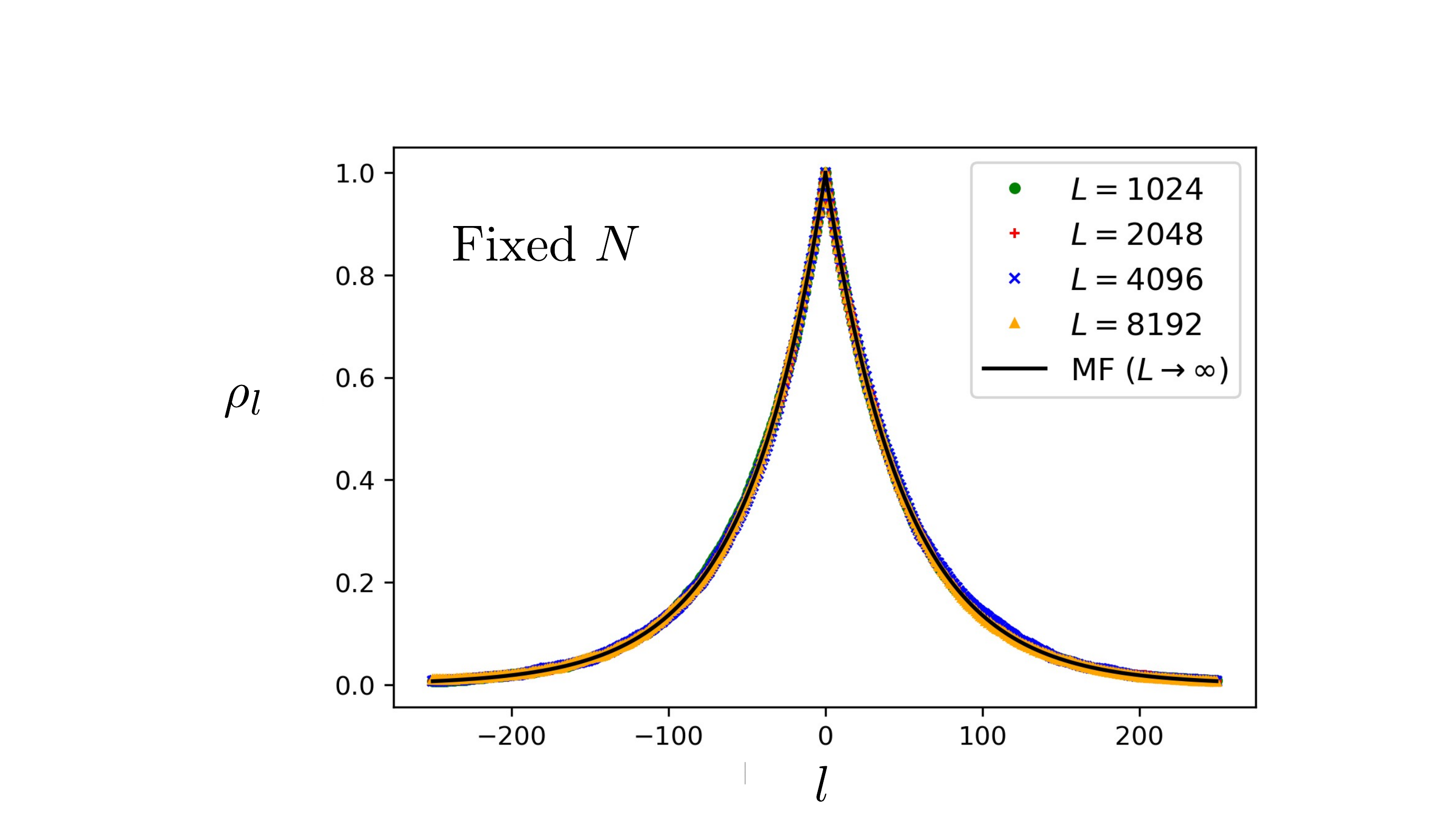}
\par\end{centering}
\caption{For the problem of symmetric simple exclusion process subject to local resetting studied in Ref.~\cite{Miron:2022} and discussed in Section~\ref{subsec:ssep-local}, the figure shows data collapse of the density profile for $r = 1$ and $\overline{\rho}L = N = 100$ plotted as a function of lattice index $l$ for $250$ sites to the left and right of the origin. The different symbols stand for various values of the system size $L$, while the solid black curve denotes the theoretical mean-field (MF) prediction in Eq.~(\ref{eq:rho_bulk_sol_2}) for fixed $N$. Here, the site indices $l$ have been shifted to $l=-L/2+1,...,0,...,L/2$. The figure is adapted from Ref.~\cite{Miron:2022}. \label{Miron-Reuveni-fig2}}
\end{figure}

\textit{Fixed particle number $N$:} Here, we consider the case where the particle number $N$ is fixed, finite, and independent of $L$. Here, the interval between the time instant the site $l=0$ is vacated and the instant it is reoccupied is finite as $L$ increases~\cite{Miron:2022}. Correspondingly, both $\rho_{0}$ and $\rho_{1}$ are asymptotically independent of $L$. We now go back to Eq.~(\ref{eq:rho_bulk_sol_2}) to relate $\rho_{l}$ at $l=0$ to $\rho_{0}$ for finite $N$ and in the limit of large $L$. Substituting $\overline{\rho}=N/L$ and $a=r\left(1-\rho_{0}\right)$ into Eq.~(\ref{eq:rho_bulk_sol_2}), setting $l=0$ and equating to $\rho_{0}$ asymptotically yields the polynomial equation~\cite{Miron:2022} 
\begin{equation}
\frac{r\left(1-\rho_{0}\right)\left(N-\rho_{0}\right)}{\sqrt{r\left(1-\rho_{0}\right)\left(r\left(1-\rho_{0}\right)+4\right)}-r\left(1-\rho_{0}\right)}\cong \rho_{0},\label{eq:rho_0_eqn}
\end{equation}
where one uses the fact that $A_{+}^{-L}$ and $A_{-}^{L}$ both vanish as $L\to \infty$, for any $L$-independent $a$.
This equation for $\rho_{0}$ has three roots, of which only one satisfies $\rho_{0}\in\left[0,1\right]$.  In the limit $1\ll N\ll L$, when the system contains many particles but the average density is low, it reduces to $\rho_{0}= 1-4r^{-1}N^{-2}+\mathcal{O}(N^{-4})$, which implies $a\cong 4/N^2$ that when substituted into Eq. ~(\ref{eq:rho_bulk_sol_2}) shows that the density profile has a width $\sim N/2$ around site $l=0$.  With $N$ being fixed, it then follows that the density profile extends only over a finite $\sim\mathcal{O}(N)$ region near the origin, as demonstrated in Fig.\ref{Miron-Reuveni-fig2}.

We now review a follow-up work, Ref.~\cite{Pelizzola:2021}. Here, the model under question has been studied in a continuum limit, assuming $L \gg 1$ and $N \gg 1 \ge \rho_0$, and introducing a scaled coordinate $x \equiv l/L \in [0,1/2]$ (the $x < 0$ portion of the profile can be obtained by symmetry; note that here, the authors have taken the lattice index to run over the range $l = -L/2 + 1, \cdots 0, \cdots L/2$, with even $L$). Their main result is a closed-form solution for the density profile:
\begin{equation}
  \rho(x) = \rho_0 \cosh \left(L \sqrt{r(1-\rho_0)} x\right) - \frac{L \sqrt{r(1-\rho_0)} N}{2 L} \sinh\left(L \sqrt{r(1-\rho_0)} x\right),
  \label{eq:profile}
\end{equation}
with  
\begin{equation}
  \rho_0 = \frac{N \sqrt{r(1-\rho_0)}}{2} \coth \frac{L \sqrt{r(1-\rho_0)}}{2}.
  \label{eq:rho0}
\end{equation}
Somewhat in analogy with Ref.~\cite{Miron:2022}, the authors considered the case of the limit $L \to \infty$ taken either at a fixed density $\overline{\rho}$ or at a vanishing density ($1 \ll N \ll L$). In the former case, one obtains $\rho_0 = \overline{\rho}L \sqrt{r(1-\rho_0)}/2 \coth(L \sqrt{r(1-\rho_0)}/2)$ and
\begin{equation}
  \rho(x) = \overline{\rho} \frac{L \sqrt{r(1-\rho_0)}/2}{\sinh(L \sqrt{r(1-\rho_0)}/2)} \cosh \left[ L \sqrt{r(1-\rho_0)} \left( \frac{1}{2} - x \right)\right],
  \label{eq:profilefixedrho}
\end{equation}
suggesting that the stationary density profile depends on $r$ and $L$ only through the combination $r L^2$  (except for the prefactor $\overline{\rho}$). Consequently, three different regimes can be found depending on the behavior of $r L^2$ in the thermodynamic limit $L \to \infty$. (i) Small resetting: $r$ tending to zero faster than $L^{-2}$ implies $\rho_0 = \overline{\rho}$ and $\rho(x) = \overline{\rho}$, which is the purely diffusive case. (ii) Large resetting: $r L^2 \to \infty$, which includes the study in Ref.\cite{Miron:2022}; here, $\rho_0$ tends to unity, and the density profile has a maximum at the origin. (iii) Intermediate resetting: $r$ tending to $0$ as $L^{-2}$, where we have $\rho_0 \in (\overline{\rho},1)$ and the profile has a maximum at the origin.
In the vanishing density case, one gets
$\rho_0 = (r N^2/8) ( \sqrt{1 + 16/(r N^2)} - 1 )$, and we have
\begin{equation}
  \rho(y) = \rho_0 \cosh (\mu y) - \frac{\mu}{2} \sinh(\mu y),
\end{equation}
with new scaled variables $\mu \equiv \sqrt{r N^2 (1 - \rho_0)}$ and $y \equiv l/N$. As $L \to \infty$, the behavior of $\rho_0$ is  determined by $r N^2$, and consequently, we find 3 different regimes. (i) $r$ tending to zero faster than $N^{-2}$ implies $\rho_0 \to 0$, the purely diffusive case. (ii) $r N^2 \to \infty$ implies $\rho_0$ tending to unity and the density profile has a maximum at the origin and covers a finite portion of the lattice (consistent with Ref.~\cite{Miron:2022}). (iii) $r N^2$ tending to a positive constant implies $\rho_0 \in (0,1)$, and the profile has again a maximum at the origin and covers a finite portion of the lattice. 

\subsection{Totally asymmetric simple exclusion process with local resetting}
\label{subsec:tasep-local}
In Ref.~\cite{Pelizzola:2021}, the authors have looked at the effect of local resetting on the SSEP as well as the TASEP. The symmetric model has already been discussed in the last portion of the section above. Here, we discuss the work on the totally asymmetric dynamics, wherein the hard-core particles occupying the sites of a one-dimensional periodic lattice of $L$ sites move unidirectionally, hopping only to the nearest-neighbour site to the right with rate unity. The local resetting dynamics takes the particles to the site at the origin: every particle resets at a constant rate $r>0$ (exponential resetting) to the origin, independently of the others. It is more convenient to let the site index $l$ take values from $0$ to $L - 1$. The asymmetry in the dynamics, combined with the resetting dynamics, causes a discontinuous behavior across the resetting site. As the average density $\overline{\rho}$ of particles is varied, the system undergoes phase transitions which can be characterized by the density as well as the current at the sites neighbouring the origin. 

As in the symmetric case, the system can be studied in various regimes, namely, large, intermediate, and small resetting. The intermediate regime is the most interesting one of these and has been studied in detail. Here, the product of resetting rate $r$ and the system size $L$ is taken to be a constant. The mean-field equations for the evolution of the local densities are  
\begin{eqnarray}
&& \hspace{-3em} \dot \rho_0(t) = - \rho_0(t) (1 - \rho_1(t)) + \rho_{L-1}(t) (1 - \rho_0(t)) + r (1 - \rho_0(t)) \sum_{l=1}^{L-1} \rho_l(t), \\
&& \hspace{-3em}  \dot \rho_l(t) = - \rho_l(t) (1 - \rho_{l+1}(t)) + \rho_{l-1}(t)(1 - \rho_l(t)) - r (1 - \rho_0(t)) \rho_l(t)  \nonumber, \\
  && \hspace{-3em} (l = 1, \ldots L-1),  
\end{eqnarray}
where the first equation is for the site at origin, to which the resetting takes place, while the second applies to all the other sites. In the stationary state, the density profile in the continuum approximation (with $x\in [0,1]$) satisfies
\begin{equation}
\frac{\mathrm{d}}{\mathrm{d}x} \left[ \rho(x) \left( 1 - \rho(x) \right) \right] = - r L (1 - \rho_0) \rho(x).
\end{equation}
The above equation can be further simplified by writing it in terms of the function $f(x) \equiv F(\rho(x))$ with $F(\rho) \equiv \rho e^{- 2 \rho}$, as
\begin{equation}
f'(x) = - r L (1 - \rho_0) f(x).
\end{equation}
If the product $rL$ goes to zero in the thermodynamic limit $L \to \infty$, then the effects of resetting become negligible, and one gets a homogeneous density $\rho(x) \equiv \rho_0$ as seen in the bare TASEP with periodic boundaries. 

However, if the quantity $rL(1-\rho_0)$ is finite in the thermodynamic limit, then the equation admits the solution  $f(x) = \mathrm{const} \cdot e^{- r L (1 - \rho_0)  x}$; This solution for $f(x)$ can then be used to determine the density $\rho(x)$ by inverting the relation $f = F(\rho) = \rho e^{-2 \rho}$. The stationary state is characterized in terms of the density values in the immediate vicinity of the origin: $\rho_+ \equiv \lim_{x \to 0^+} \rho(x)$ and $\rho_- \equiv \lim_{x \to 1^-} \rho(x)$. The current out of the origin to the right and into the origin from the left are respectively given by $J_\pm = \rho_\pm (1-\rho_\pm)$, which can then be used to obtain the current into the origin due to resetting as $J_* \equiv J_+ - J_-$, implying that the current has a finite discontinuity at the origin due to the reset dynamics that perpetually introduces particles at the origin. It can be shown that $J_*=\overline{\rho} r L (1 - \rho_0)$ in terms of the average density  
\begin{equation}
\overline{\rho} = \frac
{\left( \rho_- - \frac{1}{2} \right)^2 
-\left( \rho_+ - \frac{1}{2} \right)^2}
{rL (1-\rho_0)}.
\label{eq:averhovsrho0}
\end{equation}

It is important to note that $f=F(\rho)=\rho e^{-2\rho}$ increases for $\rho \in (0,1/2)$ and has a maximum at $\rho = 1/2$. It decreases for $\rho \in (1/2,1)$ and for $f \in (1/e^2,1/(2e))$, its inverse is not single valued. In fact, it can then be written in terms of a Lambert $W$ function as $\rho = - W(- 2 f)/2$. There are two real branches of $W$, named $W_0(z)$ and $W_{-1}(z)$, such that $W_{-1}(z) \le W_0(z)$, and the equality applies at $z = -1/e$. These branches lead to two possible solutions for the density profile. The low density (LD) phase is associated with the density $\rho_{LD}(x) = - W_0(- 2 f(x))/2 \in (0,1/2)$ for $f(x) \in (0, 1/(2 e))$, while the high-density (HD) phase has the density $\rho_{HD}(x) = - W_{-1}(- 2 f(x))/2 \in (1/2,1)$ for $f(x) \in (1/e^2, 1/(2 e))$. 

\begin{figure}
\begin{center}
\includegraphics[scale=0.5]{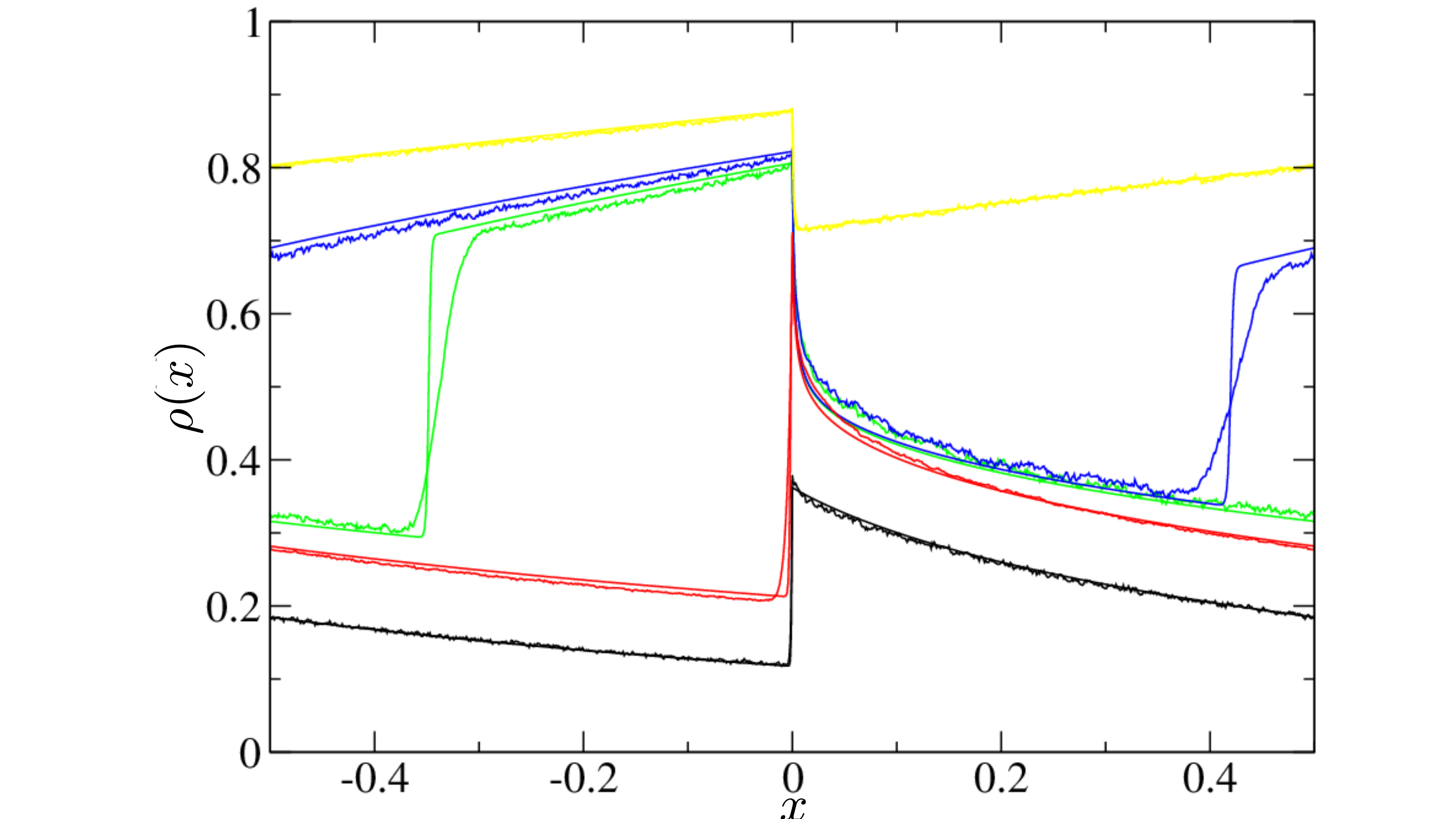}
\caption{For the totally asymmetric simple exclusion process studied in Ref.~\cite{Pelizzola:2021} and discussed in Section~\ref{subsec:tasep-local}, the figure shows the stationary density profiles in the intermediate resetting regime for $L = 10^3$, $r = 10^{-3}$ and several values of $\overline{\rho}$. Smooth lines: mean-field approximation. Noisy lines: data from kinetic Monte Carlo simulations. Here, we have $\overline{\rho} = 0.2$ (black), $0.3$ (red), $0.5$ (green), $0.6$ (blue), $0.8$ (yellow). The $x$ range has been shifted to $(-1/2,1/2)$ for clarity by taking advantage of periodic boundary conditions. The figure is adapted from Ref.~\cite{Pelizzola:2021}.}
\label{fig:TASEP-rL1}
 \end{center}
 \end{figure}

On the basis of the above analysis, the system can be shown to exist in four phases. At low values of the average density, $\overline{\rho} < \overline{\rho}_{c1}$, the system is in a low-density (LD) phase and the densities to the left and right of the origin are given by
\begin{equation}
\rho_+ = \rho_0, \qquad  
\rho_- 
= - {\textstyle \frac{1}{2}} W_0 \left( -2 F(\rho_0) e^{-rL(1-\rho_0)} \right).
\label{eq:rhopmLD}
\end{equation}
A typical density profile is shown in Fig.~\ref{fig:TASEP-rL1} ($\overline{\rho} = 0.2$ (black)). It can be seen that the numerical data agree well with the mean-field predictions.

As the density is increased, the system goes into a maximal current (M) phase, in which $J_*=1/4$ is maximal. This phase exists for average density $\overline{\rho} \in (\overline{\rho}_{c1},\overline{\rho}_{c2})$, and the densities in the vicinity of the origin are given by
\begin{equation}
\rho_+ = {\textstyle \frac{1}{2}} < \rho_0, \ 
\rho_- 
= - {\textstyle \frac{1}{2}} W_0 \left( - 2 F(\rho_+) e^{-rL(1-\rho_0)} \right). 
\label{eq:rhopmMC}
\end{equation}
An example for the density profile can be seen in Fig.~\ref{fig:TASEP-rL1} ($\overline{\rho} = 0.3$ (red)), where we observe that the shape of the profile is similar to the LD case, but the density $\rho_+$ is always half.

When the density is in the region $\overline{\rho} \in (\overline{\rho}_{c2},\overline{\rho}_{c3})$, the system exhibits phase separation into two pure phases: an M phase on the right of the origin, and a high-density (HD) phase (with $\rho'_{HD}(x)>0$) on the left. The density profile is a piecewise combination of an M portion for $x \in (0,x_s)$ and an HD one for $x \in (x_s,1)$, with 2 domain walls: one is at the origin, where the density jumps downward from $\rho_- = \rho_0$ to $\rho_+ = 1/2$ and the other  at $x = x_s$, where the density jumps upward from $\rho_s < 1/2$ to $1 - \rho_s > 1/2$.  The system is said to be in the M-HD phase, and typical density profiles can be seen in Fig.~\ref{fig:TASEP-rL1} ($\overline{\rho} = 0.5$ (green) and $\overline{\rho} = 0.6$ (blue)). The mean-field analysis gives the following piecewise expression for the density
\begin{equation}
\rho(x) 
= 
\begin{cases}
- {\textstyle \frac{1}{2}} W_0 \left( -2 F(\rho_+) e^{- r L (1 - \rho_0) x} \right); 
&  x \in (0,x_s),
\\
- {\textstyle \frac{1}{2}} W_{-1} \left( -2 F(\rho_-) e^{ r L (1 - \rho_0) (1-x)} \right); 
& x \in (x_s,1).
\end{cases}
\end{equation}

On further increase of the average density, $\overline{\rho} > \overline{\rho}_{c3}$, the system enters into a pure high density (HD) phase where the density is
greater than half everywhere and the slope of the density profile is always positive. An example of the density profile in this phase can be seen in Fig.~\ref{fig:TASEP-rL1} ($\overline{\rho} = 0.8$ (yellow)). The mean-field solution gives
\begin{equation}
\rho_- = \rho_0,
\qquad 
\rho_+ 
= - {\textstyle \frac{1}{2}} W_{-1} \left( -2 F(\rho_0) e^{rL(1-\rho_0)} \right).
\end{equation}
The densities $\overline{\rho}_{c1}$, $\overline{\rho}_{c2}$ and $\overline{\rho}_{c3}$ can be expressed in terms of transcendental equations, and can be obtained as a function of the product $rL$. 

The product $rL$ appears naturally in the expressions above and is of importance in determining the phase behavior. The small resetting regime characterized by $rL \to 0$ shows a diffusive behavior, while the intermediate regime, where $rL$ is a constant, shows the rich phase behavior described above. In the large resetting regime, $r L \to \infty$, the density at the origin $\rho_0$ is very close to one, while a pure HD profile is observed at large average density and a pure LD or an 
M profile would be observed only for a very small average density, which tends to zero as $rL \to \infty$.

\section{Resetting in quantum systems}
\label{sec:quantum}
Until now we have discussed resetting in classical systems. We now discuss briefly stochastic resets in quantum systems. This line of research has been initiated in Ref.~\cite{Sengupta}, a work that we now review briefly. 
Consider a quantum system described by a given Hamiltonian $H(t)$. The system is prepared initially at $t=0$ in
the state $|\psi(0)\rangle$. In an infinitesimal $[t,t+\mathrm{d}t]$, the state $|\psi(t)\rangle$ evolves as
\begin{eqnarray}
|\psi(t+\mathrm{d}t)\rangle =
\begin{cases} |\psi(0)\rangle, \quad\quad\quad\quad\quad\quad {\rm
with}\,\, {\rm
prob.}\,\, r\, \rmd t \\
\nonumber\\ [1-i H(t)\, \mathrm{d}t]\, |\psi(t)\rangle \,\,\,\, \mathrm{with}\,\,
\mathrm{prob.}\,\, 1-r\,\mathrm{d}t.
\end{cases}\\
\label{reset_dyn.0}
\end{eqnarray}
Here and in the following, the Planck's constant is set to unity, and the parameter $r\ge 0$ denotes
the constant rate with which the system is reset to the
initial state (exponential resetting). The density matrix of the system at time $t$ is given by ${\hat \rho}(t)= |\psi(t)\rangle \langle \psi(t)|$. We thus observe that starting with the density matrix $\rho_0 \equiv |\psi(0)\rangle \langle \psi(0)|$, the system after evolving for a random time $\tau$ distributed as the exponential~(\ref{eq:ptaur}) resets the time-evolved density matrix $U^\dagger(0,\tau)
\rho_0 U(0,\tau)$ to $\rho_0$. Here, we have the unitary operator $U(t_1, t_2) \equiv T_{t}\, \exp[- \mathrm{i}\int_{t_1}^{t_2} \rmd t'~H(t')]$, and $T_t$ denotes time ordering. The density matrix at time $t$,
obtained by averaging over all possible reset histories, is given on using Eq.~(\ref{eq:renewal-basic0}) as
\begin{eqnarray}
\rho(t) &=& e^{-r t}\, U^\dagger(0,t)\, \rho_0\, U(0,t)+r\int_0^{t} \rmd \tau~e^{-r \tau}\, U^\dagger(0,\tau)\, \rho_0\,  U(0,\tau).
 \label{denmat1}
\end{eqnarray}
The first term on the right hand side represents events for no reset and consequently unitary evolution over the interval $[0,t]$, while the second term stands for the last reset to have happened between times $\tau$ and $\tau + \rmd \tau;~\tau \in [0, t]$ and unitary evolution from the instant of last reset up to time $t$. As $t\to \infty$, one obtains a stationary state:
\begin{eqnarray}
\rho_\mathrm{st}  &=r& \int_0^{\infty} \rmd \tau~e^{-r \tau}\, U^\dagger(0,\tau)
\rho_0 U(0,\tau).\label{denmat2}
\end{eqnarray}

Consider a generic non-integrable system described by a time-independent Hamiltonian $H$  subject to a quantum quench and in absence of any reset events ($r=0$). An arbitrary initial quantum state $|\psi(0)\rangle$ is given
in the energy basis by
\begin{eqnarray}
|\psi(0) \rangle &=& \sum_{\alpha} c_{\alpha} |\alpha\rangle;  \quad
H |\alpha \rangle = \epsilon_{\alpha} |\alpha \rangle,~c_{\alpha} \equiv \langle \alpha |\psi(0)\rangle,  \label{wav1}
\end{eqnarray}
yielding the elements of
the density matrix at any time $t$ as
\begin{eqnarray}
\rho_{\alpha \beta}(t) = \langle \beta|\rho|\alpha\rangle &=& c_{\alpha}^{\ast} c_{\beta} e^{-\mathrm{i}
\omega_{\beta \alpha} t};~\omega_{\beta \alpha} \equiv (\epsilon_{\beta} -
\epsilon_{\alpha}), \label{dm1}
\end{eqnarray}
and consequently, a time average of $\rho_{\alpha \beta}$ as
\begin{eqnarray}
{\overline \rho_{\alpha \beta}}(t) &=& \lim_{T \to \infty} \frac{1}{T} \int_0^T \rmd t~ \rho_{\alpha \beta}(t)= |c_\alpha|^2 \delta_{\alpha \beta}.   \label{sdm1}
\end{eqnarray}
Correspondingly, the system has a stationary state in the limit of long times, which is described by a diagonal density matrix
$\rho_D = \rho_{\alpha \alpha} \delta_{\alpha \beta}$, and any quantum operator $\hat O$ of the system has the stationary
expectation value
\begin{eqnarray}
\langle O\rangle &=& \sum_{\alpha \beta} O_{\alpha \beta}
\rho_{\beta \alpha} = \sum_{\alpha} |c_{\alpha}|^2 O_{\alpha \alpha}
= O_D. \label{opdiag}
\end{eqnarray}

In presence of stochastic resets, using Eqs.~(\ref{dm1})
and~(\ref{denmat2}), one has
\begin{eqnarray}
{\overline \rho_{\alpha \beta}} &=& \int_0^{\infty} \rmd \tau~r e^{- (r+ i \omega_{\beta \alpha}) \tau} c_{\alpha}^{\ast} c_{\beta} \nonumber\\
&=& (\rho_0)_{\alpha \beta} \frac{r}{r+ \mathrm{i} \omega_{\beta \alpha}}  \quad  {\rm for} \, \, \beta \ne \alpha, \nonumber\\
&=& \rho_D  = (\rho_0)_{\alpha \alpha}   \quad  {\rm for}\,\,
\alpha=\beta, \label{dmsto1}
\end{eqnarray}
where we have used $(\rho_0)_{\alpha \beta}= c_{\alpha}^{\ast} c_{\beta}$. Thus, we find
that the reset-averaged stationary-state density matrix is not diagonal in
the energy basis, unlike the situation in absence of resets. We are therefore led to conclude that stochastic resets lead
to novel stationary-state density matrices. Presence of 
off-diagonal terms in the stationary-state density matrix will result in the expectation value of any generic operator to be given by
\begin{eqnarray}
\langle O \rangle &=& O_D + \sum_{\alpha \ne \beta}
c_{\alpha}^{\ast} c_{\beta} O_{\beta \alpha} \frac{r}{r+ \mathrm{i}
\omega_{\beta \alpha} }, \label{opexp}
\end{eqnarray}
implying clear deviation from diagonal
ensemble values. 

As an illustration of application of the aforementioned ideas, consider a fermion chain model consisting of free spinless
fermions with nearest-neighbour hopping described by the Hamiltonian
\begin{eqnarray}
H= -(1/2) \sum_m (c_m^{\dagger} c_{m+1} +{\rm
h.c.}),\label{chainham1}
\end{eqnarray}
where $c_m$ is the annihilation operator for a fermion on the
$m$-th site and the hopping amplitude of fermions is set to
$1/2$. The initial state $|\psi(0)\rangle$ is a step
function: the origin and the sites to the left of the origin are all occupied by a
fermion, while the sites to the right are empty, i.e., $\langle
c_{m}^{\dagger} c_n \rangle= \delta_{mn} \theta(-n)$, where $\theta$
denotes the Heaviside step
function. Hence, initially, the
density per site is on an average equal to $1/2$, and subsequent unitary
evolution will preserve the total number of particles. One can show that under Hamiltonian evolution, the expected density
$n_m(t)= \langle \psi(0)|c_{m}^{\dagger}(t)\, c_m(t)|\psi(0)\rangle$ of the
fermions at site $m>0$ at any time $t$ is
\begin{equation}
n_m(t) =  \sum_{k=m}^{\infty} J_k^2(t),
\label{fchaineq1}
\end{equation}
where $J_m(t)$ is the Bessel function. For $m<0$, the density is simply
\begin{equation}
n_m(t)= 1- n_{1-m}(t).
\label{dens_negative}
\end{equation}
As 
$t\to \infty$, we have $n_m(t)\to 1/2$ for every $m$,
i.e., the density profile is asymptotically flat with value
$1/2$. At any given finite $t$, the density approaches
asymptotically to $1$ as $m\to -\infty$ and approaches zero as $m\to \infty$. However, away from these two
boundaries, for large $t$ and large $|m|$, but with the ratio
$m/t=v$ fixed, $n_m(t)$ converges to a scaling form
\begin{equation}
n_m(t) \to S\left(\frac{m}{t}\right),
\label{dens_profile.1}
\end{equation}
where the scaling function behaves as
\begin{eqnarray}
S(v) & = & \frac{1}{\pi}\, \cos^{-1}(v) \quad {\rm for}\,\, 0<v<1, \nonumber \\
&=& 0 \quad\quad\quad\quad\quad\,\, {\rm for}\, \, v\ge 1,\nonumber \\
&=& 1-S(-v) \quad {\rm for}\,\, v<0.
\label{shape_negative}
\end{eqnarray}

In presence of stochastic resets, the average stationary density profile is obtained from Eq.~(\ref{denmat2}) as
\begin{eqnarray}
n_m(r) & = & \langle m| \rho_\mathrm{st} |m\rangle = r
\int_0^{\infty} \rmd\tau\, e^{-r \tau}\, n_m(\tau), \label{den_reset.1}
\end{eqnarray}
where $n_m(\tau)$ is the average density profile at time $\tau$
without reset, given in Eqs.\ (\ref{fchaineq1})
and (\ref{dens_negative}). We obtain for $m>0$ that 
\begin{eqnarray}
n_m(r)&=\sum_{k=m}^{\infty} C_k(r);\nonumber \\
C_k(r)&\equiv r\, \int_0^\infty \rmd \tau~J_k^2(\tau)\, e^{-r\tau}=\frac{4^m}{\pi} \Gamma^2(k+1/2) r^{-(2k)} \nonumber\\
& ~~~~\times _2 F_{1}(\frac{1}{2}+k, \frac{1}{2}+k, 1+2k;
-\frac{4}{r^2}),
\label{fchaineq1.5}
\end{eqnarray}
while for $m\le 0$, we have
\begin{equation}
n_{1-m}(r)= 1- n_m(r). 
\label{dens_negative_reset}
\end{equation}
Here, $\Gamma$ is the Gamma function, while $_{2}F_1$ is the
regularized Hypergeometric function. We thus observe that as opposed to the flat density profile under bare evolution, a nontrivial profile is induced by the dynamics of stochastic resets. 
In Ref.~\cite{Sengupta}, the authors have also studied a periodically-driven Dirac Hamiltonian in $d$ dimensions, whose unitary evolution is interrupted by a reset after a random integer number of periods. The unitary dynamics of the system is dictated by a periodic drive characterized by a time period $T$. It has been shown that the reset-averaged stationary state of such a driven system is nontrivial and corresponds to the so-called generalized Gibbsian ensemble characterized by the reset rate $r$. 

As a further illustration of effects of resetting in quantum systems, we now consider the example studied in Ref.~\cite{Sevilla:2023}, which involves a two-level atom that interacts with a single mode of a quantized electromagnetic field. The system is described by the paradigmatic Jaynes-Cummings Hamiltonian:
%%%%%%%%
\begin{equation}\label{Hatomo}
\hat H_\mathrm{JC}=\hbar \omega\Big(\hat a^{\dagger}\hat a+\frac{1}{2}\Big)-\frac{1}{2}\hbar \omega \hat \sigma_z-\mathrm{i}~\frac{1}{2}\hbar \Omega\,(\hat \sigma^{+}\hat a-\hat \sigma^{-}\hat a^{\dagger}),
\end{equation}
where $\hat a^{\dagger}$ and $\hat a$ are respectively the creation and the annihilation operator of the electromagnetic field, $\hat \sigma^{+}$ and $\hat \sigma^{-}$ are respectively the raising and the lowering operator of the atom, $\omega$ is the frequency of the electromagnetic mode, which is taken to be the same as the frequency of transition between the excited and the ground state of the atom, and the real parameter $\Omega$ is the coupling constant. In the above equation, the first two terms describe respectively the free Hamiltonian of the field and the atom, whereas the last term describes the excitation (de-excitation) of the atom and the associated absorption (emission) of a photon.

Let us consider an initial state for the system to be
\begin{equation} \label{psit0}
|\psi(0)\rangle =(a|\downarrow\rangle+b|\uparrow\rangle)|0\rangle;~~|a|^2+|b|^2=1,
\end{equation}
where $|\downarrow\rangle$ and $|\uparrow\rangle$ are respectively the ground and the excited state of the atom, which correspond to the eigenvectors of $\hat \sigma_z$ with eigenvalues $+1$ and $-1$, respectively. On the other hand, $|0\rangle$ is the vacuum state of the electromagnetic field, and $|1\rangle$ is the excited state of the field with one photon. In the interaction picture, the initial state evolves in time to yield 
\begin{equation}\label{psievolved}
|\psi(t)\rangle=a|\downarrow\rangle |0\rangle+b\sqrt{1-p(t)}~|\uparrow\rangle|0\rangle+b\sqrt{p(t)}~|\downarrow\rangle|1\rangle,
\end{equation}
where we have $p(t) \equiv \sin^2(\Omega t/2)$.

The reduced state of the two-level atom is given in terms of the reduced density operator
\begin{equation}
\hat\rho^{A}(t) \equiv \Tr_{F}\,|\psi(t)\rangle \langle \psi(t)|,
\end{equation}
where ${\rm Tr}_{F}$ indicates tracing over the degrees of freedom of the field. One obtains 
\begin{eqnarray}\label{rhoatom}
\hat \rho^{A}(t)=\hat \rho^{A}(p(t))&=&\Bigl(|a|^2+p(t)|b|^2\Bigr)|\downarrow\rangle \langle \downarrow|+|b|^2\Bigl(1-p(t)\Bigr)|\uparrow\rangle \langle \uparrow|\nonumber\\
&&+ab^*\sqrt{1-p(t)}|\downarrow\rangle \langle \uparrow|+
a^*b\sqrt{1-p(t)}|\uparrow\rangle \langle \downarrow|.
\end{eqnarray}
Assuming that $a,b\in \mathbb{R}$, and writing the operator (\ref{rhoatom}) in the form 
\begin{equation}
\hat \rho^{A}(p(t))=\frac{1}{2}\bigl[\mathbb{I}+\boldsymbol{r}^A(p(t))\cdot\op{\boldsymbol{\sigma}}\bigr],
\end{equation}
we identify the corresponding Bloch vector as
\begin{equation}\label{rhoAunit}
\boldsymbol{r}^A(p(t))=\Bigl(2ab\sqrt{(1-p)},0,1-2b^2(1-p)\Bigr).
\end{equation}
Under exponential resetting to the state $\hat \rho^{A}(p(0))$, we obtain the reduced density operator of the two-level atom on using Eq.~(\ref{eq:renewal-basic0}) as 
\begin{equation}
\hat \rho^{A}_r(p(t))=e^{-rt}\hat \rho^{A}(p(t))+r\int_0^t \rmd \tau~e^{-r\tau}\hat \rho^{A}(p(t-\tau)),
\end{equation} 
implying the corresponding equation for the Bloch vector in presence of resets:
\begin{equation}
\boldsymbol{r}^A_r(p(t))=e^{-rt}\boldsymbol{r}^A(p(t))+r\int_0^t \rmd \tau~e^{-r\tau}\boldsymbol{r}^A(p(t-\tau)).
\label{blochvectorSR}
\end{equation}

\begin{figure}[!]
\vspace{0.8cm}
\begin{center}
\includegraphics[width=0.4\textwidth]{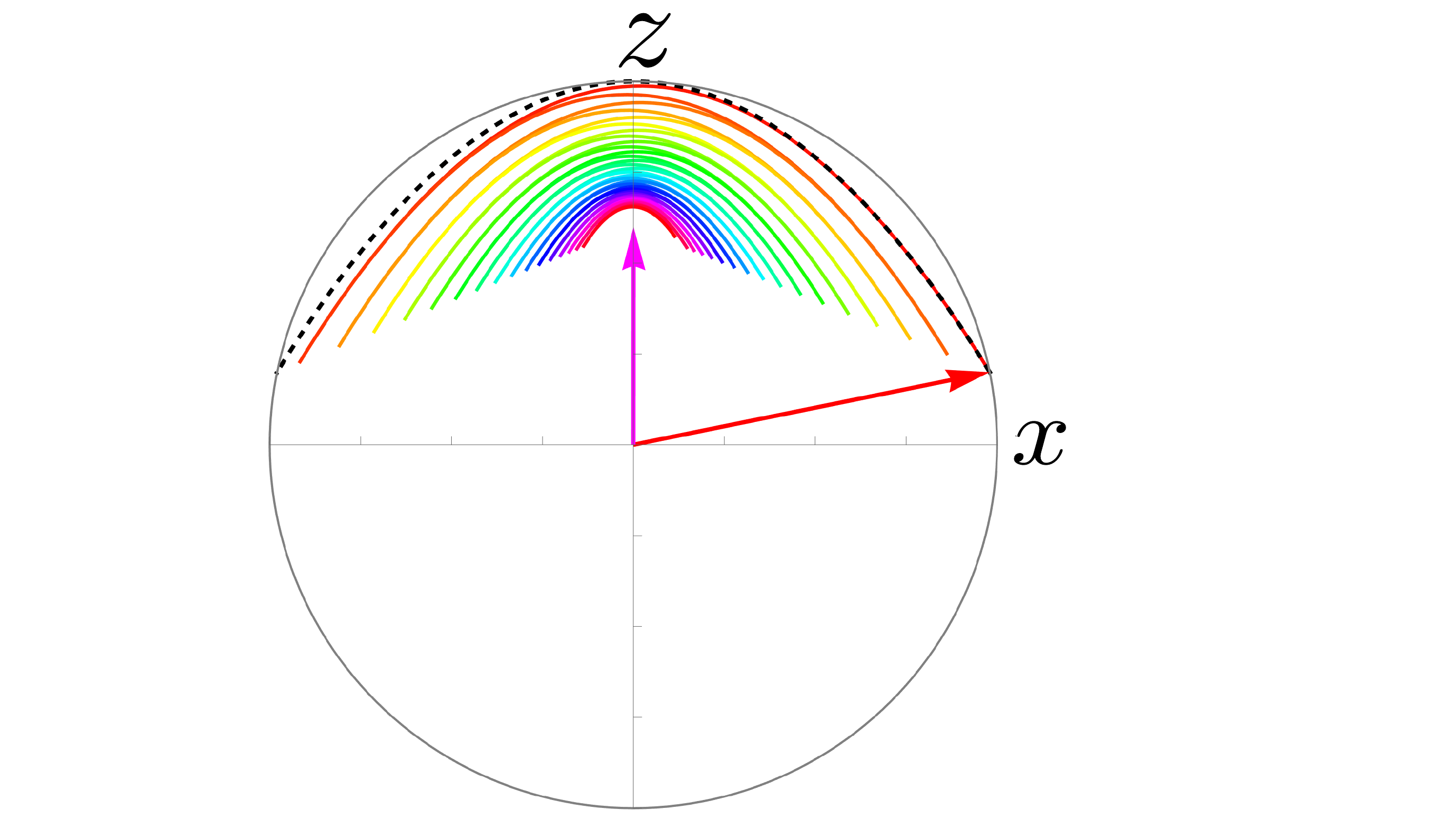}\quad\quad
\includegraphics[width=0.4\textwidth]{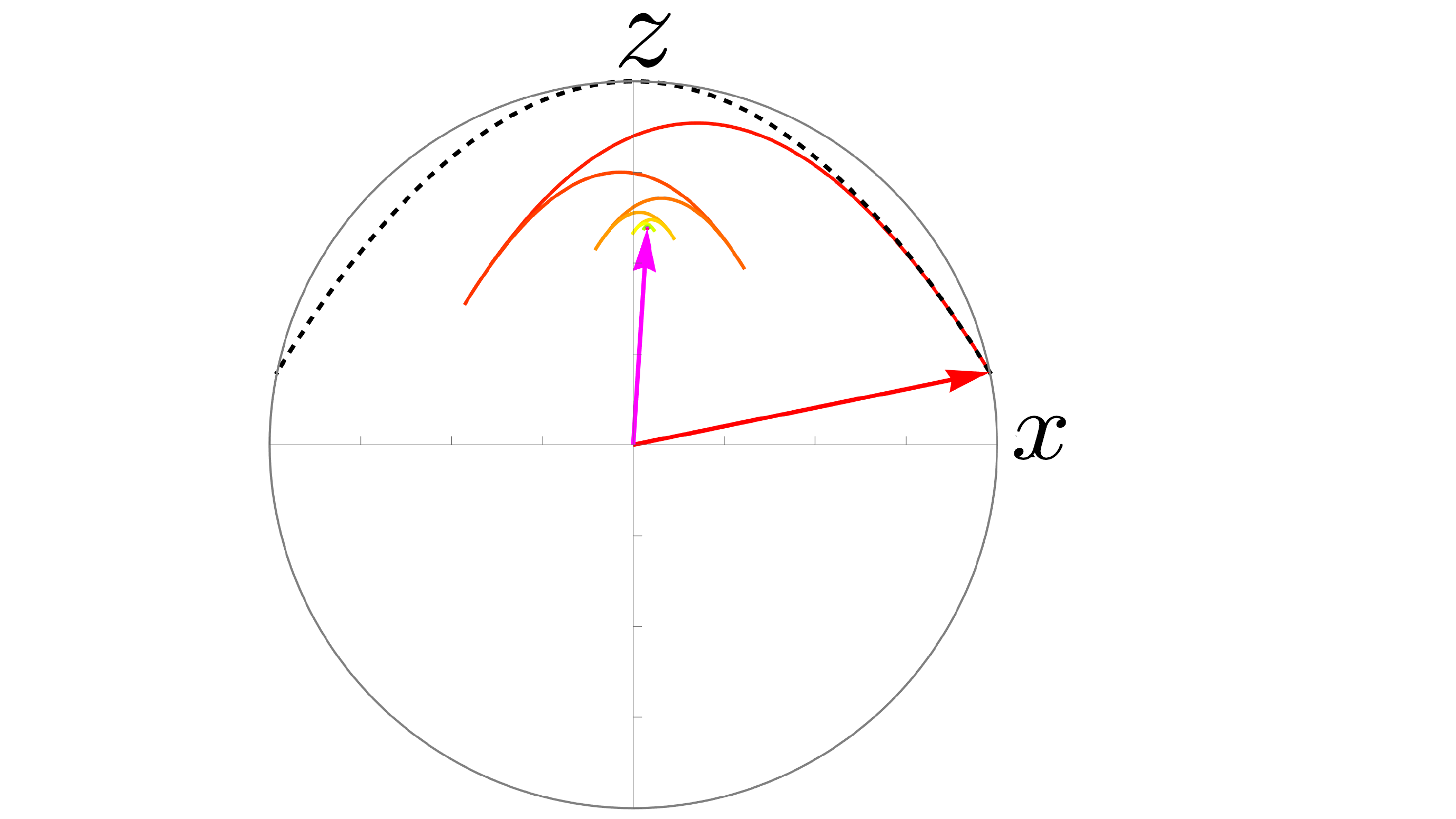}\\
\includegraphics[width=0.4\textwidth]{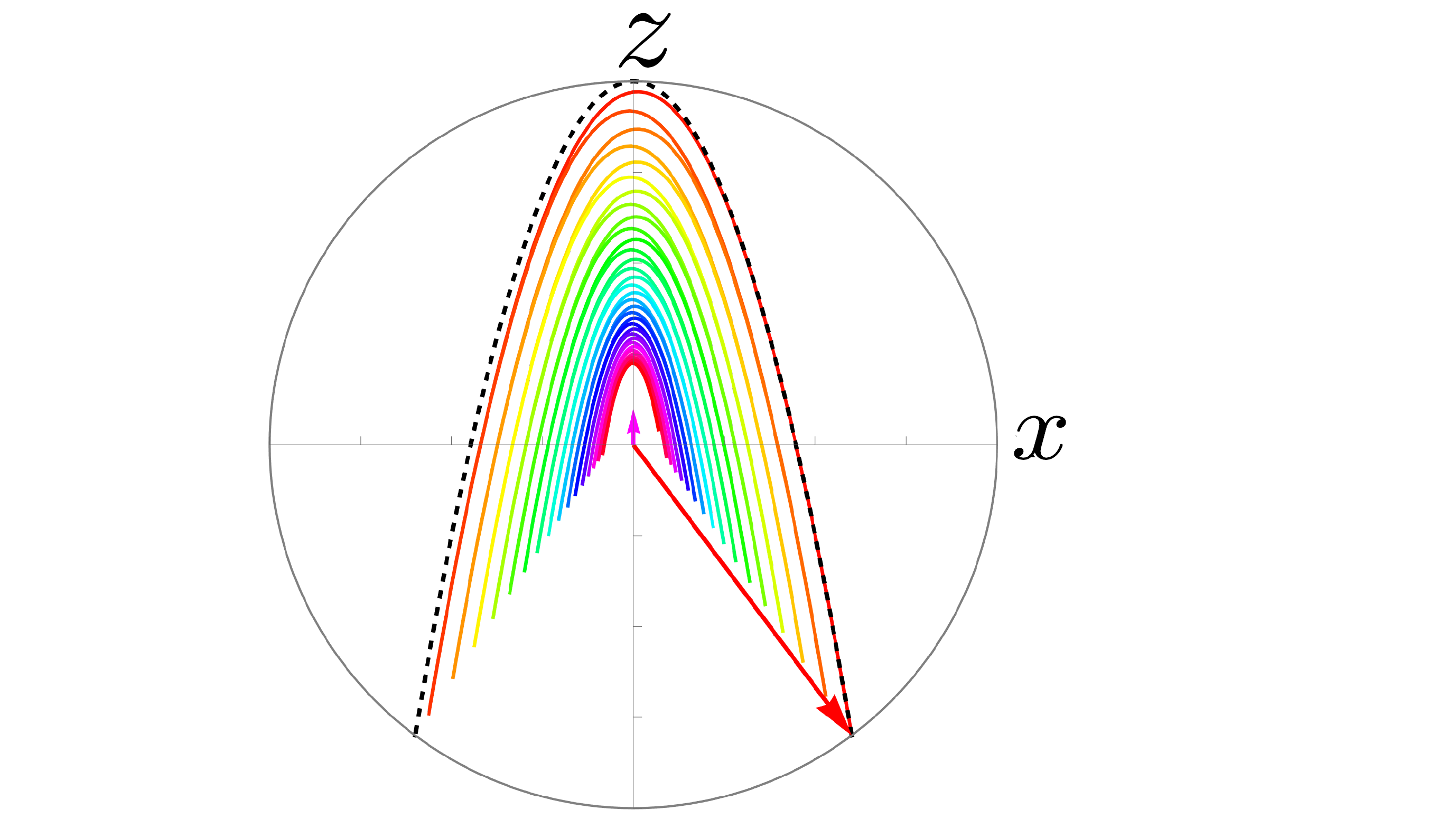}\quad\quad
\includegraphics[width=0.4\textwidth]{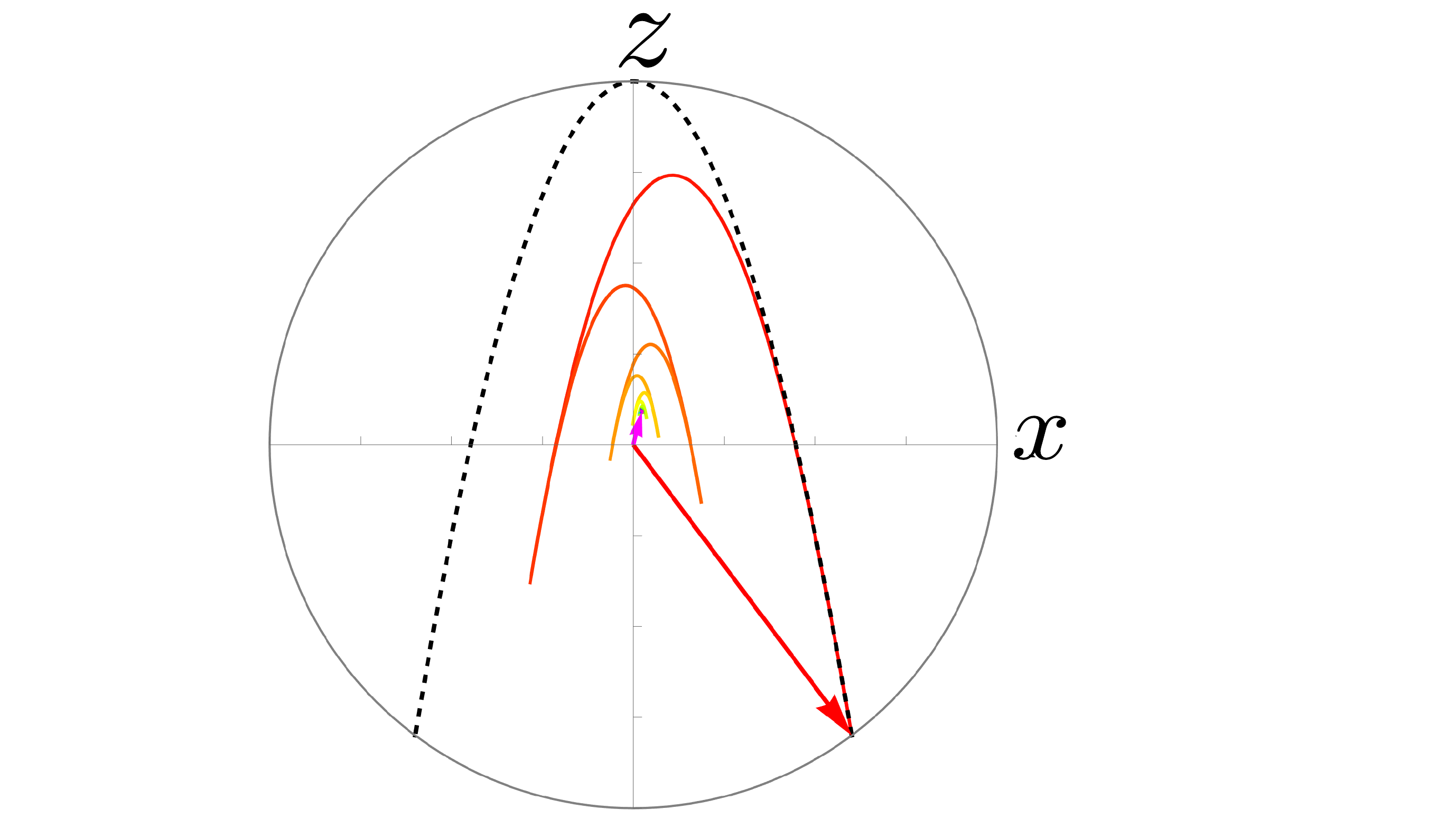}
\caption[]{\label{rhoA}For the problem of stochastic resets in a two-level atom interacting with an electromagnetic field, studied in Ref.~\cite{Sevilla:2023} and discussed in Section~\ref{sec:quantum}, the figure shows the time evolution of the Bloch vector $\boldsymbol{r}^A(p(t))=(2ab\cos{\Omega t/2},0,1-2b^2\cos^2{\Omega t/2})$ for: $a=\sqrt{3/5}$,  $b=\sqrt{2/5}$ (upper panel); $a=\sqrt{1/10}$,  $b=\sqrt{9/10}$ (bottom panel). The resetting corresponds to the exponential distribution~(\ref{eq:ptaur}) with  $r=1/100$ (left panel) and $r=1/10$ (right panel) in units of $\Omega$, and time runs from $t=0$ to $t=200$ (units of $\Omega^{-1}$), in color scale ranging from red to magenta. The initial ($t=0$) and the final ($t\rightarrow\infty$) Bloch vectors are indicated by a red and a magenta arrow, respectively. The black-dashed line depicts the periodic trajectory of the Bloch vector in absence of resetting ($r=0$).
} 
\end{center}
\end{figure}

Figure~\ref{rhoA} shows the time evolution of $\boldsymbol{r}^A_r(p(t))$ for the exponential distribution \eref{eq:ptaur} of resetting and for two different initial states (upper and lower panels) and two values of reset rate $r$ (right and left panels). Equations~(\ref{rhoAunit}) and (\ref{blochvectorSR}) imply that $\boldsymbol{r}^A_r(p(t))$ is restricted to the $x$-$z$ plane. In each plot, we show the time evolution of the tip of the vector $\boldsymbol{r}^A_r(p(t))$, and flow of time is represented by the color scale. The initial ($t=0$) and the final ($t\rightarrow \infty$) Bloch vectors are represented, respectively, by red and magenta arrows. In absence of resetting, when one has a periodic trajectory (traversed back and forth) of the vector $\boldsymbol{r}^A(p(t))$, the corresponding time evolution is depicted with the black-dashed line; this trajectory corresponds to Bloch vectors that vary their magnitude as they evolve, on account of the fact that $\op{\rho}^A(p(t))$ evolves non-unitarily in time. Figure \ref{rhoA} demonstrates that under exponential resetting, the magnitude of the Bloch vector continually reduces in time until a steady-state vector is attained. Comparison of the upper and lower panels shows that the magnitude of  $\boldsymbol{r}^A_r(p(t))$  differs more drastically from that of $\boldsymbol{r}^A(p(0))$ 
as the initial population $b^2$ of the atom's excited state $|\uparrow\rangle$ increases. The value of the resetting rate $r$ dictates the speed of relaxation to the steady state and also the precise nature of the steady state. 

Let us end this section by citing a selection of recent work on stochastic resets in quantum systems. Since this review is primarily on classical systems undergoing stochastic resetting dynamics, our aim here in citing the references is not to be exhaustive, but to only offer a flavour of the kind of work being pursued in the quantum domain. Reference~\cite{Perfetto:2021} studied closed quantum many-body systems subject to stochastic resetting, with a reset reinitializing the dynamics to a state chosen from a set of reset states conditionally on the outcome of a measurement taken immediately before resetting. In the case of the transverse field Ising model, it has been shown that the signatures of its ground-state quantum phase transition may be observed as a sharp crossover in the stationary state of the reset dynamics. 
In Ref.~\cite{Turkeshi:2022}, a phenomenological theory for entanglement dynamics in monitored quantum many-body systems with well-defined quasiparticles (such as in the transverse field Ising model) is put forward, wherein entanglement is carried by ballistically propagating non-Hermitian quasiparticles that are stochastically reset by the measurement protocol, with the rate of resetting given by the finite inverse lifetime of the quasiparticles. A renewal equation for the statistics of the entanglement entropy revealed that different entanglement scaling and even sharp entanglement phase transitions may arise depending on the spectrum of quasiparticle decay rates. 
A recent work~\cite{Debraj1} considered stochastic resets in the framework of the so-called tight-binding model relevant in various contexts in solid-state physics. In one dimension, the corresponding tight-binding chain (TBC) models the motion of a charged particle between the sites of a lattice, wherein the particle is for most times localized on the sites, but which owing to spontaneous quantum fluctuations tunnels between nearest-neighbour sites. In the case in which the density operator is at random times reset to its initial form, the particle is found to be localized on the sites at long times, leading to a time-independent mean-squared displacement of the particle about its initial location. This stands in stark contrast to the situation in absence of resets, wherein the particle does not ever get to a state with a time-independent probability to be found on the different sites. This localization of particle on the sites induced by stochastic resets is observed even when the TBC is subject to an external field that is periodic in time. Here, it is known that one may induce localization in the bare model through tuning the ratio of the strength to the frequency of the field to have a special value, namely, equal to one of the zeros of the zeroth order Bessel function of the first kind. The work~\cite{Debraj2} established that localization may be induced by a far simpler procedure of subjecting the system to stochastic resets.

\section{Conclusion}
\label{sec:conclusions}
In this work, we reviewed recent work on stochastic resetting in systems involving more than one degree of freedom that are interacting with one another in a wide variety of ways. The manner in which the interaction is modelled leaves a strong signature on the emergent long-time static and dynamic behavior, particularly when the bare dynamics of the model is in suitable competition with stochastic resetting. The study of resetting in interacting multi-particle systems not only allows us to model real-world systems, but also paves the way for developing new theoretical techniques beyond those available to study single-particle systems. This work provides a bird's eye view of the rich variety of emergent behavior resulting from stochastic resetting in many-body dynamics (primarily in the classical domain), and we hope that it will serve as an invitation to researchers to unveil many more interesting behavior in the framework of stochastic resetting in many-particle dynamics. As possible trends of future research, we may mention the following: (i) resetting in exclusion processes with either bond-wise or particle-wise disorder~\cite{Barma:2006}, (ii) resetting in exclusion processes with several class of particles~\cite{Blythe}, (iii) resetting in interfaces with radial geometry~\cite{Singha:2005}, (iv)  difference if any in emergent static and dynamical behavior due to resetting in short vs. long-range interacting systems~\cite{Campa}, and (v) resetting in quantum systems exhibiting quantum phase transitions~\cite{Sachdev}.
Note added in proof: While the review was being finalized, many (many) relevant papers of interest that deal with resetting in quantum systems have appeared, which we feel pertinent to point out here, with the purpose of demonstrating the diverse applicability of stochastic resetting dynamics. In pointing out the references, we however would not be able to be exhaustive and would have to be selective, as the literature is literally exploding at the moment with interesting papers appearing almost on a regular basis. Also, resetting in quantum systems has not been the main purpose of this review, and our focus has been primarily on resetting in classical systems involving many interacting particles. In Ref.~\cite{Garrahan1}, spectral properties of classical and quantum Markovian processes were studied, and it was shown that resetting causes a uniform shift in the eigenvalues of the Markov generator, thereby accelerating or even inducing relaxation to a stationary state. Reference~\cite{Garrahan2} considered resetting dynamics in discussing dynamical large deviations of quantum trajectories in Markovian open quantum systems.

A few relevant papers catching attention, which dealt with resetting in classical interacting systems are: Ref.~\cite{Meerson1}, addressing stationary fluctuations in systems involving Brownian particles undergoing stochastic resetting in one dimension, in which the particles reset to the origin either independently or that only the particle farthest from the origin can be reset to the origin; Ref.~\cite{Meerson2}, addressing a simple model of interagent competition by considering particles performing independent Brownian motions on the line, with two particles selected at random and at random times, and the particle closest to the origin being reset to it; Ref.~\cite{Meerson3}, studying a problem of ``Brownian bees," wherein  an ensemble of independent branching Brownian particles is studied, in which, when a particle branches into two particles, the particle farthest from the origin is eliminated so as to keep the number of particles constant; Ref.~\cite{Satya1}, studying a one-dimensional gas of Brownian particles that diffuse independently, but are simultaneously reset to the origin at a constant rate - the system approaches a nonequilibrium stationary state with long-range interactions induced by the simultaneous resetting; Ref.~\cite{Satya2}, addressing the issue of Brownian particles diffusing independently on a line in the presence of an absorbing target at the origin, with the particles undergoing stochastic resetting either independently or all together. A very interesting recent paper studying predator-prey dynamics is Ref.~\cite{RK}, in which the authors study a diffusing lion running after a diffusing lamb when both of them are allowed to get back to their homes (reset) intermittently under resetting at a constant rate. Modelling the dynamics with a pair of vicious random walkers, it was demonstrated that the distribution of the location of annihilation is composed of two parts, namely, one in which the trajectories cross without being reset and the other where trajectories are reset at least once before they cross each other. Moreover, the process of the lion running after the lamb ends in two ways: either the lamb makes it to the safe haven (success) or is captured by the lion (failure), with the conditional distribution in the absence of resetting and for both success and failure possessing a finite mean, but no higher moments exist. Under resetting at a constant rate, the probability of success exhibits a monotonic dependence on the restart parameters, with the distribution of first passage times exhibiting an exponential decay~\cite{RK2}.

We hope that this review would serve the desirable purpose of enticing readers, young and old, to discover by themselves many of the excellent papers that could not be covered in this review. We did not by any means want this review to be the end but rather the beginning of one's exploration into the kingdom of stochastic resetting. 

\section{Acknowledgements}
SG thanks A. Acharya, M. Chase, D. Das, S. Dattagupta, A. Gambassi, L. Giuggioli, S. N. Majumdar, R. Majumder, G. Morigi, \'{E}. Rold\'{a}n, S. Roy, M. Sarkar, G. Schehr, V. Tripathi, and G. Tucci for fruitful collaborations and insightful discussions on the topic of stochastic resetting. SG also thanks ICTP-Abdus Salam International Centre for Theoretical Physics, Trieste, Italy, for support under its Regular Associateship scheme, and acknowledges support from the Science and Engineering Research Board (SERB), India under SERB-MATRICS scheme Grant No. MTR/2019/000560, and SERB-CRG scheme Grant No. CRG/2020/000596.

\end{document}